\documentclass[12pt,a4paper]{book}
\usepackage[french,english]{babel}
\usepackage[latin1]{inputenc}
\usepackage[centertags]{amsmath}
\usepackage{amsfonts}
\usepackage{amssymb}
\usepackage{amsthm}
\usepackage{newlfont}
\usepackage{epigraph}
\usepackage{epsf}
\usepackage{graphicx}
\usepackage{fancyhdr}
\newlength{\defbaselineskip}
\newcommand{\setlinespacing}[1]%
           {\setlength{\baselineskip}{#1 \defbaselineskip}}

\newcommand{\singlespacing}{\setlength{\baselineskip}{\defbaselineskip}}
\newcommand{\onehalfspacing}{\setlength{\baselineskip}{1.5 \defbaselineskip}}
\theoremstyle{plain}
\newtheorem{thm}{Theorem}[chapter]
\newtheorem{cor}[thm]{Corollary}
\newtheorem{lem}[thm]{Lemma}
\newtheorem{prop}[thm]{Proposition}
\theoremstyle{definition}
\newtheorem{defn}[thm]{Definition}
\newtheorem{axiom}{Axiom}
\theoremstyle{remark}
\newtheorem{rem}[thm]{Remark}
\newcounter{axiomcounter}
\numberwithin{axiom}{section} 

\newenvironment{chsum}{\small
\begin{quote}}{\end{quote}}

\font\mybby=eufb8 at 12pt
\def\tb#1{\hbox{\mybby#1}}
\def\M{\tb M}

\def\BA{\mathcal A}
\def\BB{\mathcal B}
\def\CB{\tb B}
\def\BE{\mathcal E}
\def\BF{\mathcal F}
\def\BM{\mathcal M}
\def\BN{\mathcal N}
\def\BH{\mathcal H}
\def\BG{\mathcal G}

\long\def\symbolfootnote[#1]#2{\begingroup%
\def\thefootnote{\fnsymbol{footnote}}\footnote[#1]{#2}\endgroup}

\def\subtitle#1{\gdef\@subtitle{#1}}
\def\@subtitle{\@latex@error{No \noexpand\title given}\@ehc}

\def\maketitle{%
  \begin{titlepage}
    \vfill\vskip 0.2 true in
    \begin{center}
      {\large Th\`{e}se pr\'{e}sent\'{e}e pour obtenir le grade de}\par\vskip 0.16 true in%
      {\large\textsc{Docteur de l'Ecole Polytechnique}}\par\vskip 0.4 true in%
\begin{tabular}{rc}
  Domaine~: & Economie et Sciences sociales \\*[0.1 true in]
  Sp\'{e}cialit\'{e}~: & Sciences cognitives th\'eoriques \\
\end{tabular}
      \vskip 0.3 true in
      {\large par}
      \vskip 0.3 true in
      {\large \bf Alexei Grinbaum}\par\vskip 0.4 true in
      {\LARGE\textbf{LE R\^{O}LE DE L'INFORMATION}}
      \vskip 0.07 true in
      {\LARGE\textbf{DANS LA TH\'{E}ORIE QUANTIQUE}}
      \vskip 0.35 true in %\par\hline\par\\*[0.2 true in]
      {\large\textbf{(THE SIGNIFICANCE OF INFORMATION}}%\par
      \vskip 0.07 true in
      {\large\textbf{IN QUANTUM THEORY)}}%\par
      \vskip 0.7 true in
Soutenue le 4 octobre 2004 devant le jury compos\'{e}
de~:\par\vskip 0.1 true in
\begin{tabular}{lll}
M. Jean-Pierre Dupuy &\textit{directeur de th\`{e}se}&Professeur \`{a} l'\'{E}cole Polytechnique \& CNRS\\
M. Jeffrey Bub & \textit{rapporteur}&Professeur \`{a} l'Universit\'{e} de Maryland\\
M. Carlo Rovelli & \textit{rapporteur}&Professeur \`{a} l'Universit\'{e} de la M\'{e}diterran\'{e}e\\
M. Michel Bitbol & \textit{examinateur}&Directeur de recherche au CNRS\\
M. Jean Petitot &\textit{examinateur}&Directeur d'\'{e}tudes \`{a} l'EHESS \& CREA\\
M. Herv\'{e} Zwirn & \textit{examinateur}&Pr\'{e}sident d'\textit{Eurobios} \& CNRS\\
\end{tabular}
    \end{center}%
    \par
    \vfill\vfill
\newpage\thispagestyle{empty}\null\vskip 3.4in
\begin{center} L'\'{E}cole Polytechnique n'entend donner aucune
approbation, ni improbation aux opinions \'{e}mises dans les
th\`{e}ses. Ces opinions doivent \^{e}tre consid\'{e}r\'{e}es
comme propres \`{a} leur auteur.
\end{center}
\end{titlepage}}

\renewcommand{\headrulewidth}{0.3pt}    % Width of head rule
\makeatletter
\def\cleardoublepage{\clearpage\if@twoside \ifodd\c@page\else%
    \hbox{}%
    \thispagestyle{empty}%              % Empty header styles
    \newpage%
    \if@twocolumn\hbox{}\newpage\fi\fi\fi}
\makeatother

\makeatletter
\let\ps@plain=\ps@empty
\makeatother

\begin{document}

%\title{Le r\^{o}le de l'information dans la th\'{e}orie quantique}
%\title{The significance of information in quantum theory}

%\author{Alexei Grinbaum}
%\address{CREA, Ecole Polytechnique \\
%     1 rue Descartes 75005 Paris, France
%\\ Email grinbaum@poly.polytechnique.fr}

\pagenumbering{roman} \maketitle \setcounter{page}{1}
%\tableofcontents

\selectlanguage{french}
\chapter*{R\'{e}sum\'{e}\markboth{R\'{e}sum\'{e}}{R\'{e}sum\'{e}}}
%\addcontentsline{toc}{chapter}{R\'{e}sum\'{e}}

\noindent Les d\'{e}rivations th\'{e}or\'{e}tico-informationnelles
du formalisme de la th\'{e}orie quantique soul\`{e}vent un
int\'{e}r\^{e}t croissant depuis le d\'{e}but des ann\'{e}es 1990,
gr\^{a}ce \`{a} l'\'{e}mergence de la discipline connue sous le
nom d'information quantique et au retour des questions
\'{e}pist\'{e}mologiques dans les programmes de recherche de
nombreux physiciens-th\'{e}oriciens. Nous proposons une
axiomatique informationnelle dont nous d\'{e}rivons le formalisme
de la th\'{e}orie quantique.

\medskip\noindent
La premi\`{e}re partie de la th\`{e}se est consacr\'{e}e aux
fondements philosophiques de l'ap\-proche informationnelle. Cette
approche s'ins\`{e}re dans un cadre \'{e}pist\'{e}mologique que
nous pr\'{e}\-sen\-tons sous la forme d'une boucle entre
descriptions th\'{e}oriques, ce qui nous permet de proposer une
m\'{e}thode nouvelle d'analyse de la fronti\`{e}re entre toute
th\'{e}orie et sa m\'{e}ta-th\'{e}orie.

\medskip\noindent
La deuxi\`{e}me partie de la th\`{e}se est consacr\'{e}e \`{a} la
d\'{e}rivation du formalisme de la th\'{e}orie quantique. Nous
posons un syst\`{e}me d'axiomes formul\'{e}s dans le langage
informationnel. En conformit\'{e} avec l'argument pour la
s\'{e}paration entre th\'{e}orie et m\'{e}ta-th\'{e}orie, nous
analysons le double r\^{o}le de l'observateur qui est \`{a} la
fois un syst\`{e}me physique et un agent informationnel. Apr\`{e}s
l'introduction des techniques de la logique quantique, les axiomes
re\c{c}oivent un sens math\'{e}matique pr\'{e}cis, ce qui nous
permet d'\'{e}tablir une s\'{e}rie de th\'{e}or\`{e}mes montrant
les \'{e}tapes de la reconstruction du formalisme de la
th\'{e}orie quantique. L'un de ces th\'{e}or\`{e}mes, celui de la
reconstruction de l'espace de Hilbert, constitue un point
important o\`{u} la th\`{e}se innove par rapport aux travaux
existants. Le double r\^{o}le de l'observateur permet de retrouver
la description de la mesure par POVM, un \textit{sine qua non} de
la computation quantique.

\medskip\noindent
Dans la troisi\`{e}me partie de la th\`{e}se, nous introduisons la
th\'{e}orie des $C^*$-alg\`{e}bres et nous proposons de cette
derni\`{e}re une interpr\'{e}tation
th\'{e}or\'{e}tico-informationnelle. L'interpr\'{e}tation
informationnelle permet ensuite d'analyser sur le plan conceptuel
les questions relatives aux automorphismes modulaires et \`{a}
l'hypoth\`{e}se du temps thermodynamique de Connes-Rovelli, ainsi
qu'\`{a} la d\'{e}rivation propos\'{e}e par Clifton, But et
Halvorson.

\medskip\noindent
Nous concluons par une liste de probl\`{e}mes ouverts dans
l'approche informationnelle, y compris ceux relevant des sciences
cognitives, de la th\'{e}orie de la d\'{e}cision et des
technologies de l'information.

\bigskip\noindent \textbf{Mots cl\'{e}s}: th\'{e}orie quantique, information,
boucle des th\'{e}ories, logique quantique, espace de Hilbert,
$C^*$-alg\`{e}bre, automorphismes modulaires, condition KMS, temps

\selectlanguage{english}

\chapter*{Abstract\markboth{Abstract}{Abstract}}
%\addcontentsline{toc}{chapter}{Abstract}

Interest toward information-theoretic derivations of the formalism
of quantum theory has been growing since early 1990s thanks to the
emergence of the field of quantum computation and to the return of
epistemological questions into research programs of many
theoretical physicists. We propose a system of
information-theoretic axioms from which we derive the formalism of
quantum theory.

\medskip\noindent
Part~\ref{part1} is devoted to the conceptual foundations of the
information-theoretic approach. We argue that this approach
belongs to the epistemological framework depicted as a loop of
existences, leading to a novel view on the place of quantum theory
among other theories.

\medskip\noindent
In Part~\ref{part2} we derive the formalism of quantum theory from
information-theoretic axioms. After postulating such axioms, we
analyze the twofold role of the observer as physical system and as
informational agent. Quantum logical techniques are then
introduced, and with their help we prove a series of results
reconstructing the elements of the formalism. One of these
results, a reconstruction theorem giving rise to the Hilbert space
of the theory, marks a highlight of the dissertation. Completing
the reconstruction, the Born rule and unitary time dynamics are
obtained with the help of supplementary assumptions. We show how
the twofold role of the observer leads to a description of
measurement by POVM, an element essential in quantum computation.

\medskip\noindent
In Part~\ref{cstar}, we introduce the formalism of $C^*$-algebras
and give it an infor\-ma\-tion-theoretic interpretation. We then
analyze the conceptual underpinnings of the Tomita theory of
modular automorphisms and of the Connes-Rovelli thermodynamic time
hypothesis. We also discuss the Clifton-Bub-Halvorson derivation
program and give an information-theoretic justification for the
emergence of time in the algebraic approach.

\medskip\noindent
We conclude by giving a list of open questions and research
directions, including topics in cognitive science, decision
theory, and information technology.

\bigskip\bigskip\noindent \textbf{Keywords}: quantum theory, information,
loop of existences, quantum logic, Hilbert space, $C^*$-algebra,
modular automorphisms, KMS condition, time

\chapter*{Acknowledgements\markboth{Acknowledgements}{Acknowledgements}}
%\addcontentsline{toc}{chapter}{Acknowledgements}

The warmest thanks I address to my advisor Jean-Pierre Dupuy. To
him I owe my lifestyle in science and in philosophy.
\bigskip

\noindent I have always felt the unceasing support, both
scientifically and administratively, of Jean Petitot, the director
of CREA. Also at CREA, the discussions with Michel Bitbol have
taught me a great deal.

\bigskip

\noindent I am indebted to Professors~M.V.~Ioffe and V.A.~Franke,
members of the Chair of High Energy Physics at St. Petersburg
State University, for many years of undemanding support.

\bigskip

\noindent I have learned a lot from the valuable discussions with
Carlo Rovelli. His name is quoted often in the dissertation, but
indeed must be quoted on almost every its page.

\bigskip

\noindent Comments made by Jeffrey Bub, Chris Fuchs, Simon
Saunders, and Bas van Fraassen were at the origin of some of the
lines of argument.

\bigskip

\noindent My fellow Ph.D. students at CREA provided numerous
remarks that made me spell the ideas clearer. I thank Stefano
Osnaghi, Patricia Kauark, Adrien Barton, Mathieu Magnaudet,
Alexandre Billon, and Manuel B\"{a}chtold.

\bigskip

\noindent The results of this dissertation were presented at
conferences organized by Andrei Khrennikov at V\"{a}xjo University
and Marisa dalla Chiara in Sardinia under the auspices of the ESF
Network for Philosophical and Foundational Problems of Modern
Physics. I am grateful to the organizers and all the participants
of those conferences who took part in the discussions.

\bigskip

\noindent Financial support and travel funds over the last five
years were provided by the Centre de Recherche en
\'{E}pist\'{e}mologie Appliqu\'{e}e of the Ecole Polytechnique,
the Fondation de l'Ecole Polytechnique, the French Embassy in
Russia, and the Ministry for Education, Research and Technology of
France.

\tableofcontents
%\listoffigures

\chapter*{Notation}
\addcontentsline{toc}{chapter}{Notation}

\begin{tabular}{ll}
$\mathbb{N}\;(\mathbb{N}_0)$ & positive (nonnegative) integers\\
$\mathbb{Z}$ & integers \\
$\mathbb{R}\;(\mathbb{R}_+)$ & (positive) real numbers\\
$\mathbb{C}$ & complex numbers\\
$\mathbb{H}$ & quaternions\\
$\mathbb{D}$ & underlying field of a vector space\\
$S,O$ & physical system (in Part~\ref{part2})\\
$\M$ & fact or measurement result\\
$A,B$ & linear operator\\
$P$ & projection operator\\
$E$ & positive operator (p. \pageref{eoper})\\
$H$ & Hamiltonian \\
$\mathcal{H}$ & Hilbert space (p. \pageref{hilbspdef})\\
$\mathcal{L}$ & lattice (p. \pageref{deflattice})\\
$x,y,z$ & lattice element\\
$Q_i$ & yes-no questions\\
$W(P)$ & set of questions \\
%$\Iota$ & relevance function\\
$\CB(\BH)$ & algebra of all bounded linear operators on $\BH$\\
$\BA,\BB$ & $C^*$-algebra (p. \pageref{algdef}) \\
$\BM,\BN$ & von Neumann algebra (p. \pageref{algdef}) \\
$\omega,\rho,\sigma$ & state over an algebra (p. \pageref{defstate})\\
$\alpha _t ^\omega$ & modular automorphism (p. \pageref{eq8})
\end{tabular}

% Line spacing in the main text
%\doublespacing
\onehalfspacing

%\pagenumbering{arabic} \setcounter{page}{1}

\selectlanguage{french}

\chapter*{Note de pr\'{e}sentation synth\'{e}tique\markboth{La note de pr\'{e}sentation
synth\'{e}tique}{Note de pr\'{e}sentation synth\'{e}tique}}
\addcontentsline{toc}{chapter}{Note de pr\'{e}sentation
synth\'{e}tique}

\section*{R\'{e}sum\'{e} des resultats et plan de la th\`{e}se}
\addcontentsline{toc}{section}{R\'{e}sum\'{e} des resultats et
plan de la th\`{e}se}

Cette th\`{e}se appartient au domaine des Fondements de la
physique. Cela signifie que nous mettons ensemble une analyse des
concepts qui se trouvent \`{a} la base de diff\'{e}rentes
th\'{e}ories physiques avec des r\'{e}sultats formels rigoureux
qui permettent d'\'{e}viter toute ambigu\"{i}t\'{e} dans les
conclusions. Le r\^{o}le de la preuve math\'{e}matique dans la
justification des r\'{e}sultats est d\'{e}cisive.

Cette th\`{e}se mobilise \'{e}galement d'autres disciplines. Dans
la partie~\ref{cstar}, la t\^{a}che principale consiste \`{a}
donner une \textit{interpr\'{e}tation} et, par cons\'{e}quent, le
domaine con\-cern\'{e} est celui de la philosophie de la physique.
Dans le chapitre~\ref{philodiss}, les questions soulev\'{e}es sont
de caract\`{e}re g\'{e}n\'{e}ral plut\^{o}t que sp\'{e}cialis\'{e}
au cas de la physique, comme dans le reste du texte~; ainsi, le
domaine concern\'{e} est celui de la philosophie des sciences ou
de l'\'{e}pist\'{e}mologie. Dans la Conclusion, qui pr\'{e}sente
les probl\`{e}mes ouverts et les th\`{e}mes appartenant \`{a}
d'autres axes de recherche, nous parlons des disciplines telles
que les sciences cognitives et la th\'{e}orie de la d\'{e}cision.

Le but de cette th\`{e}se est de d\'{e}velopper une d\'{e}rivation
coh\'{e}rente de l'ensemble du formalisme de la th\'{e}orie
quantique \`{a} partir des principes
th\'{e}or\'{e}tico-informationnels. Au cours de la d\'{e}rivation,
nous \'{e}tudions les diverses questions conceptuelles et
techniques qui se posent. La r\'{e}ussite du programme de
d\'{e}rivation dans la partie~\ref{part2} permet d'avancer la
th\`{e}se suivante~: \begin{quote} \textbf{La th\'{e}orie
quantique est une th\'{e}orie g\'{e}n\'{e}rale de l'information,
dont la g\'{e}n\'{e}ralit\'{e} est toutefois restreinte par
quelques importantes contraintes
th\'{e}o\-r\'{e}\-ti\-co-infor\-ma\-tion\-nelles. Elle peut
\^{e}tre formellement d\'{e}\-riv\'{e}e d'une axiomatique
informationnelle qui correspond \`{a} ces contraintes.}\end{quote}

Il y a trois mani\`{e}res dont nous innovons en mati\`{e}re des
fondements de la phy\-sique~:
\begin{itemize} \item
Nous d\'{e}rivons le formalisme quantique \`{a} partir des axiomes
th\'{e}or\'{e}\-tico-infor\-ma\-tion\-nels de fa\c{c}on nouvelle.
\item Nous donnons une formulation de l'attitude
\'{e}pist\'{e}mologique pr\'{e}sent\'{e}e sous forme de boucle et
nous montrons \'{e}galement son utilit\'{e} pour l'analyse des
th\'{e}ories autres que les th\'{e}ories physiques. \item Nous
donnons une interpretation th\'{e}or\'{e}tico-informationnelle de
l'approche des $C^*$-alg\`{e}bres et de la th\'{e}orie des
automorphismes modulaires Tomita.
\end{itemize}

Le premier de ces r\'{e}sultats est le plus important. Il est
commun de consid\'{e}rer, m\^{e}me aujourd'hui, que la th\'{e}orie
quantique est une th\'{e}orie du micromonde, ou des objets
r\'{e}els tels que les particules et les champs, ou d'une autre
entit\'{e} fondamentale qui ait n\'{e}cessairement un statut
ontologique. La d\'{e}rivation th\'{e}or\'{e}tico-informationnelle
du formalisme quantique donne \`{a} ces questions une clart\'{e}
longtemps d\'{e}sir\'{e}e~: toutes les pr\'{e}suppositions
ontologiques sont \'{e}trang\`{e}res \`{a} la th\'{e}orie
quantique, qui est, en soi, une pure \'{e}pist\'{e}mologie. La
th\'{e}orie quantique comme th\'{e}orie de l'information doit
\^{e}tre d\'{e}barrass\'{e}e des pr\'{e}suppos\'{e}s
r\'{e}alistes, qui ne doivent leur existence qu'aux
pr\'{e}jug\'{e}s et croyances individuelles des physiciens, sans
appartenir de quelque fa\c{c}on que ce soit \`{a} la th\'{e}orie
quantique propre. Ce qui appartient \`{a} la th\'{e}orie
quantique, c'est exclusivement ce dont on a besoin pour sa
d\'{e}rivation, c'est-\`{a}-dire pour sa reconstruction dans le
contexte de l'approche th\'{e}or\'{e}tico-informationnelle. Au
cours d'une telle d\'{e}rivation nous montrons, pour la
premi\`{e}re fois dans la litt\'{e}rature, comment \`{a} partir
des axiomes informationnels on peut reconstruire l'espace de
Hilbert --- un \'{e}l\'{e}ment essentiel de la th\'{e}orie
quantique. Nous utilisons ensuite des th\'{e}or\`{e}mes
math\'{e}matiques puissants afin de reconstruire le reste du
formalisme.

Pour s\'{e}parer la th\'{e}orie quantique de l'ontologie
superficielle par le moyen de la d\'{e}rivation
th\'{e}o\-r\'{e}\-tico-information\-nelle, on doit la d\'{e}river
\`{a} partir de postulats dont la philosophie sous-jacente soit
d\'{e}nou\'{e}e d'engagements de caract\`{e}re ontologique. Cela
marque le deuxi\`{e}me point d'innovation de la th\`{e}se. Non
seulement on expose la philosophie de la physique sans se
r\'{e}f\'{e}rer \`{a} l'ontologie, mais on montre \'{e}galement
comment cette philosophie peut \^{e}tre li\'{e}e de mani\`{e}re
coh\'{e}rente au programme de d\'{e}rivation formul\'{e}
\textit{dans le langage math\'{e}matique}.

Pour passer au troisi\`{e}me point d'innovation de la th\`{e}se,
nous changeons d'attitude, passant de celle d'un scientifique qui
d\'{e}montre les th\'{e}or\`{e}mes \`{a} celle d'un philosophe de
la physique. La t\^{a}che est double~: nous donnons une
interpr\'{e}tation th\'{e}or\'{e}tico-informationnelle du
formalisme alg\'{e}brique en th\'{e}orie quantique et nous
\'{e}tudions les pr\'{e}suppos\'{e}s conceptuels de la th\'{e}orie
de Tomita et de l'hypoth\`{e}se du temps modulaire de
Connes-Rovelli. Nous continuons \`{a} suivre l'approche
informationnelle, et c'est l'interpr\'{e}tation
th\'{e}or\'{e}tico-informationnelle du formalisme des
$C^*$-alg\`{e}bres qui est innovatrice par rapport aux travaux
existants.

La th\`{e}se est compos\'{e}e de trois parties. Dans
partie~\ref{part1}, apr\`{e}s quelques remarques de caract\`{e}re
g\'{e}n\'{e}ral, le chapitre~\ref{philodiss} s'ouvre par une
section dans laquelle nous expliquons pourquoi, apr\`{e}s
plusieurs d\'{e}cennies d'oubli, s'est r\'{e}veill\'{e}
l'int\'{e}r\^{e}t des physiciens pour la philosophie. Nous passons
ensuite \`{a} la section centrale du chapitre o\`{u} nous
introduisons le concept de la boucle entre les th\'{e}ories. Dans
la derni\`{e}re section, nous montrons en quoi consiste la
r\'{e}ponse que l'on donne du point de vue ici choisi \`{a} la
question que pose tout philosophe de la physique \`{a} toute
approche dite nouvelle~: Comment est-ce que cela r\'{e}sout le
probl\`{e}me de la mesure~? La r\'{e}ponse est que notre approche
ne r\'{e}sout pas, mais dissout le probl\`{e}me.

Dans le chapitre~\ref{qcompch}, nous introduisons les notions de
la computation quantique. Elles ne seront pas directement
utilis\'{e}es dans la th\`{e}se, mais elles servent \`{a} motiver
l'int\'{e}r\^{e}t croissant pour la notion d'information. Ce
chapitre peut \^{e}tre omis par le lecteur int\'{e}ress\'{e}
exclusivement au d\'{e}veloppement de la ligne d'argumentation
principale.

La partie~\ref{part2} est consacr\'{e}e \`{a} la d\'{e}rivation
th\'{e}or\'{e}tico-informationnelle du formalisme de la
th\'{e}orie quantique. Cette d\'{e}rivation est expos\'{e}e en
trois chapitres.

Le chapitre~\ref{sect3} est d\'{e}di\'{e} aux fondements
conceptuels de l'approche
th\'{e}or\'{e}\-tico-infor\-mation\-nelle. Il s'ouvre par une
section historique o\`{u} on pr\'{e}sente un r\'{e}sum\'{e} des
tentatives d'axiomatisation en m\'{e}canique quantique. Puis le
chapitre se poursuit avec une section sur la \og M\'{e}canique
quantique relationnelle \fg~de Rovelli, qui justifie l'intuition
que nous utiliserons pour le choix des axiomes. Les
sections~\ref{defmeasurement} et~\ref{axioms12} sont au c\oe ur de
l'approche th\'{e}or\'{e}tico-informationnelle en ce que nous y
posons, respectivement, les notions fondamentales de la
th\'{e}orie et les axiomes th\'{e}or\'{e}tico-informationnels
formul\'{e}s en termes de ces notions fondamentales. Le chapitre
se conclut avec une section importante sur le double r\^{o}le de
l'observateur, qui est \`{a} la fois un syst\`{e}me physique et un
agent informationnel.

Le chapitre~\ref{chapeql} est consacr\'{e} au formalisme de la
logique quantique qui sera utilis\'{e} dans la suite. Certains
r\'{e}sultats de ce chapitre nous appartiennent, mais la plupart
sont dus \`{a} d'autres chercheurs. La derni\`{e}re section du
chapitre traite de la question cruciale~: comment caract\'{e}riser
un treillis pour que l'espace dont ce treillis est le treillis de
sous-espaces clos soit un espace de Hilbert~?

C'est dans le chapitre~\ref{chaptreconstr} que nous pr\'{e}sentons
les r\'{e}sultats les plus importants du programme de
d\'{e}rivation. Le chapitre s'ouvre par une section dans laquelle
nous nous demandons quels \'{e}l\'{e}ments du formalisme de la
th\'{e}orie quantique il faut reconstruire \`{a} partir des
axiomes th\'{e}or\'{e}tico-informationnels. La section suivante
expose l'id\'{e}e de preuve due \`{a} Rovelli. Toutefois, la vraie
preuve est d\'{e}velopp\'{e}e ind\'{e}pendamment dans la
Section~\ref{rovellirigorous} qui est le point central de la
th\`{e}se. Dans cette section, partant de l'axiomatique
th\'{e}or\'{e}tico-informationnelle, nous d\'{e}montrons le
Th\'{e}or\`{e}me~\ref{constrhilb} qui assure que l'espace de la
th\'{e}orie est un espace de Hilbert. Dans les sections qui
suivent, on traite les probl\`{e}mes du caract\`{e}re quantique de
l'espace de Hilbert~; du corps sous-jacent \`{a} l'espace de
Hilbert et du th\'{e}or\`{e}me de Sol\`{e}r~; de la reconstruction
de la r\`{e}gle de Born par le moyen du th\'{e}or\`{e}me de
Gleason justifi\'{e} par les arguments
th\'{e}or\'{e}tico-informationnels~; et de la dynamique temporelle
unitaire, d\'{e}riv\'{e}e d'un ensemble minimal de
pr\'{e}suppos\'{e}s \`{a} l'aide des th\'{e}or\`{e}mes de Wigner
et de Stone.

La partie~\ref{cstar} de la th\`{e}se est consacr\'{e}e aux
fondements conceptuels de l'approche des $C^*$-alg\`{e}bres. Elle
contient deux chapitres.

Le chapitre~\ref{cstarf} pr\'{e}sente le formalisme des
$C^*$-alg\`{e}bres. Sa premi\`{e}re section est d\'{e}di\'{e}e aux
\'{e}l\'{e}ments de base de cette approche, tandis que dans les
deux sections suivantes on traite de la th\'{e}orie des
automorphismes modulaires de Tomita, \`{a} laquelle
l'int\'{e}r\^{e}t contemporain est en grande partie d\^{u} aux
travaux d'Alain Connes, et de la condition KMS.

Dans le chapitre~\ref{itvalapp}, nous interpr\'{e}tons les
concepts de base du formalisme pr\'{e}sent\'{e} au chapitre
pr\'{e}c\'{e}dent. Le chapitre s'ouvre par une section
consacr\'{e}e \`{a} la justification
th\'{e}or\'{e}tico-informationnelle des notions premi\`{e}res de
la th\'{e}orie. Nous identifions les pr\'{e}suppos\'{e}s les plus
charg\'{e}s philosophiquement. Cela nous m\`{e}ne \`{a} faire une
parenth\`{e}se dans la section suivante, o\`{u} nous exposons la
d\'{e}rivation de la th\'{e}orie quantique par von Neumann.
Malheureusement, von Neumann s'est tromp\'{e} sur quelques points,
et dans la troisi\`{e}me section nous d\'{e}veloppons une
interpr\'{e}tation conceptuelle de l'approche moderne bas\'{e}e
sur la th\'{e}orie des alg\`{e}bres locales. Le retour au
programme de justification th\'{e}or\'{e}tico-informationnelle
sugg\`{e}re, dans la section suivante, la n\'{e}cessit\'{e}
d'analyser la seule d\'{e}rivation
th\'{e}or\'{e}tico-informationnelle de la th\'{e}orie quantique
alg\'{e}brique qui existe, \`{a} savoir celle de Clifton, Bub et
Halvorson. Nous montrons les points forts de leur d\'{e}rivation,
mais aussi ses faiblesses, qui engendrent des id\'{e}es \`{a}
propos de l'espace, le temps et la localit\'{e} qui ne sont pas
motiv\'{e}es du point de vue th\'{e}or\'{e}tico-informationnel.
Enfin, le chapitre se conclut avec une section sur le r\^{o}le du
temps dans laquelle nous analysons le probl\`{e}me de
justification th\'{e}or\'{e}tico-informationnelle du temps.

La th\`{e}se se cl\^{o}t par la Conclusion o\`{u} nous
pr\'{e}sentons les questions ouvertes et d'autres axes de
recherche concern\'{e}s par les id\'{e}es expos\'{e}es dans la
th\`{e}se, \`{a} savoir ceux des sciences cognitives et de la
th\'{e}orie de la d\'{e}cision. Dans la derni\`{e}re section, nous
sugg\'{e}rons l'hypoth\`{e}se qu'avec le d\'{e}veloppement des
technologies de l'information, le langage de l'information
deviendra non seulement \textit{le} langage de la physique, comme
nous l'argumentons dans la th\`{e}se, mais aussi celui d'autres
disciplines scientifiques.

\section*{Partie~\ref{part1}}\addcontentsline{toc}{section}{Partie~\ref{part1}}

Le premier et crucial pr\'{e}suppos\'{e} philosophique fait dans
la th\`{e}se est que le monde peut \^{e}tre d\'{e}crit comme une
\og boucle des existences \fg~(~Wheeler~). Cette expression est
d\'{e}nu\'{e}e de tout engagement ontologique : l'accent est
plac\'{e} sur le mot \og d\'{e}crit \fg~et non pas sur \og monde
\fg. Par cons\'{e}quent, notre programme est celui de
l'\'{e}pist\'{e}mologie : nous \'{e}tudions la mise en jeu des
descriptions sans se prononcer sur la r\'{e}alit\'{e} de l'objet
d\'{e}crit, une telle r\'{e}alit\'{e} pouvant exister ou ne pas
exister. Quelle que soit la r\'{e}ponse, la question n'est pas
pertinente. Afin d'\^{e}tre pr\'{e}cis et d'\'{e}viter les termes
dont la signification est vide, comme \og monde \fg~ou \og
existences \fg, nous posons que la boucle d\'{e}crit non pas les
existences comme \'{e}l\'{e}ments de la r\'{e}alit\'{e} externe,
mais les descriptions, c'est-\`{a}-dire les diff\'{e}rentes
th\'{e}ories. Ainsi, le premier pr\'{e}suppos\'{e} devient :
\textit{L'ensemble de toutes les th\'{e}ories est d\'{e}crit sous
forme cyclique comme une boucle.}

Le deuxi\`{e}me pr\'{e}suppos\'{e} philosophique consiste \`{a}
dire que chaque description th\'{e}o\-rique particuli\`{e}re peut
\^{e}tre obtenue \`{a} partir de la boucle par une op\'{e}ration
consistant en sa \textit{coupure}. Toute coupure s\'{e}pare
l'objet de la th\'{e}orie des pr\'{e}suppos\'{e}s de la m\^{e}me
th\'{e}orie. Il est impossible de donner une description
th\'{e}orique de la boucle tout enti\`{e}re, sans la couper. Une
fois la coupure donn\'{e}e, certains \'{e}l\'{e}ments de la boucle
deviennent l'objet d'\'{e}tude de la th\'{e}orie, d'autres restent
dans la m\'{e}ta-th\'{e}orie de cette th\'{e}orie. En changeant
l'endroit o\`{u} est effectu\'{e}e la coupure, il est possible
d'\'{e}changer les r\^{o}les de ces \'{e}l\'{e}ments : ceux qui
\'{e}taient \textit{explanans} deviennent \textit{explanandum} et
l'inverse. Il est important de noter que la coupure a \'{e}t\'{e}
fix\'{e}e, c'est une \textit{erreur logique} de se poser des
questions qui n'ont un sens que par rapport \`{a} une autre
coupure de la boucle. Le probl\`{e}me de la mesure se dissout
ainsi comme une simple erreur logique, puisqu'il est d\'{e}nu\'{e}
de sens dans l'approche th\'{e}or\'{e}tico-informationnelle.

Les deux pr\'{e}suppos\'{e}s que nous avons faits forment un
argument transcendantal, c'est-\`{a}-dire un argument \`{a} propos
des \textit{conditions de possibilit\'{e}}. Dans notre cas, il
s'agit de la possibilit\'{e} de th\'{e}orisation. Il n'est
possible de construire une th\'{e}orie que si la boucle a
\'{e}t\'{e} coup\'{e}e. L'absence de la coupure m\`{e}ne au cercle
vicieux et \`{a} l'inconsistance logique. La th\'{e}orie ne se
rend possible que par la mise en \'{e}vidence de ses propres
limites. La possibilit\'{e} de th\'{e}orisation est
conditionn\'{e}e par la coupure de la boucle.

La physique et l'information se trouvent dans la boucle en deux
points dia\-m\'{e}\-trale\-ment oppos\'{e}s. Il s'agit pour nous
de couper la boucle de telle sorte que l'information soit \`{a} la
base da la th\'{e}orie physique particuli\`{e}re que nous
consid\'{e}rons, \`{a} savoir la th\'{e}orie quantique.

\section*{Partie~\ref{part2}}\addcontentsline{toc}{section}{Partie~\ref{part2}}

Dans la partie~\ref{part2}, nous focalisons l'attention sur la
coupure de la boucle qui fonde la th\'{e}orie physique sur
l'information. On introduit trois notions fondamentales qui ne
peuvent pas \^{e}tre d\'{e}finies dans le cadre de la th\'{e}orie
s\'{e}lectionn\'{e}e : \textit{syst\`{e}me, information} et
\textit{fait}. La signification de ces notions n'est pas
donn\'{e}e par la th\'{e}orie quantique, et par cons\'{e}quent il
faut les consid\'{e}rer comme des notions m\'{e}ta-th\'{e}oriques.

La coupure de von Neumann entre l'observateur et le syst\`{e}me
\'{e}tant mise au niveau z\'{e}ro, tout peut \^{e}tre vu comme un
syst\`{e}me physique. La premi\`{e}re notion fondamentale, celle
de syst\`{e}me, est ainsi universelle. La deuxi\`{e}me notion
fondamentale, celle d'information, ne pr\'{e}suppose pas encore
l'un des sens math\'{e}matiques pr\'{e}cis de ce terme : les
significations math\'{e}matiques n'apparaissent qu'\`{a}
l'\'{e}tape o\`{u} les notions fondamentales seront traduites dans
les termes math\'{e}matiques de l'un des formalismes de la
th\'{e}orie quantique. Les faits se pr\'{e}\-sentent en tant
qu'actes d'engendrement de l'information ou l'information
index\'{e}e par le moment temporel o\`{u} elle a \'{e}t\'{e}
engendr\'{e}e. La nature de la temporalit\'{e} qui entre en jeu
sera \'{e}tudi\'{e}e dans la Section~\ref{nfrole}. Dans une
th\'{e}orie physique, les faits sont habituellement introduits
sous nom de r\'{e}sultats de la mesure. La question de la
repr\'{e}sentation math\'{e}matique de ces notions devient ainsi
la question de ce qu'est la mesure. Nous y r\'{e}pondons selon les
lignes du formalisme de la logique quantique. La mesure
\'{e}l\'{e}mentaire est d\'{e}finie par une question binaire,
c'est-\`{a}-dire une question qui n'admet que deux r\'{e}ponses~:
oui ou non.

Il convient maintenant de poser deux axiomes informationnels sur
lesquels sera bas\'{e}e la reconstruction du formalisme de la
th\'{e}orie quantique. Axiome~\ref{ax1}~: Il existe une
quantit\'{e} maximale de l'information pertinente qui peut
\^{e}tre extraite d'un syst\`{e}me. Axiome~\ref{ax2}~: Il est
toujours possible d'acqu\'{e}rir une information nouvelle \`{a}
propos d'un syst\`{e}me. Contrairement aux apparences, il n'y a
pas de contradiction entre les axiomes, en vertu de l'utilisation
du terme \og pertinente \fg. Le premier axiome parle non pas d'une
information quelconque, mais de l'information pertinente, tandis
que le deuxi\`{e}me axiome \'{e}nonce qu'une information nouvelle
peut toujours \^{e}tre engendr\'{e}e, m\^{e}me s'il faut pour cela
rendre une autre information, pr\'{e}c\'{e}demment disponible,
non-pertinente. La notion d'information pertinente est li\'{e}e
aux \textit{faits}, et du fait du caract\`{e}re
m\'{e}ta-th\'{e}orique de la notion fondamentale de fait, on
s'attend naturellement \`{a} ce que la notion de pertinence ne
puisse pas \'{e}merger de l'int\'{e}rieur de la th\'{e}orie, mais
qu'elle n\'{e}cessitera une d\'{e}finition externe. Ce sera le cas
dans notre approche.

Chaque syst\`{e}me \'{e}tant trait\'{e} comme syst\`{e}me
physique, mais aussi, potentiellement, comme observateur qui
obtient l'information, il est urgent de distinguer ces deux
r\^{o}les. En effet, dans chaque syst\`{e}me, nous distinguons le
P-observateur, qui est ce syst\`{e}me vu comme un syst\`{e}me
physique, et l'I-observateur, qui est l'agent informationnel.
L'I-observateur est m\'{e}ta-th\'{e}orique par rapport \`{a} la
th\'{e}orie quantique dans l'approche
th\'{e}or\'{e}tico-informationnelle. La possibilit\'{e},
donn\'{e}e par le formalisme, d'\'{e}liminer le P-observateur de
la consid\'{e}ration d'une mesure permet d'obtenir la description
de la mesure qui est essentielle pour la computation quantique,
\`{a} savoir celle par une POVM, la mesure \`{a} valeurs dans la
classe des op\'{e}rateurs positifs. Enfin, la distinction entre
P-observateur et I-observateur nous permet de poser le
troisi\`{e}me axiome de l'approche
th\'{e}or\'{e}tico-informationnelle. Si les deux premiers axiomes
t\'{e}moignent de la pr\'{e}sence de la contextualit\'{e}
m\'{e}tath\'{e}orique, le troisi\`{e}me installe la
non-contextualit\'{e} intrath\'{e}orique~: si une information I a
\'{e}t\'{e} engendr\'{e}e, alors cela s'est pass\'{e} sans
l'engendrement de l'information J \`{a} propos du fait
d'engendrement de l'information I. Cet axiome est \'{e}quivalent
\`{a} la demande d'absence de la m\'{e}ta-information.

Nous nous limitons ici \`{a} donner un seul r\'{e}sultat du
chapitre~\ref{chapeql} qui sera utilis\'{e} dans le
th\'{e}or\`{e}me principal de la th\`{e}se. Ce r\'{e}sultat
(Th\'{e}or\`{e}me~\ref{idth}), dû \`{a} Kalmbach, est le suivant~:
Soit $\BH$ un espace vectoriel de dimension infinie sur le corps
$\mathbb{D}=\mathbb{R}$, $\mathbb{C}$ ou $\mathbb{H}$ et soit
$\mathcal{L}$ un treillis complet orthomodulaire de sous-espaces
de $\BH$ qui satisfait aux conditions suivantes : tout sous-espace
de dimension finie de $\BH$ appartient \`{a} $\mathcal{L}$, et
pour tout $U\in\mathcal{L}$ et pour tout sous-espace $V$ de
dimension finie de $\BH$ la somme $U+V$ appartient \`{a}
$\mathcal{L}$. Alors il existe le produit interne $f$ sur $\BH$
tels que $\BH$ avec $f$ est un espace de Hilbert, qui a
$\mathcal{L}$ pour son treillis de sous-ensembles clos. $f$ est
d\'{e}termin\'{e} de fa\c{c}on unique \`{a} une constante positive
r\'{e}elle pr\`{e}s. Un r\'{e}sultat analogue est d\'{e}montr\'{e}
pour les espaces de dimension finie.

Nous proc\'{e}dons maintenant \`{a} la reconstruction de la
th\'{e}orie quantique \`{a} partir des axiomes
th\'{e}or\'{e}tico-informationnels \`{a} l'aide du formalisme de
la logique quantique. Le premier \'{e}l\'{e}ment \`{a}
reconstruire est l'espace de Hilbert de la th\'{e}orie. Cette
reconstruction se fait en sept \'{e}tapes.

\`{A} la premi\`{e}re \'{e}tape, on d\'{e}finit le treillis des
questions binaires qui repr\'{e}sentent la notion fondamentale
d'information. La r\'{e}ponse \`{a} une question binaire
repr\'{e}sente la notion fondamentale de fait. On postule
(Axiomes~\ref{axiii}, \ref{axiv} et \ref{axv}) la structure
requise dans la d\'{e}finition du treillis et, \'{e}galement, que
le treillis est complet. \`{A} la deuxi\`{e}me \'{e}tape, on
d\'{e}finit la compl\'{e}mentation orthogonale dans le treillis et
on d\'{e}montre que cette notion correspond bien \`{a} toutes les
conditions qui s'imposent sur le compl\'{e}ment orthogonal. \`{A}
la troisi\`{e}me \'{e}tape, on utilise la compl\'{e}mentation
orthogonale pour d\'{e}finir la pertinence d'une question par
rapport \`{a} une autre. \`{A} l'aide de l'Axiome~\ref{ax1}, on
prouve un lemme d\'{e}cisif d\'{e}montrant que le treillis ainsi
construit est orthomodulaire.

\`{A} la quatri\`{e}me \'{e}tape, on introduit un espace de Banach
arbitraire dont le treillis de sous-espaces clos est isomorphe au
treillis que nous avons construit. \`{A} la cinqui\`{e}me
\'{e}tape, on \'{e}tudie les propri\'{e}t\'{e}s de cet espace et
on montre, en particulier, que les conditions ci-mentionn\'{e}es
\`{a} propos des sous-espaces de dimension finie sont
valid\'{e}es. \`{A} la sixi\`{e}me \'{e}tape, on introduit
axiomatiquement le type du corps sous-jacent \`{a} l'espace en
question. Enfin, \`{a} la septi\`{e}me \'{e}tape, on prouve que
cet espace est un espace de Hilbert.

\`{A} l'aide de l'Axiome~\ref{ax2}, et en supposant l'absence des
r\`{e}gles de supers\'{e}lection dans l'espace de Hilbert
construit, nous montrons le caract\`{e}re quantique, et non pas
classique, de cet espace. Pour cela, nous prouvons que toute
sous-alg\`{e}bre bool\'{e}enne du treillis orthomodulaire que nous
avons construit est sa sous-alg\`{e}bre propre. Par
cons\'{e}quent, le treillis lui-m\^{e}me est non-distributif.

Nous discutons ensuite d'une alternative \`{a}
l'Axiome~\ref{contaxiom} qui porte sur le type du corps
num\'{e}rique sous-jacent \`{a} l'espace de la th\'{e}orie. Au
lieu de postuler que c'est un corps simple, on pouvait utiliser le
th\'{e}or\`{e}me de Sol\`{e}r qui engendre ce r\'{e}sultat au prix
de pr\'{e}supposer l'existence, dans l'espace de la th\'{e}orie,
d'une s\'{e}quence infinie orthonormale. \`{A} cause de
l'obscurit\'{e} de justification
th\'{e}or\'{e}tico-informationnelle potentielle de l'existence
d'une telle s\'{e}quence, nous choisissons de ne pas suivre la
voie alternative sugg\'{e}r\'{e}e par le th\'{e}or\`{e}me de
Sol\`{e}r.

Une fois que l'espace de Hilbert a \'{e}t\'{e} construit, il est
n\'{e}cessaire de reconstruire les deux autres \'{e}l\'{e}ments du
formalisme de la th\'{e}orie quantique : la r\`{e}gle de Born avec
l'espace des \'{e}tats et la dynamique temporelle unitaire. En
utilisant le th\'{e}or\`{e}me de Gleason, justifi\'{e} par
l'Axiome~\ref{ax3}, on retrouve la r\`{e}gle de Born. Pour obtenir
la dynamique temporelle, on postule que les ensembles de questions
index\'{e}s par la variable du temps sont tous isomorphes. \`{A}
l'aide des th\'{e}or\`{e}mes de Wigner et de Stone, on obtient
ensuite la description hamiltonienne du d\'{e}veloppement du
syst\`{e}me physique dans le temps et l'\'{e}quation de Heisenberg
pour l'op\'{e}rateur de l'\'{e}volution.

Nous concluons la partie~\ref{part2} par une d\'{e}monstration de
la description de la mesure en tant que POVM, gr\^{a}ce \`{a}
notre argument concernant le temps et \`{a} la s\'{e}paration
entre I-observateur et P-observateur.

La liste compl\`{e}te des axiomes qui ont \'{e}t\'{e} utilis\'{e}s
pour la reconstruction du formalisme de la th\'{e}orie quantique
est ainsi comme suit~:
\begin{description}
\item[Axiome~\ref{ax1}.]Il existe une quantit\'{e} maximale de
l'information pertinente qui peut \^{e}tre extraite d'un
syst\`{e}me. \item[Axiom~\ref{ax2}.]Il est toujours possible
d'acqu\'{e}rir une information nouvelle \`{a} propos d'un
syst\`{e}me.\item[Axiome~\ref{ax3}.]Si information I à propos d'un
syst\`{e}me a \'{e}t\'{e} engendr\'{e}e, alors cela s'est
pass\'{e} sans l'engendrement de l'information J \`{a} propos du
fait d'engendrement de l'information
I.\item[Axiome~\ref{axiii}.]Pour toute paire de questions
binaires, il existe une question binaire \`{a} laquelle la
r\'{e}ponse est positive si et seulement si la r\'{e}ponse \`{a}
au moins une des questions initiales est
positive.\item[Axiome~\ref{axiv}.]Pour toute paire de questions
binaires, il existe une question binaire \`{a} laquelle la
r\'{e}ponse est positive si et seulement si la r\'{e}ponse \`{a}
chacune des questions initiales est
positive.\item[Axiome~\ref{axv}.]Le treillis des questions
binaires est complet.\item[Axiome~\ref{contaxiom}.]Le corps
num\'{e}rique sous-jacent à l'espace de la th\'{e}orie est l'un
des corps $\mathbb{R}$, $\mathbb{C}$ ou $\mathbb{H}$ et
l'anti-automorphisme involutif dans ce corps est
continu.\end{description} De ces axiomes on d\'{e}duit que,
premi\`{e}rement, la th\'{e}orie est d\'{e}crite par un espace de
Hilbert qui est de caract\`{e}re quantique; deuxi\`{e}mement, sur
cet espace de Hilbert on construit l'espace des \'{e}tats, puis on
d\'{e}rive la r\`{e}gle de Born et on d\'{e}rive aussi, avec
quelques pr\'{e}suppos\'{e}s suppl\'{e}mentaires, la dynamique
temporelle unitaire sous la forme classique de l'\'{e}volution
hamiltonienne.

\section*{Partie~\ref{cstar}}\addcontentsline{toc}{section}{Partie~\ref{cstar}}

Dans la partie~\ref{part2}, \`{a} l'aide de l'approche de la
logique quantique, nous avons d\'{e}riv\'{e} le formalisme de la
th\'{e}orie quantique. Dans la partie~\ref{cstar}, nous
consid\'{e}rons une approche diff\'{e}rente, celle de la
th\'{e}orie des $C^*$-alg\`{e}bres. Dans ce cadre, le programme de
d\'{e}rivation sera r\'{e}duit au probl\`{e}me de
l'interpr\'{e}tation th\'{e}or\'{e}tico-informationnelle de
l'approche alg\'{e}brique. Une fois ladite interpr\'{e}tation sera
achev\'{e}e, les th\'{e}or\`{e}mes des $C^*$-alg\`{e}bres
permettront de retrouver le formalisme de la th\'{e}orie quantique
sous la forme pr\'{e}cise du formalisme de la th\'{e}orie des
alg\`{e}bres locales.

Le chapitre~\ref{cstarf} est consacr\'{e} \`{a} la
pr\'{e}sentation de quelques \'{e}l\'{e}ments math\'{e}matiques du
formalisme alg\'{e}brique. Nous introduisons les notions de
$C^*$-alg\`{e}bre et d'alg\`{e}bre de von Neumann concr\`{e}tes et
abstraites. Nous d\'{e}finissons ensuite ce qu'est un \'{e}tat sur
une alg\`{e}bre et nous donnons la premi\`{e}re classification des
facteurs de von Neumann.

Dans la section~\ref{modusect}, les concepts de la th\'{e}orie de
Tomita sur les automorphismes modulaires sont introduits, ce qui
m\`{e}ne \`{a} la deuxi\`{e}me classification des facteurs de von
Neumann, due \`{a} Connes, et aux th\'{e}or\`{e}mes montrant
l'unicit\'{e} des alg\`{e}bres hyperfinies de type $II_1$ et
$III_1$.

Dans la Section~\ref{skms}, il s'agit de la th\'{e}orie KMS et du
lien avec la thermodynamique. Le th\'{e}or\`{e}me principal est
celui de Tomita et Takesaki, qui dit que tout \'{e}tat fid\`{e}le
sur une alg\`{e}bre est un \'{e}tat KMS \`{a} la temp\'{e}rature
inverse $\beta =1$, par rapport \`{a} l'automorphisme modulaire
qu'il g\'{e}n\`{e}re lui-m\^{e}me. Ainsi, exactement de la
m\^{e}me fa\c{c}on que dans le cas de la m\'{e}canique classique,
un \'{e}tat \`{a} l'\'{e}quilibre contient toute l'information sur
la dynamique du syst\`{e}me qui peut \^{e}tre d\'{e}finie par
l'hamiltonien, sauf la constante $\beta$. Cela signifie que
l'information sur la dynamique peut \^{e}tre enti\`{e}rement
remplac\'{e}e par l'information sur l'\'{e}tat thermique. Le fait
que $\beta$ soit constante et non-modifiable de l'int\'{e}rieur de
la th\'{e}orie quantique dans l'approche
th\'{e}or\'{e}tico-informationnelle m\`{e}ne \`{a} placer la
thermodynamique, comme une science qui \'{e}tudie les variations
de la temp\'{e}rature et, par cons\'{e}quent, de $\beta$, dans la
coupure de la boucle des th\'{e}ories diff\'{e}rente de celle
o\`{u} se trouve la th\'{e}orie quantique. La thermodynamique
appartient ainsi, dans l'approche
th\'{e}or\'{e}tico-informationnelle, \`{a} la m\'{e}ta-th\'{e}orie
de la th\'{e}orie quantique.

C'est dans le chapitre~\ref{itvalapp} que nous donnons une
interpr\'{e}tation th\'{e}or\'{e}\-tico-infor\-ma\-tion\-nelle de
l'approche alg\'{e}brique. Les notions fondamentales sont
traduites par des notions math\'{e}matiques de $C^*$-alg\`{e}bre
et d'\'{e}tat sur cette alg\`{e}bre. Une alg\`{e}bre correspond
\`{a} un syst\`{e}me, tandis que l'\'{e}tat, en tant que
l'\'{e}tat informationnel, d\'{e}crit l'information \`{a} propos
de ce syst\`{e}me. Cela nous m\`{e}ne \`{a} consid\'{e}rer la
notion de pr\'{e}paration comme catalogue de \textit{toute}
l'information que l'observateur a \`{a} propos d'un syst\`{e}me,
et, \`{a} son tour, l'analyse de la notion de pr\'{e}paration est
intrins\`{e}quement li\'{e}e \`{a} l'id\'{e}e initiale de von
Neumann concernant la m\'{e}thode de d\'{e}rivation du formalisme
de la th\'{e}orie quantique. Von Neumann se pr\'{e}occupait de la
notion d'ensemble \'{e}l\'{e}mentaire non-ordonn\'{e}, qui lui a
servi pour fonder l'Ansatz statistique -- le premier jalon de la
m\'{e}canique quantique. Von Neumann a utilis\'{e} son programme
de d\'{e}rivation, que nous exposons dans la
Section~\ref{vNdisill}, pour argumenter le passage de la
m\'{e}canique quantique bas\'{e}e sur l'espace de Hilbert, \`{a}
la m\'{e}canique quantique bas\'{e}e sur un facteur de type $II$.
Malheureusement, les facteurs de ce type, dans la th\'{e}orie
quantique moderne, se sont r\'{e}v\'{e}l\'{e}s inutiles, et c'est
\`{a} l'interpr\'{e}tation des concepts de cette derni\`{e}re que
nous proc\'{e}dons.

Il s'agit dans la Section~\ref{partchoi} de justifier le choix
particulier qui est fait par la th\'{e}orie des alg\`{e}bres
locales, qui donne la pr\'{e}f\'{e}rence \`{a} l'alg\`{e}bre
hyperfinie de type $III_1$. Toutefois, nous commen\c{c}ons par une
analyse des pr\'{e}suppos\'{e}s cach\'{e}s dans le choix d'une
$C^*$-alg\`{e}bre et d'un \'{e}tat sur elle comme
repr\'{e}sentants des notions de syst\`{e}me et d'information. Le
deuxi\`{e}me choix, celui d'un fonctionnel positif lin\'{e}aire
comme repr\'{e}sentant de la notion d'information, est lourd de
postulats implicites. En effet, toute la d\'{e}rivation \`{a}
l'aide de la logique quantique avait pour but l'obtention de la
structure de l'espace de Hilbert, et ceci au prix d'une seule
d\'{e}finition m\'{e}ta-th\'{e}orique, \`{a} savoir celle de la
notion d'information pertinente. Avec la traduction de la notion
d'information sous forme de la notion d'\'{e}tat, le nombre de
pr\'{e}suppos\'{e}s m\'{e}ta-th\'{e}oriques augmente~: ils sont
deux -- lin\'{e}arit\'{e} et positivit\'{e}, tandis que, dans ce
cadre, pour d\'{e}river l'espace de Hilbert il suffit de se
r\'{e}f\`{e}rer \`{a} la construction GNS sans rentrer dans
l'explicitation des d\'{e}tails comme on l'a fait dans le cas de
la logique quantique.

Une fois que les pr\'{e}suppos\'{e}s cach\'{e}s ont \'{e}t\'{e}
d\'{e}gag\'{e}s, il convient de passer \`{a} l'interpr\'{e}tation
de la th\'{e}orie des alg\`{e}bres locales par les
Axiomes~\ref{ax1} et \ref{ax2}. Il est sugg\'{e}r\'{e} et
argument\'{e} que ces deux axiomes correspondent \`{a} la demande
que l'alg\`{e}bre en question soit hyperfinie. L'argumentation
pr\'{e}cise est donn\'{e} dans le texte de la th\`{e}se.

Ayant donn\'{e} l'interpr\'{e}tation
th\'{e}or\'{e}tico-informationnelle de l'approche alg\'{e}brique
\`{a} l'aide des axiomes pos\'{e}s dans le chapitre~\ref{sect3},
nous nous posons maintenant la m\^{e}me question que dans la
Section~\ref{quantsect}, \`{a} savoir celle du caract\`{e}re
quantique vs. classique de la th\'{e}orie. Il est n\'{e}cessaire
de se restreindre, par le moyen des pr\'{e}suppos\'{e}s
th\'{e}or\'{e}tico-informationnels, au cas quantique. La solution
a \'{e}t\'{e} propos\'{e}e par Clifton, Bub et Halvorson dans un
article o\`{u} ils op\`{e}rent une d\'{e}rivation de la
th\'{e}orie quantique \`{a} partir des th\'{e}or\`{e}mes de la
computation quantique. Les trois th\'{e}or\`{e}mes qu'ils
utilisent sont~: l'absence de transfert supralumineux de
l'information via la mesure (`no superluminal information transfer
via measurement'), l'absence de \og t\'{e}l\'{e}diffusion \fg~des
\'{e}tats (`no broadcasting') et l'impossibilit\'{e} d'engager un
octet de mani\`{e}re d\'{e}cisive dans un processus de
transmission (`no bit commitment'). Nous analysons les d\'{e}tails
de leur d\'{e}rivation et, tout en l'endossant sur le plan formel,
sauf en une seule occasion, nous la critiquons sur le plan
conceptuel, en rapport avec l'utilisation d'un vocabulaire
non-pertinent pour ce qui est de l'approche alg\'{e}brique. Nous
la reformulons ensuite pour donner un crit\`{e}re
th\'{e}or\'{e}tico-informationnel des syst\`{e}mes physiques
\textit{distincts}. \`{A} l'aide de ce crit\`{e}re et en utilisant
les th\'{e}or\`{e}mes d\'{e}montr\'{e}s par Clifton, Bub et
Halvorson, on retrouve le caract\`{e}re quantique de
l'alg\`{e}bre.

L'une des critiques que nous adressons \`{a} Clifton, Bub et
Halvorson consiste \`{a} mettre en question l'utilisation qu'ils
font des concepts d'espace et de temps. Dans l'approche
th\'{e}or\'{e}\-tico-infor\-ma\-tion\-nelle, ces notions
n'appartiennent pas \`{a} l'ensemble des notions fondamentales et
elles doivent, par cons\'{e}quent, \^{e}tre d\'{e}riv\'{e}es des
notions fondamentales et des axiomes. Nous y consacrons la
Section~\ref{nfrole}. En vertu de la th\'{e}orie KMS, chaque
\'{e}tat sur une alg\`{e}bre acquiert son courant modulaire de
Tomita, et c'est ce courant que nous appelons le temps
d\'{e}pendant de l'\'{e}tat. Il faut souligner trois
cons\'{e}quences importantes de la r\'{e}f\'{e}rence \`{a} la
th\'{e}orie KMS pour la d\'{e}finition du temps~:
\begin{itemize}
\item Le temps est un concept qui d\'{e}pend de l'\'{e}tat. Si
l'\'{e}tat ne change pas, le temps ne change pas non plus. Un
changement dans le temps signifie un changement de l'information.
Ce dernier peut \^{e}tre engendr\'{e} dans un nouveau fait. Alors,
\`{a} chaque fait, le temps d\'{e}pendant de l'\'{e}tat \og
red\'{e}marre \fg. On observe que la temporalit\'{e} des faits (la
variable $t$ qui indexe les faits) n'a rien \`{a} voir avec la
notion du temps qui d\'{e}pend de l'\'{e}tat. \item La
thermodynamique ne jou\'{e} pas de r\^{o}le. Pour voir un \'{e}tat
comme un \'{e}tat KMS \`{a} $\beta =1$ et pour d\'{e}finir le
courant temporel, il n'est pas n\'{e}cessaire de dire que
l'\'{e}tat sur une $C^*$-alg\`{e}bre est un concept
thermodynamique. Par cons\'{e}quent, cela permet d'identifier la
thermodynamique comme m\'{e}ta-th\'{e}orie dans l'approche
th\'{e}or\'{e}tico-informationnelle. Pour faire ainsi, il suffit
de consid\'{e}rer le temps modulaire et d'ex\'{e}cuter la rotation
de Wick, en appelant temp\'{e}rature le r\'{e}sultat. Si l'on veut
modifier la temp\'{e}rature ind\'{e}pendamment du temps modulaire,
il est in\'{e}vitable d'introduire un degr\'{e} de libert\'{e}
nouveau par rapport \`{a} la th\'{e}orie quantique dans l'approche
th\'{e}or\'{e}tico-informationnelle. \item Dans le cadre de
l'interpr\'{e}tation th\'{e}or\'{e}tico-informationnelle de la
th\'{e}orie des alg\`{e}bres locales, on justifie le caract\`{e}re
hyperfini de la $C^*$-alg\`{e}bre du syst\`{e}me. Par
cons\'{e}quent, s'il n'y a pas eu d'engendrement de l'information
nouvelle, et si l'alg\`{e}bre est un facteur de von Neumann de
type $III_1$, le spectre du temps varie de $0$ jusqu'\`{a}
$+\infty$. Ce r\'{e}sultat correspond \`{a} notre intuition sur la
fa\c{c}on dont le temps se comporte.
\end{itemize}

Le temps est une notion d\'{e}pendante de l'\'{e}tat, mais l'on
voudrait aussi avoir dans la th\'{e}orie un temps qui ne
d\'{e}pend pas de l'\'{e}tat. Pourquoi~? Parce que nous sommes
habitu\'{e}s au temps lin\'{e}aire newtonien qui ne d\'{e}pend pas
de l'\'{e}tat informationnel. Pour obtenir ce temps
non-d\'{e}pendant de l'\'{e}tat, nous factorisons les
automorphismes modulaires par les automorphismes internes et nous
choisissons toute une classe de ces derniers qui correspond \`{a}
un seul automorphisme externe. En effectuant cette op\'{e}ration,
nous n\'{e}gligeons une certaine information, \`{a} savoir celle
qui distinguait entre eux les automorphismes modulaires, ceux qui
ont tous \'{e}t\'{e} projet\'{e}s sur un seul automorphisme
externe. Ainsi l'\'{e}mergence du temps devient la question du
rejet d'une certaine information comme non-pertinente. Cela
\'{e}voque le mot de Bohr qui disait, \og Les concepts d'espace et
de temps, par leur nature m\^{e}me, n'acqui\`{e}rent un sens que
gr\^{a}ce \`{a} la possibilit\'{e} de n\'{e}gliger les
interactions avec les moyens de la mesure \fg. Nous concluons le
chapitre en d\'{e}montrant comment ces propos de Bohr
acqui\`{e}rent un sens th\'{e}or\'{e}tico-informationnel gr\^{a}ce
\`{a} la division entre I-observateur et P-observateur.

%\section*{Partie~\ref{part4}}\addcontentsline{toc}{section}{Partie~\ref{part4}}

\selectlanguage{english}\cleardoublepage\pagenumbering{arabic}\setcounter{page}{1}

\part{Introduction}\label{part1}

\chapter{General remarks}

\section{Disciplinary identity of the dissertation}

This dissertation belongs to the field of Foundations of Physics.
It means that we aim at combining the analysis of concepts
underlying physical theories with rigorous \textit{formal} results
that allow to avoid ambiguity in conclusions. Role of mathematical
proof in the justification of conclusions is a deciding factor.

This dissertation also reaches out to other disciplines. In
Part~\ref{cstar} our task is to give an interpretation and the
area concerned is closer to the philosophy of physics. In
Chapter~\ref{philodiss} questions that are raised are general
rather than specialized to the case of physics: the area, then, is
the one of the philosophy of science or epistemology. In the
Conclusion, speaking about open topics and the application of the
ideas of the dissertation, we discuss disciplines such as
cognitive science and decision theory.

\section{Goals and results}

The goal of this dissertation is to give a consistent derivation
of the formalism of quantum theory from information-theoretic
principles. We also study a variety of issues that arise in the
process of derivation. Successful accomplishment of the derivation
program in Part~\ref{part2} allows us to advance the following
thesis:
\begin{quote} \textbf{Quantum theory is a general theory of information
constrained by several important information-theoretic principles.
It can be formally derived from the corresponding
information-theoretic axiomatic system.}\end{quote}

In three ways we innovate in the field of the foundations of
physics:
\begin{itemize} \item We derive the quantum formalism
from information-theoretic axioms in a novel way. \item We
formulate an epistemological attitude presented in the form of a
loop and we demonstrate its utility for the analysis of theories
other than physics. \item We give an information-theoretic
interpretation to the $C^*$-algebraic approach, including the
Tomita theory of modular automorphisms and the issue of time
emergence.
\end{itemize}

The first of these three goals remains the most important one. It
is commonplace to think, even nowadays, that quantum theory is a
theory of the microworld, or of real objects like particles and
fields, or of some other ``first matter'' that necessarily has the
ontological status. Information-theoretic derivation of the
quantum formalism installs the long lusted clarity: all the
ontological assumptions are alien to quantum theory which is, in
and of itself, a pure epistemology. Quantum theory as a theory of
information must be cleared from the realist ideas which are
merely brought in by the physicists working in quantum theory,
with all their individual prejudices and personal beliefs, rather
than belong to the quantum theory proper. What belongs to quantum
theory is no more than what is needed for its derivation, i.e. for
a reconstruction of the quantum theoretic formalism. In the
process of such derivation we for the first time demonstrate how,
from information-theoretic axioms, one can reconstruct the Hilbert
space---a crucial element of quantum theory. We then use powerful
mathematical results to reconstruct the remainder of the
formalism.

In order to separate it from the superficial ontology by means of
the informa\-tion-theoretic derivation, quantum theory must be
derived from such postulates of which the underlying philosophy is
devoid of ontological commitments. This is the role of the second
point on which innovates this dissertation. Not only it gives an
exposition of the philosophy of physics that is disconnected from
ontology, but it also shows how such a philosophy can be
consistently linked to the derivation program \textit{formulated
in the mathematical language}.

To move to the third point of innovation, we change the attitude
from the one of the scientist proving theorems to the attitude of
the philosopher of physics. The task is now to give an
information-theoretic \textit{interpretation} of the algebraic
formalism in quantum theory and to study philosophical
underpinnings of the Tomita theory and of the Connes-Rovelli
modular time hypothesis. What links this field to the previous
parts of the dissertation is that we continue to follow the
information-theoretic approach; what innovates with respect to the
currently existing work is that, even if there were a few
specialists in the foundations of physics who worked on the
conceptual basis of the $C^*$-algebraic approach, there is
virtually no published work on the conceptual foundations of the
Tomita theory of modular automorphisms in connection with the KMS
condition and the modular time hypothesis. We bring together
various mathematical results in an attempt to give a
philosophically sound exposition of the key ideas in this field.

\section{Outline}

The remainder of this introduction will be devoted to two needs:
presentation in Chapter~\ref{philodiss} of the philosophy in which
will be rooted the dissertation; and presentation in
Chapter~\ref{qcompch} of the few elements of quantum computation.

Chapter~\ref{philodiss} opens with a section in which we explain
why interest for philosophy has reemerged in the community of
physicists after the many decades of \textit{oubli}. We then move
to the highlight of the chapter, where we introduce the philosophy
of the loop of existences. In the concluding section, we explain
how this point of view responds to the question that any
philosopher of physics immediately asks when he hears of a new
approach: How does that solve the measurement problem? Our answer
is that it does not solve, but rather dissolves the problem.

Chapter~\ref{qcompch} introduces the ideas of quantum computation.
They will not be used in the thesis but serve to motivate the
rising interest toward the notion of information. A reader solely
interested in following the main line of the dissertation can skip
this chapter.

In Part~\ref{part2} we present the information-theoretic
derivation of the formalism of quantum theory. It is exposed in
three chapters.

Chapter~\ref{sect3} is devoted to laying out the conceptual
foundations of the information-theoretic approach. It opens with a
historic section about axiomatization attempts in quantum
mechanics. It then continues with a section on Rovelli's
Relational Quantum Mechanics that justifies the intuition which we
use for selection of information-theoretic axioms.
Sections~\ref{defmeasurement} and \ref{axioms12} form the core of
the information-theoretic approach by postulating, respectively,
the fundamental notions of the theory and information-theoretic
axioms formulated in the language of these fundamental notions.
The chapter then concludes with an important section on the
twofold role of the observer as physical system and as
informational agent.

Chapter~\ref{chapeql} is devoted to exposition of the quantum
logical formalism that will be used in the sequel. A few results
belong to us but most are taken from the literature. The last
section of the chapter treats the crucial question of how to
characterize a lattice so that it will force the space of which
this lattice is the lattice of closed subspaces to be a Hilbert
space.

It is in Chapter~\ref{chaptreconstr} that we present the most
important results of the derivation program. The chapter opens
with a section in which we ask ourselves what are the elements of
the formalism of quantum theory that we have to reconstruct from
information-theoretic axioms. The next section gives a sketch of
Rovelli's idea of derivation. The actual proof, however, is
independently developed in Section~\ref{rovellirigorous} which is
the highlight of the whole dissertation. In this section, based on
the information-theoretic axiomatic system, we prove
Theorem~\ref{constrhilb} which shows that the space of the theory
is a Hilbert space. Consequent sections address the problems of
quantumness versus classicality of the theory; of the field
underlying the Hilbert space and the Sol\`{e}r theorem; of
reconstruction of the Born rule by means of Gleason's theorem
justified information-theoretically; and of the unitary time
dynamics derived from the allegedly minimal set of assumptions
with the help of Wigner's and Stone's theorems.

Part~\ref{cstar} is devoted to the conceptual foundations of the
$C^*$-algebraic approach. It consists of two chapters.

Chapter~\ref{cstarf} presents the $C^*$-algebraic formalism. Its
first section is dedicated to the basic elements of this approach,
while the two subsequent sections treat of the Tomita theory of
modular automorphisms, much of the contemporary interest in which
is due to Alain Connes's work, and of the KMS condition.

In Chapter~\ref{itvalapp} we analyze the concepts underlying the
formalism presented in the previous chapter. The opening section
is devoted to information-theoretic interpretation of the basic
notions of the theory. We uncover the assumptions that have a
maximal philosophical weight. This leads us to a digression in the
next section in which we expose von Neumann's derivation of
quantum theory. Unfortunately, von Neumann was wrong on certain
points, and in the third section we develop a conceptual
interpretation of the modern approach based on the theory of local
algebras. This return to the program of information-theoretic
justification suggests, in the following section, a necessity to
discuss the only available information-theoretic derivation of the
algebraic quantum theory due to Clifton, Bub and Halvorson. We
show the strong points of this derivation but also its weaknesses
that lead to informationally unmotivated assumptions concerning
space, time, and locality. Finally, we conclude with a section on
the role of time where we address the problem of its
information-theoretic justification.

The dissertation ends with the Conclusion in which we address
questions that were left open and apply the ideas of the
dissertation to theories other than physics: cognitive science and
decision theory. The last section advances a hypothesis that, with
the development of information technology, the language of
information will become not only a language of physics, the
possibility of which we demonstrate in the dissertation, but also
of other scientific disciplines.

\chapter{Philosophy of this dissertation}\label{philodiss}

\section{``The Return of the Queen''}

The conceptual revolution brought to science by quantum theory is
now almost a century old. Despite this old age, the theory's full
significance has not yet been appreciated outside a limited circle
of physicists and philosophers of science. Although terms like
``uncertainty principle'' or ``quantum jumps'' have been
incorporated into the everyday, common language, they are often
used to convey ideas which have no relation with the physical
meaning of these terms. One could say that the wider public took
note of the metaphorical powers of the quantum theory, while the
essence of the quantum revolution remains largely unknown, even
more so because of the slow reform of the educational system.

The situation is somewhat different for another great physical
revolution, the one of relativity. Ideas of relativity have much
better penetrated in the mainstream culture. Terms like ``black
holes'' and ``spacetime'' are a familiar occurrence in popular
scientific journals. Such a relative success of relativity
compared to quantum theory may be due to two reasons.

First, quantum theory's rupture with the preceding classical
paradigm, although, as we argue in Section~\ref{rovhilbspace}, due
to a similar shift in understanding, is more radical than the
rupture of relativity with Galilean and Newtonian physics. A
non-scientist can easier understand that at high velocities
unusual effects occur or that black holes absorb matter and light,
than that the very notions of velocity, position, particle or wave
must be questioned. Interpretation of quantum theory has always
been a motive for argument even among professional physicists,
leave alone the general public.

Second, the discussion of foundations of quantum theory always
remained away from practical applications of the theory, and
therefore away from a wider audience fascinated by the
breathtaking technical development. Educational systems nowadays
do little or nothing to explain that computers, mobile phones, and
many other everyday devices work thanks to quantum mechanics, and
even if educational systems did explain this, they would probably
avoid referring explicitly to any particular interpretation of
quantum theory. Working applications and problems of
interpretation have long been isolated from each other.

This situation has evolved in the last ten years with the
appearance of the new field of quantum information. Practical
quantum information applications are perhaps around the corner,
with prototypes of quantum cryptographic devices and the
teleportation of structures as large as atoms already realized in
laboratories~\cite{teleatom2,teleatom1}. These applications, for
the first time in history, illustrate highly counter-intuitive
features of quantum theory at the level of everyday utility. One
sign of the growing importance of quantum information methods and
results is the increasing use of them in introductory courses of
quantum mechanics. In a broader context, we see the public
excitement by research in quantum information, through mass media
and governmental action.

We shall see that applications of quantum theory to quantum
information often suggest what is essential and what is accessory
in quantum theory itself, highlighting features which may be of
practical and theoretical importance. It appears that taking
seriously the role of information in quantum theory might be
unavoidable for the future major developments.

Yet another change in the circumstances occurred due to which the
foundations of quantum mechanics receive now more attention.
Echoing what we said in the discussion of the first reason, this
change has to do with the ongoing effort to unite the quantum
mechanical ideas with the ideas of general theory of relativity.
Unlike the founding fathers of modern physics, most of their
followers of the second half of the XXth century viewed questions
like ``What is space? What is time? What is motion? What is being
somewhere? What is the role of the observer?'' as irrelevant. This
view was appropriate for the problems they were facing: one does
need to worry about first principles in order to solve a problem
in semiconductor physics or to write down the symmetry group of
strong interactions. Physicists, working pragmatically, lost
interest in general issues. They kept developing the theory and
adjusting it for particular tasks that they had to solve; when the
basis of problem-solving is given, there is no need to worry about
foundations. The period in the history of physics from 1960s till
the end of 1980s was dominated by the technical attitude. However,
to understand quantum spacetime and the unification of quantum
mechanics with gravity, physicists need to come back to the
thinking of Einstein, Bohr, Heisenberg, Boltzmann and many others:
to unite the two great scientific revolutions in one, one ought to
think as generally as did the great masterminds of these
revolutions. The questions that we enlisted above all reemerged at
the front line of the scientific interest. Queen Philosophy
returned to her kingdom of physics.

\section{Loop of existences}\label{loopsect}

Before we start laying down the foundations of the
information-theoretic approach to quantum theory, it is necessary
to say what role this approach plays in our general view of the
scientific venture. This section presents a philosophy in which
will be rooted all of the dissertation.

A first and crucial philosophical assumption is that the world is
\textit{best described} as a \textit{loop of existences} or, as
Wheeler called it, a ``self-synthesizing system of existences''
\cite{WheIBM}. This phrase is devoid of any ontological
commitments; the accent is placed on the word ``described'' and
not on ``world.'' The program therefore is the one of
epistemology: we are studying the interplay of descriptions
without saying anything on the reality of the object described, if
there is any such reality. Perhaps there is none: the question is
irrelevant. To be precise and to remove pure placeholders like
``world'' or ``existences,'' we say that the loop of existences
describes not the existences as elements of external reality, but
the descriptions, the various theories. The first assumption then
becomes: The ensemble of all theories is best described in a
cyclic form as a loop.

The second philosophical assumption is that any particular
theoretical description is achieved by cutting the loop at some
point and thus separating the target object of the theory from the
theory's presuppositions. It is impossible to give a theoretical
description of the loop of existences as a whole. Bohr said about
the necessity of a cut, although from a somewhat different
philosophical position, that ``there must be, so to speak, a
\textit{partition} between the subject which communicates and the
object which is the content of the
communication''\symbolfootnote[2]{Our emphasis.} \cite{aagebohr}.
With the position of the cut being fixed, some elements of the
loop will be object of the theory, while other elements will fall
into the domain of meta-theory of this theory. At another loop cut
these elements may exchange roles: those that were explanans
become explanandum and those that were explanandum become
explanans. The reason why one cannot get rid of the loop cut and
build a theory of the full loop is that the human venture of
knowing needs a basis on which it can rely; at another time, this
basis itself becomes the object of scientific inquiry, but then a
new basis is unavoidably chosen. It is not the case that these
bases form a pyramid which is reduced to yet more and more
primitive elements; on the contrary, for the study of one part of
the world-picture, another its part must be postulated and vice
versa. Employing a notion characteristic of Wittgensteinian
philosophy \cite{wittphilinvest}, Wheeler calls this endeavour a
\textit{mutual illumination}. Francisco Varela, in the context of
phenomenology and cognitive science, spoke about \textit{mutual
constraints} \cite{varelaCC}.

\begin{figure}[htbp]
\begin{center}
\epsfysize=2.5in \epsfbox{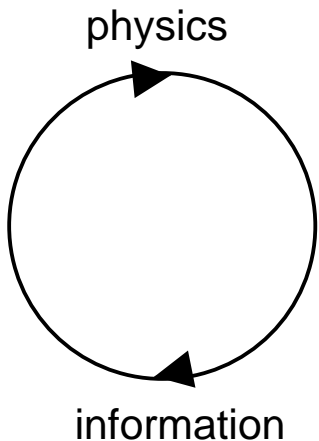}
\end{center}
\caption{The loop of existences between physics and
information}\label{loop0}
\end{figure}

Consider the loop between physical theory and information
(Figure~\ref{loop0}). Arrows depict possible assignment of the
roles of explanans and explananda, of what falls into the
meta-theory and what will be object of the theory. Physics and
information mutually constrain each other, and every theory will
give an account of but a part of the circle, leaving the other
part for meta-theoretic assumptions. For long time physicists have
lacked the understanding of this epistemological limitation. Thus
historically quantum physics has been predominantly conceived as
theory of non-classic waves and particles. Einstein, for instance,
believed that the postulate of existence of a particle or a
quantum is a basic axiom of the physics. In a letter to Born as
late as 1948 he writes \cite[p.~164]{einborn}:
\begin{quote}
We all of us have some idea of what the basic axioms in physics
will turn out to be. The quantum or the particle will surely be
one amongst them; the field, in Faraday's or Maxwell's sense,
could possibly be, but it is not certain.
\end{quote}
We part radically with this view. The venture of physics is now to
be seen as an attempt to produce a structured, comprehensible
theory based on information. Physical theory, quantum theory
including, is a general theory of information constrained by
several information-theoretic principles. As Andrew Steane puts it
\cite{steane},
\begin{quote}
Historically, much of fundamental physics has been concerned with
discovering the fundamental particles of nature and the equations
which describe their motions and interactions. It now appears that
a different programme may be equally important: to discover the
ways that nature allows, and prevents, information to be expressed
and manipulated, rather than particles to move.
\end{quote}
If one removes from this quotation the reference to nature, which
bears the undesired ontological flavor, what remains is the
program of giving physics an information-theoretic foundation.
This is what we achieve by cutting the loop: We treat quantum
theory as theory of information. This is a no small change in the
aim of physics. Bub \cite{bubstudies} argues that information must
be recognized as ``a new sort of \emph{physical} entity, not
reducible to the motion of particles and
fields''\symbolfootnote[2]{Our emphasis.}. Although we fully
endorse the second part of this phrase, we are forced into a
different attitude concerning the first one. In the loop
epistemology, information is not a physical entity or object of
physical theory like particles or fields are. Were it physical,
information would be fully reducible to the intratheoretic
physical analysis. This, then, would do nothing to approach the
problem of giving quantum physics a \textit{foundation}. The only
way to give an information-theoretic foundation to quantum physics
is through putting information in the domain of metatheoretic
concepts. When one does so consistently, conventional physical
concepts such as particles and fields are reduced to information,
not put along with it on equal grounds. Then the physical theory
will fully and truly be a theory of information.

\begin{figure}[htbp]
\begin{center}
\epsfysize=2.5in \epsfbox{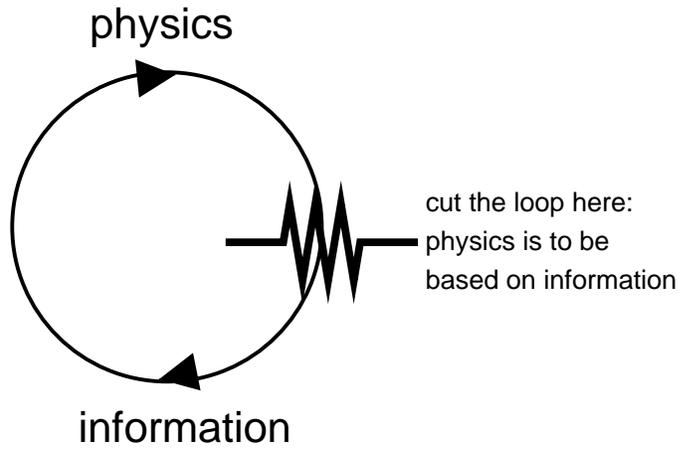}
\end{center}
\caption{Loop cut: physics is informational}\label{loop01}
\end{figure}

In the loop cut shown on Figure~\ref{loop01} information lies in
the meta-theory of the physical theory, and physics is therefore
based on information. The next step is to derive physics from
information-theoretic postulates. In this dissertation such a
derivation will be developed for the part of physics which is
quantum theory.

\begin{figure}[htbp]
\begin{center}
\epsfysize=2.5in \epsfbox{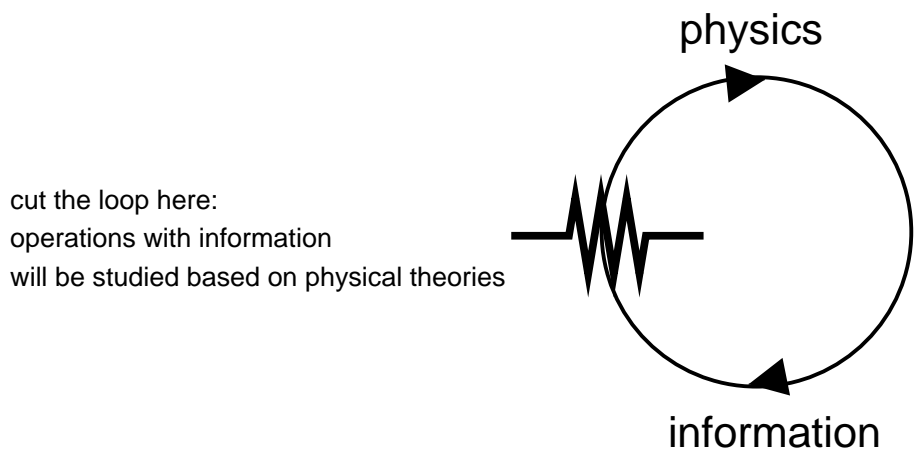}
\end{center}
\caption{Loop cut: information is physical}\label{loop1}
\end{figure}

\begin{figure}[htbp]
\begin{center}
\epsfysize=2.5in \epsfbox{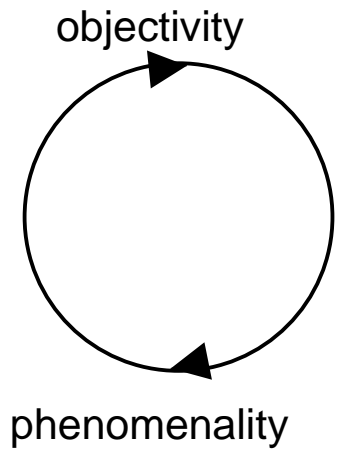}
\end{center}
\caption{The loop of existences between objectivity and
phenomenality}\label{loop02}
\end{figure}

In a different loop cut (Figure~\ref{loop1}), informational agents
are physical beings, and one can describe their storage of, and
operation with, information by means of effective theories that
are reduced, or reducible in principle, to physical theories.
Cognitive science is a vast area of science dealing with this
task; but informational agents can also be non-human systems such
as computers. In this case, the underlying physical theory is
assumed without questioning its origin and validity. Physics now
has itself the status of meta-theory and it is
\textit{postulated}, i.e. it lies in the very foundation of the
theoretical effort to describe the storage of information and no
result of the new theory of information can alter the physical
theory. Therefore, in the loop cut of Figure~\ref{loop1},
particularly in the context of cognitive science, the question of
derivation or explanation of physics is meaningless. Once a
particular loop cut is assumed, it is a \textit{logical error} to
ask questions that only make sense in a different loop cut. To
make this last assertion clearer, let us look at the loop of
existences formed by the two notions different from information
and physics (Figure~\ref{loop02}). We return to the study of the
loop cut of Figure~\ref{loop1} in Section~\ref{cognsc}.

Loop between the phenomenal and the objective is important for
understanding Husserl's phenomenology and his denunciation of
science \cite{husskrisis}. He argued that the only foundation of
science is the phenomenality, and therefore no science can claim
to explain the phenomenality, as, in his view, physics did.
Husserl was right and wrong at the same time: if one assumes his
premise about the universal primary role of phenomena, then
neither physics nor any other science can explain phenomena;
otherwise it would amount to a theory of the loop uncut. It then
becomes a \textit{logical error} to consider physics as explanans
for phenomena. However, if one considers Husserl's premise about
phenomena not as a universal---sort of ontological---claim, but as
an epistemological one: for the purposes of a \textit{given}
description it is necessary to treat the phenomenality as
meta-theoretical, then nothing precludes from treating physics as
meta-theoretical for the purposes of a description of
phenomenality. At the very moment Husserl's premise is transferred
to the sphere of epistemology, the necessity of loop cut removes
the cause for Husserl's critique of physics.

Our two assumptions: viewing the ensemble of theories as a loop
and postulating the necessity of loop cut for any particular
theory, form a transcendental argument. Here we meet the
conclusion of the paper by Michel Bitbol \cite{bitbolcercles} in
which what he calls ``epistemological circles'' also receive a
transcendental treatment. By definition, a transcendental argument
is an argument from the conditions of possibility. In our case,
one is concerned with the conditions of possibility of theorizing,
of building \textit{a} theory, of course irrespective of the
content of the theory. Theorizing is only possible if the loop is
cut; uncut loop, i.e. no separation between theory and
meta-theory, as in the example of Husserl's critique, is a logical
error. In order to avoid the error and together with it a vicious
circle, thus meeting the necessary condition of logical
consistency, one must cut the loop. A theory is only possible when
it knows its limits. The possibility of theorizing is conditioned
by cutting the loop prior to building a theory.

\section{Dissolution of the measurement problem}\label{dissom}

As a digression from the main line of development of the
dissertation, in this section we address the question of how the
epistemology of the loop of existences shapes the purported
solution of the measurement problem. The latter is formulated as
follows. In quantum mechanics a physical system can be in a
superposition state, which corresponds to a certain linear
combination of the eigenvectors of some observable. Temporal
evolution is unitary and linear, and therefore initial
superpositions of vector states are mapped onto corresponding
superpositions of image vector states. Consequently, any
measurement instrument will generally be entangled with the
quantum system it measures. The theory dictates that there shall
be no breakdown of such entanglement. So, at the end of what we
take for a measurement, neither the measuring instrument nor the
system measured will have separable properties. On the other hand,
our commonsense understanding of the phenomenon is that the
instrument registers a definite measurement outcome. The problem
is then to explain how a passage is possible, from the
superposition to a definite outcome.

A classical way to tackle the measurement problem is by
introducing a ``wavefunction collapse.'' This amounts to
suspending the unitary dynamics whenever there is a measurement
and saying that the quantum state collapses to one of the states
in the superposition that corresponds to a definite measurement
outcome. Then the final state at the end is represented as a
statistical mixture of different outcomes with weights equal to
probabilities defined by the entangled state. The difference
between statistical mixture and entangled state is the same to
which d'Espagnat refers to as proper and improper mixtures
\cite{desp}.

Other solutions to the problem of measurement include collapse
theories that modify the unitary dynamics \cite{GRW}; many-worlds
interpretation \cite{everett,Wallace}; or subscribing to some form
of modal interpretation, although it remains to be seen how this
can help to solve the problem. All these theories are empirically
equivalent and can be distinguished from one another on
non-experimental grounds only. Apparent underdetermination of
quantum theory is expressed in the fact that it allows for all the
various equivalent theories to exist. We argue that this only
happens if quantum theory is viewed in the usual way physical
theories are looked at: namely, as a theory about physical
entities that really exist, such as particles or waves, and aiming
at describing these entities. Now, if one changes the stance and
adopts our view of the physical theory being the theory of
information, the problem of choice between various answers to the
measurement problem and, indeed, the measurement problem itself
are, not solved but \textit{dissolved}. Because the loop must be
cut in construction of any particular theory, the measurement
problem is a mere logical error, a consequence of the failure to
distinguish between theory and meta-theory.

Indeed, if we identify the measuring system with the one which
stores and manipulates information, it follows from the discussion
of the two possible loop cuts that the measuring system
\textit{must} remain unaccounted for by the physical theory based
on information. A new, separate theory of measuring systems is
possible, but in order to construe it, one ought to choose a new
cut of the loop and thereby be swayed away from the theory that
had information as primary notion. A purported solution of the
quantum mechanical measurement problem belongs to the loop cut of
Figure~\ref{loop1}, while quantum mechanics as physical theory
belongs to the loop cut of Figure~\ref{loop01}.

The quantum mechanical measurement problem is then equivalent to
the non-existence of cut in the loop, to merely confusing
questions that make sense in one loop cut with questions that make
sense in the opposite loop cut. Assumption of necessity of the
loop cut, grounded in the transcendental argument, with its origin
in the structure of the human venture of theorizing, dissolves the
measurement problem: at the very moment a cut appears in the loop,
the problem disappears.

\chapter{Quantum computation}\label{qcompch}

This chapter is a brief introduction to the ideas of quantum
computation, a domain whose rapid development in 1990s motivated
the increase of interest toward the notion of information. The
chapter is not essential for following the main argument of the
dissertation and a reader only interested in the latter may go
directly to Part~\ref{part2}.

\section{Computers and physical devices}

Since ever humanity has been seeking tools to help to solve
problems and tasks, and with growth of complexity of these tasks,
the tools became needed for solving the problem of calculation.
One needs to calculate the area of land, stress on rods in
bridges, or the shortest way from one place to another. Simple
calculation evolved in a complicated computation. A common feature
of all these tasks, however, was that they follow the pattern: $$
\textrm{Input}\rightarrow \textrm{Computation}\rightarrow
\textrm{Output.}$$

The computational part of the process is inevitably performed by a
dynamical physical system, evolving in time. In this sense, the
question of what can be computed is connected to the question of
what systems can be physically realized. If one wants to perform a
certain computational task, one must seek the appropriate physical
system, such that the evolution of the system in time corresponds
to the desired computation. If such a system is initialized
according to the input, its final state will correspond to the
output.

An example \cite{doritaha} of interconnection between physical
systems and computation was invented by Gaudi, the great Spanish
architect. The plan of his Sagrada Familia church in Barcelona is
very complicated, with towers and arcs emerging from unexpected
places, leaning to other towers and arcs, and so forth. It was
practically impossible to solve the set of equations which
correspond to the requirement of equilibrium of this complex.
Instead of solving the equations Gaudi did the following: for each
arc he took a rope, of length proportional to the length of the
arc. Where arcs were supposed to lean on each other, he tied the
end of one rope to the middle of another rope. Then he tied the
edges of the lowest ropes, which must correspond to the lower
arcs, to the ceiling. All computation was thus instantaneously
done by gravity. Angles between the arcs and radii of the arcs
could be easily read from this analog computer, and the whole
church could be seen by simply putting a mirror on the floor under
the rope construction.

Many examples of analog computers exist, which were devised to
solve a specific computation task; but we do not want to build a
completely different machine for each task that we have to
compute. We would rather have a general purpose machine, which is
``universal.'' A mathematical model for such machine is Turing
machine, which consists of an infinite tape, a head that reads and
writes on the tape, a machine with finitely many possible states,
and a transition function $\delta$. Given what the head reads at
time $t$ and the machine's state at time $t$, function $\delta$
determines what the head will write, to which direction it will
move and what will be the new machine's state at time $t+1$. The
Turing machine defines a concept of computability, according to
the Church-Turing thesis in a very broad formulation:

\begin{description}\item [Church-Turing thesis:] A Turing machine can compute any
function computable by a reasonable physical device.
\end{description}

What does ``reasonable physical device'' mean? The Church-Turing
thesis is a statement about universal qualities of the physical
world and not a formal mathematical statement; therefore it cannot
be rigorously proven. However, up to now all physical systems used
for computation seem to have a simulation by a Turing machine,
although often only in principle.

It is an astonishing fact that there exist families of functions
which cannot be computed. In fact, most of the functions cannot be
computed: there are more functions than there are ways to compute
them. The reason for this is that the set of Turing machines is
countable, whereas the set of families of functions is not. In
spite of the simplicity of this argument (which can be formalized
using the diagonal argument, as did G\"{o}del), the observation
itself came as a big surprise when it was discovered in 1930s. The
subject of computability of functions is the cornerstone of
\textit{computational complexity}. Often we are interested not
only in which functions can be computed but in the \textit{cost}
of such computation. The cost, or computational complexity, is
measured naturally by the physical resources invested in solving
the problem, such as time, energy, space, etc. A fundamental
question in computational complexity is how the cost of computing
a function varies as a function of the input size, $n$, and in
particular whether it is polynomial or exponential in $n$. In
computer science problems which can only be solved in exponential
cost are regarded as intractable. The class of tractable problems
consists of problems which have solutions with polynomial cost.

It is worth reconsidering what it means to \textit{solve} a
problem. An important conceptual breakthrough was the
understanding \cite{probalgor} that sometimes it is advantageous
to relax the requirements that a solution be always correct, and
allow some (negligible) probability of error. This gives rise to a
much more rapid solutions of different problems, which make use of
random coin flips, such as an algorithm to test whether an integer
is prime or not \cite{primealg}. The class of tractable problems
is now considered as problems solvable with a negligible
probability for error in polynomial time. These solutions will be
computed by a deterministic Turing machine, except that the
transition function can change the configuration in one of several
possible ways, randomly. The modern Church thesis refines the
Church Turing thesis and asserts that the probabilistic Turing
machine captures the entire concept of computational complexity:

\begin{description}\item [The Modern Church thesis:] A probabilistic Turing machine
can compute any function computable by a reasonable physical
device in polynomial cost.\end{description}

Again, this thesis cannot be proven because it is not a
mathematical statement. It is worthwhile mentioning a few models
which at the first sight might seem to contradict the modern
Church thesis, such as the DNA computer \cite{DNAcomp}. Most of
these models, which are currently a subject of growing interest,
\textit{do not rely on classical physics}.

\section{Basics of quantum computation}\label{qcompsect}

In the beginning of 1980s Feynman \cite{feyn82,feyn86} and Benioff
\cite{benioff80,benioff82} started to discuss the question of
whether computation can be done on the scale of quantum physics.
In classical computers, the elementary information unit is a bit,
the value of which is either 0 or 1. The quantum analog of a bit
would be a two-state particle, called a \textit{qubit}. A
two-state system is described by a unit vector in the Hilbert
space isomorphic to $\mathbb{C}^2$. Zero state of a bit
corresponds to vector
$1\times|0\rangle+0\times|1\rangle=|0\rangle$, state one of the
bit corresponds to the state $|1\rangle$. These two states
constitute a orthogonal basis in the two-dimensional Hilbert
space, and the general state of a qubit is described as their
normalized linear combination. To build a computer, we need to use
a large number of qubits. Then the Hilbert space is a product of
$n$ spaces $\mathbb{C}^2$. Naturally classical strings will
correspond to quantum states:
\begin{equation}i_1i_2\ldots i_n\leftrightarrow |i_1\rangle\otimes
|i_2\rangle\otimes\ldots\otimes |i_n\rangle\equiv|i_1\ldots
i_n\rangle.\end{equation}

How to perform calculation using qubits? Suppose that we want to
compute the function $f: i_1\ldots i_n\mapsto f(i_1\ldots i_n)$,
from $n$ bits to $n$ bits. We would like the system to evolve
according to the time evolution operator $U$:
\begin{equation}\label{unitevol}
|i_1\ldots i_n\rangle\mapsto U|i_1\ldots i_n\rangle=|f(i_1\ldots
i_n)\rangle.
\end{equation}
We therefore have to find a Hamiltonian $H$ which generates this
evolution. According to the Schr\"odinger equation, this means
that we have to solve for $H$:
\begin{equation}
|\Psi_f\rangle=\exp\left(-\frac{i}{\hbar}\int H
dt\right)|\Psi_0\rangle=U|\Psi_0\rangle
\end{equation}
A solution for $H$ always exists, as long as the linear operator
$U$ is unitary. Unitarity is an important restriction. Note that
the quantum analog of a classical operation will be unitary only
if $f$ is one-to-one, or bijective. Hence, a reversible classical
function can be implemented by a physical Hamiltonian. It turns
out that any classical function can be represented as a reversible
function on a larger number of bits \cite{Bennett73}, and that
computation of $f$ can be made reversible without losing much in
efficiency. Moreover, if $f$ can be computed classically by
polynomially many elementary reversible steps, the corresponding
$U$ is also decomposable into a sequence of polynomially many
elementary unitary operations. We see that quantum systems can
imitate all computations which can be done by classical systems,
and do so without losing much in efficiency.

Quantum computation is interesting not because it can imitate
classical computation, but because it can probably do much more.
Feynman pointed out the fact that quantum systems of $n$ particles
seem to be hard to simulate by classical devices, and this
exponentially in $n$. In other words, quantum systems do not seem
to be polynomially equivalent to classical systems, including
classical computational devices, which violates the modern Church
thesis. This provides an insight on why as computational devices
quantum systems may be much more powerful than classical systems.

How to use ``quantumness''? Consider, for example, the
Green\-ber\-ger-Horne-Zei\-lin\-ger (GHZ) triparticle state
\cite{GHZ}:
\begin{equation}
  \frac{1}{\sqrt{2}}\left( |000\rangle+|111\rangle \right)
.\end{equation} What is the superposition described by the first
qubit? The answer is that is no such superposition. Each of the
three qubits does not have a state of its own, and the state of
the system is not a tensor product of states of each particle.
Such states are called entangled. Entanglement is used in the
Einstein-Podolski-Rosen ``paradox'' \cite{EPR} and Bell
inequalities both in the original formulation by Bell
\cite{Bell1,Bell2} and the one proposed by Clauser, Holt, Horne
and Shimony \cite{CHHS}. Because of entanglement, the state of the
system can only be described as a superposition of all $2^n$ basis
states, and, consequently, $2^n$ coefficients are needed. This
exponentiality of resources in the Hilbert space is the crucial
property needed for quantum computation. To take another example,
consider a uniform superposition of all basis states:
\begin{equation}
  \frac{1}{\sqrt{2^n}} \sum_{i_1,i_2,\ldots,i_n=0}^{1} |i_1 i_2 \ldots i_n\rangle
.\end{equation} Now apply to it the unitary operation which
computes $f$, as in Equation \ref{unitevol}. From the linearity of
quantum mechanics we get:
\begin{equation}
  \frac{1}{\sqrt{2^n}} \sum_{i_1,i_2,\ldots,i_n=0}^{1} |i_1, i_2, \ldots, i_n\rangle\mapsto
  \frac{1}{\sqrt{2^n}} \sum_{i_1,i_2,\ldots,i_n=0}^{1} |f(i_1, i_2, \ldots, i_n)\rangle
.\end{equation} The conclusion is that by applying $U$ one
computes $f$ simultaneously on all the $2^n$ possible inputs $i$,
which is an enormous gain in parallelism.

In fact, such an exponential gain in parallelism does not imply
exponential increase in computational power. The problem lies in
the question of how to extract information out of the system. In
order to do this, one has to \textit{observe} the quantum system.
In a standard interpretation of quantum mechanics, after the
measurement, the state is projected on one of the exponentially
many possible states, and all information appears to be lost. To
gain advantage, one therefore needs to combine parallelism with
another feature, which is interference. The goal is to arrange the
cancellation by interference so that only the interesting
computations remain and all the rest cancel out. If one expresses
this operation in the initial basis, rearrangement will take the
form of a POVM measurement, i.e. of a measurement represented as a
positive operator-valued measure \cite{davieslewis}. This explains
why POVM are an essential tool in the science of quantum
information. Formal development of this idea will follow in
Chapter~\ref{chaptreconstr}.

Combination of parallelism and interference plays an important
role in quantum algorithms. A quantum algorithm is a sequence of
elementary unitary steps, which manipulate the initial quantum
state $|i\rangle$ for an input $i$ so that a measurement of the
final state of the system yields the correct output. The first
quantum algorithm, which combines parallelism and interference to
solve a problem faster than a classical computer, was discovered
by Deutsch and Jozsa \cite{deutschjozsa}. The algorithm must
distinguish between ``constant'' (all items are the same) and
``balanced'' databases. The quantum algorithm solves this problem
exactly in polynomial cost. Classical computers cannot do better
than to check all items in the database, which is exponentially
long, and in polynomial cost they can only solve the problem
approximately. Deutsch and Jozsa's algorithm provides an exact
solution in virtue of using the Fourier transform. A similar
technique also gave rise to the most important quantum algorithm
that we know today, Shor's algorithm \cite{Shor}.

Shor's algorithm is a polynomial quantum algorithm for factoring
integers and for finding a logarithm over a finite field. For both
problems the best known classical algorithms are exponential.
However, there is no proof that a classical polynomial algorithm
is impossible. Shor's result is extremely important both
theoretically and practically, due mainly to the fact that the
factorization task is a cornerstone of the RSA cryptographic
system, which is used almost everywhere in our life, to start with
internet browsers. A cryptographic system must be secure; this
means that an eavesdropper will not be able to learn in reasonable
time significant information about the message that has been sent.
For RSA system, to be successful in cracking the system, the
eavesdropper needs to have an efficient algorithm for factoring
big numbers. It is therefore understandable why Shor's result is
viewed as the first potential implication of quantum information
science that will prove to be of great practical significance.

It is important to note that quantum computation does not rely on
unreasonable precision of measurement, but a polynomial precision
is enough. This means that the new model requires physically
reachable resources, in terms of time, space, and precision; yet
it is exponentially stronger than the ordinary model of a
probabilistic Turing machine. Currently, quantum computer is the
only model which threatens the modern Church thesis.

There are several major directions of research in the area of
quantum computation. Introduction and comprehensive analysis can
be found in a number of recent monographs
\cite{boum,heiss,nielsenchuang}.

\section{Why quantum theory and information?}

The remarkable achievements of the science of quantum information
and computation allow one to take it as the viewpoint from which
to look at all of quantum physics. Still, it is obvious that
quantum computation only uses a tiny fraction of the results of
quantum physics, although conceptually the most profound ones.
From this viewpoint one must ask, as we do in this dissertation,
if information-theoretic axioms can serve as a foundation for
quantum physics and if not fully, then to what extent. We close
the introduction by giving four arguments why in our opinion such
program deserves attention.

\medskip\noindent \textbf{Argument for a specialist in quantum
computation}. A researcher in quantum computation would like to
view quantum computation as an autonomous scientific area, which
merits its own development from its first principles, without
bringing in much from other disciplines. Such a project would try
to establish ``axiomatic closure'' or self-sufficiency of this
discipline, i.e. all information-theoretic results in quantum
computation will be derived from information-theoretic axioms.
With this idea in mind, a researcher in quantum computation would
like to see which parts of quantum mechanics he or she needs
\textit{prima facie}, and which parts it is possible to deduce
from information-theoretic axioms. The result then will show to
what extent the science of quantum computation can be treated as
autonomous discipline.

\medskip\noindent \textbf{Argument for a theoretical physicist}. Working
physicists seldom address problems in the foundations of quantum
theory and are often unprepared to talk about the role of this or
that of the bricks that compose it. To understand better the
structure of the theory, the origin of its first principles and
their interconnections, it is challenging to attempt a
reconstruction of the quantum theory from information-theoretic
axioms: a reconstruction implies derivation, and the mathematical
language of the derivation program is familiar to physicists.
Still, one must be from the very start aware that such a
derivation will not fully replace any of the usual ways of
introducing quantum theory in physics, as it would be too
ambitious to expect that all of modern quantum theory, including
field theory and unification attempts, can be reconstructed from
the few information-theoretic axioms; additional features often
need additional assumptions. In the derivation proposed in this
dissertation, we only justify the algebraic structure of the
theory, and with regard to issues not directly linked to algebra,
such as time dependence, reconstruction from information-theoretic
principles demands more assumptions (see Section~\ref{sect44}).

\medskip\noindent \textbf{Argument for a laboratory physicist}. The best
method to decide in which way to give a foundation to a scientific
discipline lays, perhaps, in looking at how the theory is applied,
i.e. at the technology that it generates. As Fuchs puts it, ``If
one is looking for something `real' in quantum theory, what more
direct tack could one take than to look to its technologies?
People may argue about the objective reality of the wave function
\textit{ad infinitum}, but few would argue about the existence of
quantum cryptography as a solid prediction of the theory. Why not
take that or a similar effect as the grounding for what quantum
mechanics is trying to tell us about nature?''~\cite{FuchsInLight}
Some steps have already been made in the direction of studying
quantum mechanics in the light of the technology to which it gave
birth \cite{nakajima}, and the program of deriving quantum
mechanics from information-theoretic principles can be viewed as a
development of this project.

\medskip\noindent \textbf{Argument for an educator}. The world is nowadays
facing a rapid development of nanotechnology
\cite{dupuygrinbaumnano} and, perhaps in the near future, of the
technology of quantum information. This means that the society
will soon need to educate quantum engineers, whose specialization
will be in quantum computers and other quantum technological
devices. As any engineer, quantum engineer will not be a scientist
doing fundamental research in physics, and thus will only need to
be given as much of physical education as he ought to have in
order to master his profession. The future educator of quantum
engineers will be interested in finding out, how much of quantum
physics the engineer needs to be taught and whether this much of
physics can be taught by being derived from the
information-theoretic principles, which, in their turn, will be a
part of the engineer's basic training.

\part{Information-theoretic derivation of quantum theory}\label{part2}

\chapter{Conceptual background}\label{sect3}

\section{Axiomatic approach to quantum mechanics}

In Part~\ref{part2} of the dissertation we demonstrate a
derivation of quantum theory from infor\-ma\-tion-theoretic
axioms. Attempts at axiomatization of quantum mechanics have been
made ever since von Neumann's early work, and we start by
presenting the idea of axiomatic approach.

As such, the axiomatic method can be traced back to the Greeks.
The XIXth century revolutionized this approach by bringing in the
idea that an axiom can no longer be considered as an ultimate
truth about reality, but a structural element---an assumption that
lies in the foundation of a certain theoretical structure.
Therefore ``not only geometry, but many other, even very abstract,
mathematical theories have been axiomatized, and the axiomatic
method has become a powerful tool for mathematical research, as
well as a means of organizing the immense field of mathematical
knowledge which thereby can be made more
surveyable''~\cite{heyting}.

The first paper where quantum mechanics was treated as a
\textit{principle theory} appeared very shortly after the creation
of quantum mechanics itself. To quote from Hilbert, von Neumann
and Nordheim \cite{HvNN}:
\begin{quote}
The recent development of quantum mechanics, stemming on the one
hand from the papers of Heisenberg, Born and Jordan and those of
Schr\"odinger on the other hand, has put us in a position to
subsume the whole domain of atomic phenomena from a single point
of view... In view of the great significance of quantum mechanics
it is an urgent requirement to formulate its principles as clearly
and generally as possible.

\ldots

The route leading to the theory is the following: we make certain
physical requirements of the probabilities, suggested by our
previous experience and developments, and whose fulfillment
entails certain relations between the probabilities. Secondly, we
look for an analytical apparatus, in which quantities occur
satisfying exactly the same relations. This analytical apparatus,
and the arithmetic quantities occurring in it, receives now
\textit{on the basis of the physical postulates} a physical
interpretation. Here, the aim is to formulate the physical
requirements so completely that the analytical apparatus is just
uniquely determined. Thus the route is of axiomatization.

\ldots

The process of axiomatization indicated above is not as a rule
exactly followed through in physics, but the route to the
establishment of a new theory is here, as elsewhere, the
following. One conjectures the analytical apparatus, before
establishing a complete axiom system and only then, by
interpretation of the formalism, obtains the basic physical
relations. It is difficult to understand such a theory if one does
not make a sharp distinction between these two things, the
formalism and its physical interpretation.\symbolfootnote[2]{Our
emphasis.}
\end{quote}

Such a standpoint led von Neumann, in collaboration with Birkhoff,
to the first study of the logic of quantum mechanics \cite{bvn}.
Later, via the theory of projective geometries, this had led to
the creation of the theory of orthomodular lattices
\cite{kalmbach83}. On the way to lattices von Neumann created the
algebraic theory of what was later called von Neumann algebras,
which further led to the explosion of algebraic techniques in
quantum mechanics, field theory, and unified theories.

Since Kolmogoroff's axioms for the probability theory \cite{kolmo}
and Birkhoff's and von Neumann's quantum logic \cite{bvn} many
axiomatic systems were proposed for quantum mechanics. On the side
of quantum logic a partial list includes Mackey
\cite{mackey57,mackey63}, Zieler \cite{zieler}, Varadarajan
\cite{varada,varada2}, Piron \cite{piron,piron72}, Kochen and
Specker \cite{kochspeck65}, Guenin \cite{guenin}, Gunson
\cite{Gunson}, Jauch \cite{jauch}, Pool \cite{pool,pool2},  Plymen
\cite{plymen}, Marlow \cite{marlow}, Beltrametti and Casinelli
\cite{bc}, Holland \cite{holland}. We propose a quantum logical
axiomatic derivation in Chapter~\ref{chaptreconstr}. Probabilistic
sets of axioms were introduced by Ludwig and his followers
\cite{ludwig}; they will not be studied in the dissertation.
Another branch of axiomatic quantum theory, the algebraic approach
was first conceived by Jordan, von Neumann and Wigner \cite{jvNw}
and developed by Segal \cite{segal1,segal2}, Haag and Kastler
\cite{hk}, Plymen \cite{plymen2}, Emch \cite{Emch}.
Information-theoretic interpretation of the algebraic approach
will be the subject of Part~\ref{cstar}.

We close this section with an illuminating passage about
axiomatization in physics due to Jean Ullmo, one of the founders
of CREA~\cite[p.~121]{ullmo}:
\begin{quote}
{\selectlanguage{french} La th\'{e}orie physique moderne manifeste
une tendance certaine \`{a} rechercher une pr\'{e}sentation
\textit{axiomatique}, sur le mod\`{e}le des axiomatiques
math\'{e}\-ma\-tiques. L'id\'{e}al axiomatique, emprunt\'{e} \`{a}
la g\'{e}om\'{e}trie, revient \`{a} d\'{e}finir tous les \og
objets \fg~initiaux d'une th\'{e}orie uniquement par des
\textit{relations}, nullement par des qualit\'{e}s
substantielles.}\symbolfootnote[2]{Modern physical theory shows a
certain tendency for one to look for an \textit{axiomatic}
representation of the theory, modelled on axiomatic systems in
mathematics. The ideal of axiomatization, borrowed from geometry,
consists in defining all the initial ``objects'' of the theory
only by \textit{relations} and not at all by some substantial
qualities.}\end{quote} Our way, thus, goes from the discussion of
axioms in this section to a discussion of relations in the next
one.

\section{Relational quantum mechanics}\label{rovsect}

\epigraph{A quantum mechanical description of an object by means
of a wave function corresponds to the relativity requirement with
respect to the means of observation. This extends the concept of
relativity with respect to the reference system familiar in
classical physics.}{Vladimir Fock \cite{fock}}

This section prepares the key two sections that follow it. It
serves to explain the motivation behind the choices made in those
sections, i.e. its goal is to communicate to the reader the
physical intuition that the author believes to have.

Any attempt at formal derivation of quantum mechanics requires a
definite conceptual background on which the derivation will
further operate. It is commonplace to say that it is not easy to
exhibit an axiomatic system that could supply such a background.
Before one starts making judgements about plausibility of axioms,
one must develop an intuition of what is plausible about quantum
theory and what is not. This can be only achieved by
\textit{practicing} the quantum theory itself, i.e. by taking its
prescriptions at face value, applying them, getting results, and
then asking questions of what these results mean. However, it is
important to notice that undertaking all actions on this list will
not yet make things clear about quantum mechanics. It purely
serves as a tool for developing intuition about what is a
plausible claim and what must be cut off. The reasons why
implausibility may arise are of various nature: from Occam's razor
to direct contradiction with observation. We discussed it in
Ref.~\cite{grinbzvezda}.

Once the intuition has been developed, a scientist who wishes to
follow the axiomatic approach must select axioms which he believes
plausible; and then the whole remaining part of the building will
be constructed ``mechanically,'' by means of the formalism. The
choice of axioms must be the only external freedom of the theory.
We argue that such a program is the exclusive way to make things
\textit{clear} about quantum mechanics. To quote from Rovelli
\cite{RovRQM},
\begin{quote}
Quantum mechanics will cease to look puzzling only when we will be
able to \textit{derive} the formalism of the theory from a set of
simple physical assertions (``postulates,'' ``principles'') about
the world. Therefore, we should not try to append a reasonable
interpretation to the quantum mechanical formalism, but rather to
\textit{derive} the formalism from a set of experimentally
motivated postulates.
\end{quote}

As the aforementioned experimentally motivated postulates we
choose informa\-tion-theoretic principles. Initially formulated by
John Wheeler \cite{WheIBM,WheFeynman}, the program of deriving
quantum formalism from information-theore\-tic principles has been
receiving lately much attention. Thus, Jozsa promotes a viewpoint
which ``attempts to place a notion of information at a primary
fundamental level in the formulation of quantum physics''
\cite{Jozsa}. Fuchs presents his program as follows: ``The task is
not to make sense of the quantum axioms by heaping more structure,
more definitions... on top of them, but to throw them away
wholesale and start afresh. We should be relentless in asking
ourselves: From what deep \emph{physical} principles might we
\emph{derive} this exquisite mathematical structure?.. I myself
see no alternative but to contemplate deep and hard the tasks, the
techniques, and the implications of quantum information
theory.''~\cite{FuchsLittleMore}

However, before we start selecting concrete information-theoretic
axioms, we must say why our intuition developed so that we believe
that precisely this kind of axioms, namely information-theoretic
ones, are a plausible set of axioms for quantum mechanics. The
intuition here is due to the relational approach to quantum
mechanics \cite{RovRQM}.

The word ``relational'' has been used by different philosophers of
quantum physics, most notably by Everett \cite{everett} and by
Mermin \cite{mermin}. Our sense of using this word, along the
lines indicated by Rovelli, goes back to the special relativity.
Special relativity is a well-understood physical theory,
appropriately credited to Einstein in 1905. But it is equally
well-known that the formal content of special relativity, i.e.
Lorentz transformations, were written by Lorentz and Poincar\'{e}
and not by Einstein, and this several years before 1905. So what
was Einstein's contribution?

Lorentz transformations were heavily debated in the years
preceding 1905 and were often called ``unacceptable,''
``unreasonable'' and so forth. Many interpretations of what the
transformations mean were offered, and among them quite a
plausible one about interactions between bodies and the ether.
This reminds of some of the modern discussion of quantum
mechanics. However, when Einstein came, things suddenly became
clear and the debate stopped. This was because Einstein gave a few
simple physical principles from which he \textit{derived} Lorentz
transformations, therefore closing the attempts to heap philosophy
\textit{a posteriori}, above the formal structure itself.
Einstein's idea was single and ingenious: he assumed that there is
no absolute notion of simultaneity. Simultaneity, said Einstein,
is relative. Once the notion of absolute simultaneity has been
removed, the physical meaning of Lorentz transformations stood
clear, and special relativity has not raised any controversy ever
since.

Vladimir Fock, as cited in the epigraph to this section, was among
the first to say that quantum mechanics generalizes Einstein's
principle of relativity. We argue that what becomes relative in
quantum theory is the notion of state.

Consider an observer $O$ that makes measurement of a system $S$.
Assume that the quantity being measured, say $x$, takes two
values, 1 and 2; and let the states of the system $S$ be described
by vectors in a two dimensional complex Hilbert space $\BH _S$.
Let the two eigenstates of the operator corresponding to the
measurement of $x$ be $|1\rangle$ and $|2\rangle$. As follows from
the standard quantum mechanics, if $S$ is in a generic normalized
state $|\psi\rangle=\alpha |1\rangle +\beta |2 \rangle$, where
$\alpha$ and $\beta$ are complex numbers and $|\alpha |^2 +|\beta
|^2 =1$, then $O$ can measure either one of the two values $1$ and
$2$ with respective probabilities $|\alpha |^2$ and $|\beta |^2$.

Assume that in a given specific measurement at time $t_1$ the
outcome is $1$. Denote this specific measurement as $\M$. The
system $S$ is affected by the measurement, and at time $t_1$ the
state of the system is $|1\rangle$. In the sequence of
descriptions, the states of $S$ at some time $t=t_0<t_1$ and
$t=t_1$ are thus
\begin{equation}\begin{array}{rcl}
t_0 & \rightarrow & t_1\\
 \alpha |1\rangle +\beta |2
\rangle & \rightarrow & |1\rangle
\end{array}\label{descr1}\end{equation}

Let us now consider the same fact $\M$ as described by a second
observer, who we call $O^\prime$. $O^\prime$ describes the system
formed by $S$ and $O$. Again, assume that $O^\prime$ uses the
conventional quantum mechanics and assume that $O^\prime$ does not
perform any measurement between $t_0$ and $t_1$ but that
$O^\prime$ knows the initial states of $S$ and $O$ and is
therefore able to give a quantum mechanical description of the
fact $\M$. Observer $O^\prime$ describes the system $S$ by means
of the Hilbert space $\BH _S$ and the system $O$ by means of a
Hilbert space $\BH _O$. The $S-O$ system is then described by
means of the product space $\BH _{SO}=\BH _S\otimes \BH _O$. Let
us denote the vector in $\BH _O$ that describes the state of $O$
at time $t_0$ at $|init\rangle$. The physical process implies
interaction between $S$ and $O$. In the course of this
interaction, the state of $O$ changes. If the initial state of $S$
is $|1\rangle$ (respectively $|2\rangle$), then $|init\rangle$
evolves into a state which we denote as $|O1\rangle$ (respectively
$|O2\rangle$). One can think of states $|O1\rangle$ and
$|O2\rangle$ as states in which ``the hand of the measuring
apparatus points at 1'' (respectively at 2). One can write down
the Hamiltonian that produces evolution of this kind, and such
Hamiltonian can be taken as a model for the physical interaction
which produces measurement.

Let us now consider the actual case of the experiment $\M$ in
which the initial state of $S$ is $|\psi\rangle=\alpha |1\rangle
+\beta |2 \rangle$. The initial full state of the $S-O$ system is
then $|\psi\rangle\otimes |init\rangle=(\alpha |1\rangle +\beta |2
\rangle)\otimes |init\rangle$. Linearity of quantum mechanics
implies
\begin{equation}
\begin{array}{rcl}
t_0 & \rightarrow & t_1 \\
(\alpha |1\rangle +\beta |2 \rangle)\otimes |init\rangle
& \rightarrow & \alpha |1\rangle\otimes |O1\rangle + \beta
|2\rangle\otimes |O2\rangle
\end{array}\label{descr2}\end{equation}
Thus at $t=t_1$ the system $S-O$ is in the state $\alpha
|1\rangle\otimes |O1\rangle + \beta |2\rangle\otimes |O2\rangle$.
This is the conventional description of the measurement as a
physical process \cite{vN32}.

We have described the actual physical process $\M$ taking place in
the laboratory. Standard quantum mechanics requires that we
distinguish the system from the observer, but it also allows us
freedom in drawing the line between the two. In the above analysis
this freedom has been exploited in order to describe the same
temporal development in terms of two different observers. In
Equation~(\ref{descr1}) the line that distinguished the observed
system from the observer was set between $S$ and $O$. In
Equation~(\ref{descr2}) this line was set between $S-O$ and
$O^\prime$. Recall that we have assumed that $O^\prime$ is not
making any measurement between $t_0$ and $t_1$. There is no
physical interaction between $O^\prime$ and $S-O$ during the
interval $t_0-t_1$. However, $O^\prime$ may make a measurement at
some later moment $t_2>t_1$; result of such measurement will agree
with the description (\ref{descr2}) that $O^\prime$ gave to the
$S-O$ system at time $t_1$. Thus, we have two different
descriptions of the state at $t_1$: the one given by $O$ and the
one given by $O^\prime$. \textit{Both are correct}. Therefore, we
conclude that
\begin{rem}\label{obsdepend}
In quantum mechanics the state is an observer-dependent concept.
\end{rem}

Observer-dependency is a crucial observation that marks
fundamentally our intuition on how to make judgements about
plausibility of postulates or principles for quantum mechanics. It
is by doing research motivated by Remark \ref{obsdepend} that we
developed the information-theoretic derivation of quantum theory.

We now advance a thesis that the argument about relative states is
in agreement with the philosophy of the loop of existences
presented in Section~\ref{loopsect}. In relational quantum
mechanics, any system is treated as physical and, consequently,
the observer is a physical system as any. Therefore the special
status of the observer only manifests itself in the asymmetry of
the relation ``$O$ has information about $S$.'' Physical states
are then seen as a manifestation of this relation, and asymmetry
of the latter makes any state a relative state: the state of $S$
is defined with respect to $O$, which is a system that has
information about $S$. Now, in some other act of bringing about
information $O$ itself can stand in the place of $S$, i.e.
information will be \textit{about} $S$. It will then be defined
with respect to some other system $O^\prime$. If one iterates such
a chain, one will never run into a contradiction barring the
question of physical nature of the systems $S$, $O$, $O^\prime$,
etc. If, for example, $S$ are light rays, $O$ is the retina of a
human eye, $O^\prime$ are the visual neurons, and then come yet
other brain systems and so forth, we naturally expect that such a
reduction of the observing physical systems ultimately stops, as
they become closer and closer to the fundamental layers of
apprehension. Rovelli denies the validity, or even relevance, of
this argument as having nothing to do with the formal construction
of his theory.
%Still, a philosopher cannot move this quickly and
%has to question the relation between the view that anything is a
%physical system with the rights equal to the rights of all other
%physical systems, and the phenomenal reality.

To safeguard a sound philosophical ground for Rovelli's point of
view, we propose to treat it in the spirit of our transcendental
argument as follows. Each recourse to brain or other physical
structures as observing systems (systems of type $O$ in the above
discussion) need not lead to questioning the applicability of
quantum mechanics or, for that matter, of any given physical
theory. This is because applications of the physical theory and
the problem of its foundation lie in the different parts of the
loop of existences and, in order to be theoretically analyzed,
require different loop cuts. The best tactics for a partisan of
the relational quantum mechanics is to loop a chain of relations
``$O$ has information about $S$,'' ``$O^\prime$ has information
about $O$'' and so on, on itself. Having become circular, the
chain will fully imitate the circle of the loop of existences as
it appears on Figure~\ref{loop0}.
%this chain will give rise to the condition of consistency that can be
%met in the context of quantum theory.
This will not, however, lead to possibly contradictory questions
concerning the method of storage and manipulation of information
by the systems concerned, because such questions are meaningful
only in a different loop cut. Therefore, with a loop cut being
fixed so that it makes physics an information-based theory,
physical theory will obtain a consistent foundation and at the
same one will be aware of the explanatory limitations of the
theory and one will know how to tackle these limitations at a
future, separate stage of reflection: it will be necessary to pass
to a different loop cut.

\section{Fundamental notions}\label{defmeasurement}

We now focus attention on the loop cut in which physical theory is
based on information (Figure~\ref{loop01}). Our task in the
remainder of this chapter is to give definitions and postulates
necessary for the formal development of this view. In this section
we choose the language of the axiomatic system to be given in the
next section.

Three notions we do not define. Their meaning is not explained by
the theory and they stand in the information-based physical theory
as meta-theoretic like the notion of ensemble stands in the set
theory. These are: \textit{system}, \textit{information},
\textit{fact}.

Without defining what these words mean, we can however explain
\textit{how} they are used in the theory. Systems are fundamental
entities of the theoretic description. Any thing \textit{distinct}
from another thing can be treated as a system. It cannot be
defined by means of other systems or of any functions of systems.
This corresponds to the neo-Platonic notion of thing as explained
by the great Russian philosopher Alexei Fedorovich Losev
\cite{losev}.

The von Neumann cut between system and observer \cite{vN32} that
we already evoked in Section~\ref{rovsect} requires that no
particular system be given preferential treatment within the
theory. All systems are \textit{a priori} equivalent. In the
context of conventional quantum mechanics, this means that only
the descriptive purposes distinguish observers from physical
systems, and any observer is a physical system as well. The von
Neumann cut between observer and system is moved to position zero,
i.e. everything is a system and all systems are viewed on equal
grounds.

It is relatively easy to comprehend that the notion of system is
chosen as meta-theoretic, whereas it is all the more difficult to
accept intuitively the same choice for the notion of information.
However, it is a requirement of the loop cut. Let us first say
which information is \textit{not} under consideration here: the
primary notion of information does not mean the quantified,
measured, calculated information that we have, for example, in the
Shannon theory \cite{shannon}. All these aspects of information
come afterwards, when one attempts a translation of the
fundamental notion of information into mathematical terms of this
or that formalism. Information in question is the primary
substrate, which serves the purposes of interaction or
communication between systems. A neo-Leibnitzian could view the
system as the ontic monad and information as the epistemic monad.

Facts are acts of bringing about information, or information
indexed by the temporal moment of it being brought about. In the
second formulation enters the notion of time, which we did not
select as fundamental. Instead of doing so, we say that facts are
fundamental, and the facts that give rise to time. The latter
statement will be explained in the generally covariant context of
Part~\ref{cstar}. Prior to that, the theory will be
non-relativistic, and time as well as facts will be treated as
coming from outside the theory, therefore introduced by way of
additional axioms. The understanding of facts as acts of bringing
about information indexed by the moments of time brings this
notion close to the notion of phenomenon, and thus the loop of
existences between physics and information is not unlike one
between phenomenality and objectivity (Figure~\ref{loop02}). If
information that is brought about in a given fact refers to some
system (it is information \textit{about} the such and such
system), then this can be viewed as an instantiation of
\textit{intentionality} of the fact. Quite naturally, once a
particular philosophical system has been chosen, as in
Section~\ref{loopsect}, then all the key conventional
epistemological concepts, like intentionality for example, find
their counterparts in the language of this philosophical system.

In the physical theory fundamental notions ought to be given a
formal treatment. This is equivalent to saying that physical
theory is always built by means of a certain mathematics, as
illuminated by Wigner in the famous paper Ref.~\cite{unreas}. In
the formalism of Chapter~\ref{chaptreconstr}, systems will be
understood as physical systems $S$, $O$, etc., that are entities
of the theory. This is a translation of the fundamental notion of
system into a mathematical notion, and it comes at almost no
price. Not so with information. Information will be translated in
a very precise mathematical manner, namely the one introduced by
Shannon. According to Shannon, information is understood as
correlation between facts about systems. Correlation is a
mathematical term that involves registration of statistical
sequences of facts and later analyzing these sequences on the
subject of finding dependencies between the facts. None of the
latter things are relevant in the information-theoretic derivation
program: neither registration, nor the analysis which is made
backward-in-time or from an Archimedes's extra-theoretic point of
view \cite{tarchi}. A theory of these processes requires a
different loop cut. Therefore, to say that information is
correlation remains a pure translation of the fundamental notion
of information into mathematical terms, and not a definition of
information.

Facts are presented in the physical theory under the name of
measurement results. The question of their mathematical
representation thus becomes the question of what is measurement
and what is its result. We treat it along the lines of quantum
logic. We understand an elementary measurement as a binary, or a
yes-no, question. Result of the elementary measurement is a
particular answer to the yes-no question. This agrees with
Rovelli's idea in Ref. \cite{RovRQM}. A detailed argument for the
choice of yes-no questions as primitive measurements is provided
in chapter 13 of Beltrametti's and Casinelli's seminal book
\cite{bc} and will not be repeated here. We limit ourselves to
postulating this choice.

To compare with a different wording that exists in the literature,
take the approach proposed by \v{C}aslav Brukner and Anton
Zeilinger \cite{BZ}. Brukner and Zeilinger use the term
``proposition.'' In search of their motivation, Timpson
\cite{timpson} compares two formulations of Zeilinger's
fundamental principle for quantum mechanics expressed in an
article different from the above~\cite{ZeilFound}:
\begin{description}
    \item[FP1] An elementary system represents the truth value of
    one proposition.
    \item[FP2] An elementary system carries one bit of
    information.
\end{description}
By referring to bits (``binary units'') and to propositions at the
same time, Zeilinger implicitly suggests that, in his and
Brukner's derivation of the quantum formalism, one should treat
the following phrase as the postulate of what is measurement:
``Yes-no alternatives are representatives of basic fundamental
units of all systems.'' Although the initial wording making use of
the notion of proposition which appears to be different from
language of yes-no measurements, we see that this appearance is
misleading: Zeilinger in fact adopts the same choice of binary
questions as elementary measurements.

\section{First and second axioms}\label{axioms12}

\epigraph{Digo que no es il\'{o}gico pensar que el mundo es
infinito. Quienes lo juzgan limitado, postulan que en lugares
remotos los corredores y escaleras y hex\'{a}gonos pueden
inconcebiblemente cesar -- lo cual es absurdo. Quienes lo imaginan
sin l\'{i}mites, olvidan que los tiene el n\'{u}mero posible de
libros. Yo me atrevo a insinuar esta soluci\'{o}n del antiguo
problema: \textit{La biblioteca es ilimitada y
peri\'{o}dica}.}{J.L.~Borges \selectlanguage{french}\og La
Biblioteca de Babel\fg}\epigraph{I say that it is not illogical to
think that the world is infinite. Those who judge it to be limited
postulate that in remote places the corridors and stairways and
hexagons can conceivably come to an end---which is absurd. Those
who imagine it to be without limit forget that the possible number
of books does have such a limit. I venture to suggest this
solution to the ancient problem: \textit{The library is unlimited
and cyclical}.}{J.L.~Borges \og The Library of Babel\fg}

After the selection of the fundamental notions in the previous
section that provides the language in which one can formulate the
axiomatic system, the time is ripe to give the
information-theoretic axioms themselves. It is the purpose of this
section.

The axioms must be such as to permit a clear and unambiguous
translation of themselves into formal terms, and this translation
must then lead to reconstruction of the structure of quantum
theory. However, first we formulate the axioms without making
reference to any particular formalism.\begin{quote}
\begin{axiom} There is a maximum amount of relevant information that can be
extracted from a system. \label{ax1}\end{axiom}
\addtocounter{axiomcounter}{1}
\begin{axiom} It is always possible to acquire new information about a system.
\label{ax2}\end{axiom}
\end{quote}
It seems that the axioms contradict each other. Indeed, at the
first sight a paradox is straightforward: Axiom~\ref{ax1} says
that the quantity of information is finite, while from
Axiom~\ref{ax2} follows that it must be infinite, because we can
always obtain some new information. But there is no contradiction:
the key is hidden in the use of the term ``relevant.'' There is no
valuation on the set of questions that would assign to each
question the amount of information that it brings about without
taking into account other questions which have been asked and
which create the context for the definition of relevance. In other
words, the amount of information is not a function of one
argument. Let us explain this in more detail.

In the conventional quantum mechanics it is from the past or the
future of a given experiment, in particular from the intentions of
the experimenter, that one can learn which information about the
experiment is relevant and which is not. What is relevant can
either be encoded in the preparation of the experiment or selected
by the experimenter later on. Both the preparation and the
posterior selection require \textit{memory}: the experimenter
compares information that was brought about in facts indexed by
different values of the time variable and decides which
information to keep and which to throw away as irrelevant. Fact is
a fundamental notion belonging to the meta-theory, and it is
therefore natural to expect, because relevance is related to
facts, that what is relevant and what is irrelevant cannot be
deduced within the theory. In every formalization of the
{axioms\label{revrefm}}, we need to give a separate definition of
relevance and the justification of such a definition will be
meta-theoretic. This is indeed what we do in the case of
Definition~\ref{revd}.

Let us repeat: the experimenter, as someone who imposes a
criterion of relevance, needs to be supplied with memory. In other
terms, his decisions are \textit{contextual}: the context here is
the sequence of facts given to the experimenter. Because facts are
meta-theoretic, we call this contextuality \textit{meta-theoretic
contextuality}. The notion of meta-theoretic contextuality must be
distinguished from the notion of intratheoretic non-contextuality
discussed in the next section.

Axioms \ref{ax1} and \ref{ax2} therefore refer, the first one to
the amount of relevant information, the second one to the fact
that new information as such can always be generated, perhaps at
the price of rendering some other available information obsolete
and thus irrelevant. In this interpretation there is no
contradiction between the axioms.

%To translate this into the language of our theory by using the
%fundamental notions only, note that acts of bringing about
%information are indexed by the time variable, and this we called
%\textit{facts}. It is from the temporal sequence of facts that one
%learns which information that had been brought about will be used
%and which will be discarded. Information that had been brought
%about and later used for description of other facts is referred to
%as relevant. Importantly, neither the temporality of facts, nor
%usage or non-usage of previous facts can be deduced by the
%theoretical means. The notion of relevance therefore does not
%arise from the formalism which solely describes, conventionally,
%the measurement within the context of the experiment, or, in our
%language, describes interconnections between facts and
%information, but not the temporality of facts or usage of the
%information. All that can be deduced from the theory is that there
%is \textit{some} relevant information. What is relevant is
%reflected in the choice of questions that are asked by I-observer
%and requires a separate definition that can be put in formal
%terms. This problem will be discussed again in
%Section~\ref{rovellirigorous} and its connection with time in
%Section~\ref{sect44}.

To give an illustration, imagine for a moment the actual
experimenter. He first makes a measurement with some fixed
measuring apparatus, then changes the apparatus and make another
measurement with another apparatus. Clearly, one would say that he
obtained some new information about the system, so Axiom~\ref{ax2}
is meaningful and justified. What about Axiom~\ref{ax1}?
Axiom~\ref{ax1} forbids the setting in which the experimenter
could change measuring apparata endlessly and each time get some
new information, also \textit{keeping all the old one}.
Axiom~\ref{ax1} tells that the information obtained in earlier
measurements must now become irrelevant. This axiom, therefore,
has two implications: first, it says that in any one act of
bringing about information, only a finite quantity of information
can be generated; second, it says that information may ``decay''
with time, in such a way that one can never have infinite relevant
information about the system, although one can still always learn
something new about it according to Axiom~\ref{ax2}.

To conclude this section, compare our axioms with a set of two
axioms proposed by Brukner and Zeilinger \cite{BZ}.

\setcounter{axiomcounter}{1}\begin{axiom}[Brukner and Zeilinger]
The information content of a quantum system is finite.
\end{axiom}\addtocounter{axiomcounter}{1}
\begin{axiom}[Brukner and Zeilinger]
Introduce the notion of total information content of the system;
state that there exist mutually complementary propositions; state
that total information content of the system is invariant under a
change of the set of mutually complementary propositions.
\end{axiom}\addtocounter{axiomcounter}{1}

Observe a telling analogy between the first axioms, apart from
Brukner's and Zeilinger's use of the term ``information content''
which suggests that they consider it as a property of the system
in itself, without bringing in the relation with another system
that plays the role of observer. Therefore, the term ``information
content'' has ontological connotations, unlike our formulation
that underlines the relational character of information and stays
within the boundaries of epistemology. Also, in spite of the
analogy between the ideas, as for the derivation of quantum
mechanics which follows the choice of axioms, Brukner and
Zeilinger opt for a technique different from ours. Following
Rovelli, we have the ambition to derive the formalism of quantum
theory from the axioms by the methods of quantum logic; and to go
further than Rovelli because he does not show a way to deduce most
of the structure, for instance the superposition principle, apart
from introducing it as a supplementary axiom. Embarking on where
Rovelli leaves, Christopher Fuchs
\cite{FuchsInLight,FuchsLittleMore} uses a decision-theoretic
Bayesian approach to derive the superposition principle. He refers
to Rovelli's paper in his own, and one is left free to suggest
that many of his axiomatic assumptions, on which he does not
clearly comment, might be similar to Rovelli's, apart from the key
issue of how to define measurement. Fuchs insists on the
fundamental character of positive operator-valued measures (POVM)
and postulates that POVM formally describe measurement. This
contradicts our choice in Section~\ref{defmeasurement} and,
indeed, may not seem intuitively evident. One is then tempted to
look for ways to avoid making this assumption; thus, even if we
dismiss the necessity to \textit{define} measurement as POVM,
there still remains an opportunity to introduce POVM in the
theory, which in itself is a virtue since it permits to establish
theorems of quantum computation following the guidelines presented
in Section~\ref{qcompsect}. POVM have a natural description as
conventional von Neumann measurements on ancillary system
\cite{Peres}, and thus to Rovelli's axiomatic derivation of the
Hilbert space structure one may try to add an account of
inevitability of ancillary systems and naturally obtain from this
the POVM description, which, in turn, will allow to follow Fuchs's
derivation. This will indeed be our plan in Section \ref{sect44}.

Brukner and Zeilinger proceed differently. If information is
primary, they argue, then any formalism must deal with information
and not with some other notions. We find it difficult to disagree
with this. Then Brukner and Zeilinger choose not to reconstruct
the physical theory, but instead to build an \textit{information
space} where they apply their axioms and use the formalism to
deduce testable predictions. Brukner and Zeilinger do not refer in
their derivation to the Hilbert space nor to the physical state
space. In part because of their choice to build a completely new
theory, Brukner and Zeilinger are forced into postulating
properties of mutually complementary propositions that are hardly
apprehensible in the conventional quantum mechanical language.
Namely, they postulate the ``homogeneity of parameter space,''
while---as we shall see---in the formalism of orthomodular
lattices one must postulate continuity of a certain well-defined
function. Of course, to Brukner's and Zeilinger's notions one can
always find counterparts in the conventional language of numeric
fields, Hilbert spaces and states; but their restriction to the
terminology of abstract information space leads to the
complications of language and renders the formalism less
transparent in use. For the reason of clarity of language we
reconstruct quantum theory in its standard form instead of giving
new names to objects that are essentially the same as all the
conventional ones.

\section{I-observer and P-observer}\label{ipobservers}

At this stage we have introduced two axioms and three fundamental
notions. We have also discussed the notion of relevance. A
question arises: Are all the terms used in the formulation of the
axioms covered by the three fundamental notions or in order to
understand the axioms one needs to employ some other notions? This
is a crucial stage where consistency of the theory is at stake.

Let us reread the formulations of Axioms~\ref{ax1} and~\ref{ax2}.
The concept of \textit{amount of information} refers to the
mathematical representation of the fundamental notion of
information as Shannon information. Consequently, this does not
raise any questions due to the commonly accepted mathematical
definition given by Shannon, where information is understood as a
measure of the number of possibly occupied states against the
total number of states. Admittedly, in our approach this latter
phrase is not a definition \textit{per se}, but it gives an
unequivocal mathematical meaning to the notion of information.

Axiom~\ref{ax1} also contains a reference implicit for someone who
reads (correctly) this axiom as a statement of the ordinary
language rather than a mathematical statement: the reference in
question is to the subject who extracts information, and it
appears in the clause ``can be extracted from.'' Note that this
reading belongs to the ordinary language, and we are therefore
obliged to analyze it in the context of the loop cut and
separation between theory and meta-theory. The same reference is
contained in Axiom~\ref{ax2} which says ``it is always possible
\textit{to acquire}\ldots'' Then the question is: Who is the one
who acquires information? By one half the answer to this has
already been given. As we said in the discussion of the notion of
system, everything is a system (apart from the whole
{Universe\label{unino}} which cannot be distinct from something).
Then the ``subject who acquires information'' is also a system in
the sense of quantum theory. With von Neumann's cut being put at
level zero, such situation is nothing else but to claim that
quantum theory is universal. Next, from the point of view of the
ordinary language, we still see a difference between the ``subject
acquiring information'' and a system that this information is
about. Language introduces an apparent dissymmetry between the
two. Where does this dissymmetry arise from in the theory and what
role does it play?

In Rovelli's phrase quoted in Section~\ref{rovsect} we stated that
``there is no physical interaction between $O^\prime$ and $S-O$
during the interval $t_0-t_1$.'' Then, if $O^\prime$ is a system
as any and translates into the language of physics as a physical
system as any, its status becomes unclear. Indeed, were $O^\prime$
a physical system, then it must have interacted with other systems
just as all physical systems do. But there is precisely no such
interaction. It means that for the purposes of description of the
interval $t_0-t_1$ and of the physical system $S-O$, the system
$O^\prime$ is not treated as physical system obeying the laws of
physics. We shall say that the system $O^\prime$ is
\textit{effectively meta-theoretic}. It means that we have chosen
to move the von Neumann cut to the position between $S-O$ and
$O^\prime$, and this only for the purposes of description of the
fact $\M$. The only function of the system $O^\prime$ which is
left after we have removed its physical function is that it is an
\textit{informational agent}, i.e. an accumulator of information
or, to match the language of the axiom, the system which
\textit{acquires} information from the fact $\M$. Because we chose
$O^\prime$ at random among all systems, we conclude that any
system can become effectively meta-theoretic for the purposes of a
fixed descriptive act of bringing about information. Therefore,
any system can be represented as a purely physical system plus an
informational agent. By definition, this distinction does not
interfere with any physical processes, because acquisition of
information is, not a theoretic but a meta-theoretic concept. Thus
the distinction, too, is meta-theoretic and bears, in the case of
each particular system, on one given fact only. In a description
of another fact, the system which has been previously treated as
an informational agent only must now again be treated as a
physical system as any.

Let us give another motivation to the distinction that we have
just made and then introduce the terminology. As one can expect,
our observation that systems are sometimes effectively
meta-theoretic will lead, not only to novel terminology, but also
to tangible theoretic results that will directly bear upon the
physical content of the theory. The reason why it happens so is
that the way in which we construct quantum theory is based on
information, and \textit{a priori} any restriction imposed on the
functioning of the concept of information must lead to constraints
on the content of the theory.

In the everyday work of a physicist who uses conventional quantum
mechanics, one is usually interested in information about
(knowledge of) the chosen system and one disregards particular
ways in which this information has been obtained. This is a
manifestation of the cut of the loop that we discussed in
Section~\ref{loopsect}. All that counts is relevant knowledge and
relevant information. Because of this, one usually pays no
attention to the very process of interaction between the system
being measured and the measuring system, and one treats the
measuring system as a meta-theoretic, i.e. non-physical,
apparatus. Correspondingly, the loop cut is the one on
Figure~\ref{loop01}. To give an example, for some experiment a
physicist may need to know the proton mass but he will not at all
be interested in how this quantity had been measured (unless he is
a narrow specialist whose interest is in measuring particle
masses). Particular ways to gain knowledge are irrelevant, while
knowledge itself is highly relevant and useful. Some of the
experiments where one is interested in the measurement as a
physical process, thus falling in the domain of the loop cut on
Figure~\ref{loop1}, are discussed in Ref.~\cite{Mensky}. In the
present derivation of quantum theory we assume a loop cut such
that physics is viewed as based on information, therefore
rendering the measurement details irrelevant.

An experimenter, though, always operates in both loop cuts at
once, i.e. he uses physics which is an information-based theory of
the first loop cut, but he also keeps in mind that ``information
is physical'' \cite{Landauer}. The last phrase means that there
always is some physical support of information, some hardware. The
necessity of the physical support requires that we carefully
justify the division between theory and meta-theory in the
selected loop cut: we first abstain from disregarding the
measurement interaction and then show how one can neglect the fact
that the measuring system is physical. This will allow to leave to
the observer solely the role of informational agent and to
formulate the physical theory only in terms of information.

The statement that any system is a physical system but also an
informational agent corresponds to making a formal distinction,
for each system, between these two roles. Call \textit{any} system
$O$ an observer. Then the observer consists of an informational
agent (``I-observer'') and of the physical realization of the
observer (``P-observer''). In the uncut loop, there is no
I-observer without P-observer. Reciprocally, there is no sense in
calling P-observer an observer unless there is I-observer
(otherwise P-observer is just a physical system as any). Two
components of $O$ are not in any way separate from each other; on
the contrary, these are merely two viewpoints that one adopts for
the needs of a given theoretical description. One has to select
the viewpoint before describing any given fact $\M$: if the
selection is for I-observer, then $O$ is treated as
meta-theoretic; if for P-observer, then $O$ is a physical system,
object of study in the physical theory.

The key point of making the distinction between I-observer and
P-\-observ\-er is that only measurement results, or the
information brought about in facts, count. We transform this
principle into an axiom that will be further discussed in
Section~\ref{bornrule}.

\begin{axiom}[``no metainformation'']\label{ax3}
If information $I$ about a system has been brought about, then it
happened without bringing in information $J$ about the fact of
bringing about information $I$.\end{axiom}

So formulated, Axiom~\ref{ax3} states that information, when it is
brought about in a fact, is ``self-sufficient,'' meaning that it
does not entail bringing about metainformation about how this
particular fact occurred. Facts bring about information that is
clearly demarcated (`\textit{this} information') and thus is
independent from other information that may be brought about in
some other facts, but \textit{a fortiori} not in the same one.

Looking at the same axiom from a different angle, let us
reformulate it in the language of measurements. It then states
that the details of measurement as physical process do not count
in making this process a measurement. This is a form of
non-contextuality that we call intratheoretic: information does
not depend on the context that belongs to the physical theory. As
we said in Section~\ref{axioms12}, intratheoretic
non-contextuality must be distinguished from meta-theoretic
contextuality, which holds in virtue of Axioms~\ref{ax1} and
\ref{ax2}. A reformulation of Axiom~\ref{ax3} then goes:

\begin{axiom}[``intratheoretic non-contextuality'']
If information is obtained by an observer, then it is obtained
independently of \textit{how} the measurement was conducted
physically, i.e. independent of the measurement's context internal
to physical theory.
\label{noncontext}\end{axiom}\addtocounter{axiomcounter}{1}

\chapter{Elements of quantum logic}\label{chapeql}

In this chapter we introduce the quantum logical formalism of the
theory of orthomodular lattices in a way suited for the program of
deriving the formalism of quantum theory from
information-theoretic principles. Most of the following exposition
is based on \cite{kalmbach83}. Several results are taken from the
seminal book on lattice theory \cite{maeda}. Each section opens
with a brief non-technical summary.

\section{Orthomodular lattices}

\begin{chsum}
This section introduces a key concept of the orthodox quantum
logic: orthomodular lattice. A lattice can be viewed as a set of
logical statements such that, for any two elements of the lattice,
two new elements formed by putting between the two old ones the
conjunction \emph{and} or the conjunction \emph{or}, also belong
to the lattice. Lattices can be distributive or Boolean, like in
classical logic; modular, which is weaker than distributive; and
orthomodular, which is yet weaker than modular. Orthomodularity is
a property defined with the help of the notion of orthogonality:
to each element corresponds a unique other element that
``complements'' it in the lattice in the sense of, roughly
speaking, having all the properties opposite to the properties of
the original element.
\end{chsum}

\begin{defn}
A \textbf{\textrm{lattice}} $\mathcal{L}$ is a partially ordered
set in which \textit{any} two elements $x,y$ have a supremum
$x\vee y$ and an infimum $x\wedge y$.\label{deflattice}
\end{defn}
Equivalently, one can require that a set $\mathcal{L}$ be equipped
with two idempotent, commutative, and associative operations
$\vee,\wedge$ : $\mathcal{L}\times\mathcal{L}\rightarrow
\mathcal{L}$, which satisfy $x\vee(y\wedge x)=x$ and $x\wedge
(y\vee x)=x$. The partial ordering is then defined by $x\leq y$ if
$x\wedge y =x$. The largest element in the lattice, if it exists,
is denoted by $1$, and the smallest one (if exists) by $0$.

\begin{defn}A lattice is called \textbf{complete} when every subset of
$\mathcal{L}$ has a supremum as well as an infimum.\label{complete}
\end{defn}
\begin{lem}
Complete lattice always contains elements $0$ and $1$.
\end{lem}
\begin{proof}
Element $0$ can be defined as infimum of all elements of
$\mathcal{L}$ and element $1$ as their supremum. Both are well
defined in virtue of completeness of the lattice.\end{proof}
\begin{defn}An \textbf{atom} of lattice $\mathcal{L}$ is an element
$a$ for which $0\leq x\leq a$ implies that $x=0$ or $x=a$. A
lattice with $0$ is called \textbf{atomic} if for every $x\neq 0$
in $\mathcal{L}$ there is an atom $a\neq 0$ such that $a\leq x$.
\end{defn}

\begin{defn}The lattice is said to be \textbf{distributive} if
\begin{equation}x\vee(y\wedge z)=(x\vee y)\wedge(x\vee z).\label{distrib}\end{equation}
\end{defn}

One can weaken the distributivity condition by requiring
(\ref{distrib}) only if $x\leq z$. This leads to the property of
modularity.

\begin{defn}The lattice is said to be \textbf{modular} if
\begin{equation}x\leq z \Rightarrow x\vee (y\wedge z)=(x\vee
y)\wedge z \quad\forall y.\label{modular}\end{equation}
\end{defn}
A canonical example of a modular lattice is the collection $L(V)$
of all linear subspaces of a vector space $V$ over an arbitrary
field $\mathbb{D}$ \cite{land}. The lattice operations are
$x\wedge y\equiv x\cap y$ and $x\vee y\equiv x + y$, where $x+y$
is the linear span of $x$ and $y$ for all linear subspaces
$x,y\subset V$. Equivalently, one can say that the partial order
is given by inclusion. Evidently, lattice elements $1=V$ and
$0=\emptyset$, the empty set.

\begin{defn} An \textbf{orthocomplementation} on lattice $\mathcal{L}$
is a map $x\mapsto x^{\bot}$, satisfying for all $x,y\in
\mathcal{L}$\label{oc}
\begin{description}
  \item[(i)] $x^{\bot\bot}=x,$
  \item[(ii)] $x\leq y \Leftrightarrow y^{\bot}\leq x^{\bot},$
  \item[(iii)] $x\wedge x^\bot =0,$
  \item[(iv)] $x\vee x^\bot =1.$
\end{description}
\end{defn}
\noindent A lattice with orthocomplementation is called an
\textbf{orthocomplemented lattice}. Two lattices $\mathcal{L}_1$
and $\mathcal{L}_2$ are isomorphic if there exists an isomorphism
between them that preserves the lattice structure. If
$\mathcal{L}_1$ and $\mathcal{L}_1$ are orthocomplemented
lattices, then they are isomorphic if the isomorphism respects the
orthocomplementarity relation.

From the definition immediately follow de Morgan laws
\begin{equation} 1^\bot=0;\quad 0^\bot=1;\quad
(x\vee y)^\bot=x^\bot\wedge y^\bot;\quad (x\wedge
y)^\bot=x^\bot\vee y^\bot.
\label{demorgan}\end{equation}

By imposing on an orthocomplemented lattice respectively the
distributive law (\ref{distrib}) and the modular law
(\ref{modular}), which is weaker than the distributive law, one
arrives at the following definitions.

\begin{defn}
A distributive orthocomplemented lattice is called a
\textbf{Bool\-ean algebra}.
\end{defn}

\begin{defn} An orthocomplemented lattice $\mathcal{L}$ is called
\textbf{orthomodular} if condition (\ref{modular}) holds for $y=x^\bot$,
that is,
\begin{equation}x\leq z \Rightarrow x\vee (x^\bot \wedge
z)=z.\label{orthomodular}
\end{equation}
\label{deforthomodular}\end{defn}

It is useful to give the following reformulation of the condition
of orthomodularity.

\begin{lem}
An orthocomplemented lattice $\mathcal{L}$ is orthomodular if and
only if $x\leq z$ and $x^\bot \wedge z=0$ imply
$x=z$.\label{lemorthomod}
\end{lem}
%\begin{proof}
\begin{proof}If the lattice is orthomodular, i.e. Equation~(\ref{orthomodular})
holds, and if $x^\bot \wedge z=0$, then $z=x \vee 0=x$. To prove
the converse statement, it suffices to show that if the lattice is
not orthomodular then there exist elements $x$ and $z$ such that
\begin{equation}
x\leq z,\quad x^\bot \wedge z=0,\quad x\neq z.\label{plusik1}
\end{equation}
Let us use the notation $x<z$ if $x\leq z$ and $x \neq z$. We can
then rewrite Equation~(\ref{plusik1}) as
\begin{equation}
x< z,\quad x^\bot \wedge z=0.\label{palka}
\end{equation}
Assume that the lattice is not orthomodular. According to the
Definition~\ref{deforthomodular} there exist elements $y$ and $z$
such that
\begin{equation}
y\leq z,\quad y\vee(y^\bot \wedge z)\neq z.\label{goriz}
\end{equation}
Now recall that on any lattice holds \cite[Chapter 2, Section
4]{cohn}\symbolfootnote[2]{We thank Prof.~V.A.~Franke for this
reference and the idea of proof.}
\begin{equation}
a\leq b\Rightarrow (c\wedge b)\vee a\leq (c\vee a)\wedge b \quad\forall c.
\label{multipl}
\end{equation}
Put in (\ref{multipl}) $a=y$, $b=z$, $c=y^\bot$. Follows that
\begin{equation}
(y^\bot \wedge z)\vee y\leq (y^\bot \vee y)\wedge z.\label{multik1}
\end{equation}
In the right-hand side replace $y^\bot \vee y$ by $1$, and
$1\wedge z=z$. Equation (\ref{multik1}) then takes the form
\begin{equation}
(y^\bot \wedge z)\vee y\leq z.\label{multik2}
\end{equation}
From equations (\ref{multik2}) and (\ref{goriz}) one obtains that
\begin{equation}
(y^\bot \wedge z)\vee y< z.\label{multik3}
\end{equation}
On the other hand, from de Morgan laws (\ref{demorgan}) one has
\begin{eqnarray}
z\wedge(y\vee(y^\bot \wedge z))^\bot=z\wedge (y^\bot \wedge(y^\bot\wedge z)^\bot)=\nonumber\\
z\wedge (y^\bot\wedge(y\vee z^\bot))=(z\wedge y^\bot)\wedge(y\wedge z^\bot)=\nonumber\\
(z\wedge y^\bot)\wedge(z\wedge y^\bot)^\bot=0.\label{multik4}
\end{eqnarray}
Now put $x=y\vee(y^\bot \wedge z)$. Equations (\ref{multik3}) and (\ref{multik4})
can be rewritten as
\begin{equation}
x<z,\quad x^\bot z=0.
\end{equation}
This is exactly the condition (\ref{palka}) that we need to
obtain.\end{proof}
%\end{proof}
%Conversely, if $x\leq z$, using a tautology $z \vee (x^\bot \wedge
%z)=z$, we obtain $x\vee (x^\bot \wedge z)\leq z$. One infers
%$(x\vee(x^\bot\wedge z))^\bot\wedge z=0$.
%Now apply the condition stated in lemma with $x$ replaced by
%$x\vee (x^\bot \wedge z)$. For this latter expression it follows from
%the lemma conditions that $x\vee (x^\bot \wedge z)=z$, and it is exactly precisely
%the required result.
We close this section with a definition of reducibility of
lattices.
\begin{defn}
The center of an orthocomplemented lattice $\mathcal{L}$ is
$$
C(\mathcal{L})=\{c\in \mathcal{L}\: | \: x=(x\wedge c)\vee
(x\wedge c^\bot)\: \forall x\in \mathcal{L}\}.
$$
\end{defn}

\begin{defn}A lattice is called \textbf{reducible} if it is (isomorphic to)
a nontrivial Cartesian product
$\mathcal{L}=\mathcal{L}_1\times\mathcal{L}_2$ with lattice
operations defined componentwise. If not, it is called
\textbf{irreducible}.
\end{defn}

\begin{lem}
The center $C(\mathcal{L})$ of an orthomodular lattice
$\mathcal{L}$ is its Boolean subalgebra.
\end{lem}

\begin{lem}
An orthocomplemented lattice is irreducible if and only if its
center is trivial, i.e. $C(\mathcal{L})=\{0,1\}$.
\end{lem}

\section{Field operations and spaces}\label{sectfield}

\begin{chsum}
This section introduces the notion of Hilbert space. We first
define automorphisms in a field, be it a numeric field like real
numbers or an abstract algebraic structure with the same
properties. The Hilbert space is then a space which is supplied
with an internal product that behaves ``rationally'', in a certain
mathematically defined way, with respect to the automorphism of
the underlying field.
\end{chsum}

Let $\mathbb{D}$ be a field, i.e. a commutative ring with addition
and multiplication such that, bar the unity element of the
additive group, one obtains a multiplicative group. A bijective
map $\theta:\mathbb{D}\mapsto\mathbb{D}$ is an
\textbf{anti-\-automorphism} if $\forall a,b\in \mathbb{D}$
\begin{equation}\theta(a+b)=\theta(a)+\theta(b)\;\mathrm{and}\;\theta(a\cdot
b)=\theta(b)\cdot\theta(a).\end{equation} The map $\theta$ is
\textbf{involutory} if $\theta^2$ is the identity. Let $\theta$ be
an involutory anti-\-auto\-morphism of the field $\mathbb{D}$ and
$V$ a vector space over $\mathbb{D}$. A map $f:V\times V\mapsto
\mathbb{D}$ is called a $\theta$\textbf{-sesquilinear} form on $V$
if $\forall x,x_1,x_2,y,y_1,y_2\in
V\;\mathrm{and}\;\alpha_1,\alpha_2,\beta_1,\beta_2\in \mathbb{D}$
one has
\begin{eqnarray}f(\alpha_1 x_1+\alpha_2 x_2,y)=\alpha_1
f(x_1,y)+\alpha_2 f(x_2,y),\nonumber\\ f(x,\beta_1 y_1+\beta_2
y_2)=f(x,y_1)\theta(\beta_1)+f(x,y_2)\theta(\beta_2).
\end{eqnarray}
Let $f$ be a $\theta$-sesquilinear form on $V$.
Then $f$ is called \textbf{Hermitian} if \begin{equation}\theta(f(x,y))=f(y,x)\end{equation}
and \textbf{definite} if $f(x,x)=0$ implies $x=0$. A Hermitian,
definite $\theta$-sesqui\-linear form is called a
$\theta$\textbf{-product}.

Now recall the definitions of Banach and Hilbert spaces.

\begin{defn}
A \textbf{Banach space} is a vector space $V$ over the field
$\mathbb{D}$ with a norm which is complete with respect to the
metric $d(x,y)\equiv \| x-y\|$ on $V$.
\end{defn}
\begin{defn}
A \textbf{Hilbert space} $\BH$ over $\mathbb{D}$ is a Banach space
whose norm comes from a $\theta$-product $f(x,y)$ for $x,y\in
\BH$, which has the following properties:
\begin{description}
    \item[(i)] $f(\alpha x+\beta y,z)=\alpha f(x,z)+\beta f(y,z),$
    \item[(ii)] $\theta(f(x,y))=f(y,x),$
    \item[(iii)] $\| x\| ^2=f(x,x).$
\end{description}
\label{hilbspdef}\end{defn}
\begin{defn}
A pre-Hilbert space is a normed linear space over $\mathbb{D}$
with its norm satisfying the parallelogram law: \begin{equation}\|
x+y\|^2+\| x-y\|^2=2(\| x\|^2+\|
y\|^2).\label{parallelogram}\end{equation}
\end{defn}
A pre-Hilbert space carries a natural $\theta$-product $f(x,y)$
defined as
\begin{equation}f(x,y)=\frac{\| x+y\|^2-\|
x-y\|^2}{4}\end{equation} and can be completed with respect to its
norm topology up to a Hilbert space that will contain the initial
pre-Hilbert space as a dense subspace. All Hilbert spaces satisfy
the parallelogram law (\ref{parallelogram}) and therefore are
pre-Hilbert spaces as well.

%\section{Irreducibility of lattices}

\section{From spaces to orthomodular lattices}

\begin{chsum}
In this section we characterize the lattice of closed subspaces of
the Hilbert space. It is found to be complete, atomic and
orthomodular.
\end{chsum}

The \textit{raison d'\^{e}tre} of the theory of orthomodular
lattices is to answer the double-way question of, firstly, how to
characterize a lattice built of subspaces of the Hilbert space;
secondly, how to characterize a lattice built of subspaces of a
vector space $V$ so that this space be a Hilbert space. One would
also like to find such a characterization of the lattice that, the
space $V$ being built upon a coordinatizing field $\mathbb{D}$,
$\mathbb{D}$ will equal either $\mathbb{R}$, $\mathbb{C}$, or
$\mathbb{H}$.

We start with characterizing the lattice $\mathcal{L}(\BH)$ that
one obtains given the Hilbert space $\BH$. Let $(\,\cdot\,
,\,\cdot\,)$~: $V\times V$ be a Hermitian form on a Banach space
$V$, defined relative to an involution $\theta$, and $L(V)$ the
lattice of all subspaces of $V$. For each $x\in L(V)$ one defines
$x^\bot\equiv\{\Psi\in V\,|\,(\Psi,\Phi)=0\; \forall\,\Phi\in x
\}$. $x^\bot$ is an element of $L(V)$ as well. One can easily see
that $x^{\bot\bot\bot}=x^\bot$ but in general $x\leq
x^{\bot\bot}$, rather than the equality required for
orthocomplementation in Definition~\ref{oc}. Therefore $L(V)$ is
not an orthocomplemented lattice.

As a remedy, consider the lattice $\mathcal{L}(V)$ of orthoclosed
subspaces of $V$, i.e. $x\in L(V)$ lies in $\mathcal{L}(V)$ if and
only if $x= x^{\bot\bot}$. The lattice operation $\wedge$ is the
same as in $L(V)$, but $\vee$ in $\mathcal{L}(V)$ is defined by
$x\vee y=(x+y)^{\bot\bot}$, which is the smallest orthoclosed
subspace containing $x$ and $y$. The symbol $+$ designates a
linear sum of subspaces. Lattice $\mathcal{L}(V)$ is complete
independently of the dimension of $V$ and is modular if and only
if $V$ is finite-dimensional.  Even in the finite-dimensional
case, $^\bot$ need not be an orthocomplementation on
$\mathcal{L}(V)$. It is straightforward, however, to check the
following necessary and sufficient condition.

\begin{prop} The map $x\mapsto x^\bot$ is an orthocomplementation
on $\mathcal{L}(V)$ if and only if $(x+x^\bot)^\bot=0$ for all
$x\in\mathcal{L}(V)$, which is equivalent to the property
$(\Psi,\Psi)=0\Leftrightarrow\Psi=0$ or to requiring that
$(\,\cdot\, ,\,\cdot\,)$ be a $\theta$-product. If in addition
$x+x^\bot$ is orthoclosed (implying $x+x^\bot=V$) for all $x\in
\mathcal{L}(V)$, then $\mathcal{L}(V)$ is orthomodular.
\label{prehilblat}\end{prop} \begin{proof} We shall prove only the
second clause of the lemma in the finite-dimensional case. For
infinite dimension we prove directly Lemma~\ref{hilblatt}.

To show that the additional assumption implies orthomodularity,
note that on this assumption, for any $x$ one has $z=z\wedge
1=z\wedge(x+x^\bot)$. If $x \leq z$, this equals $x+z\wedge
x^\bot$ by the modular law (\ref{modular}) in $\mathcal{L}(V)$,
with $y=x^\bot$. Taking the double orthocomplement of the equation
$z=x+z\wedge x^\bot $ thus found yields $z^{\bot\bot}=z$ for the
left-hand side (since $z\in \mathcal{L}(V)$ by assumption) and
$(x+z\wedge x^\bot)^{\bot\bot}=x\vee(z\wedge x^\bot)$ by the
definition of $\vee$ in $\mathcal{L}(V)$. This proves the
orthomodular law (\ref{orthomodular}).\end{proof}

\begin{lem}\label{hilblatt}
The lattice $\mathcal{L}(\BH)$ of all closed subspaces of a
Hilbert space is complete, atomic, and orthomodular.
\end{lem}
\begin{proof} In the finite-dimensional case proof follows
directly from Proposition~\ref{prehilblat}. We now give a general
proof that can also be applied in the infinite-dimensional case.

Recall the following properties of Hilbert spaces:
\begin{description}
  \item[1)] Any closed subspace of the Hilbert space is itself a Hilbert space.
  \item[2)] In every Hilbert space there exists a complete orthonormal basis.
  \item[3)] If in a Hilbert space one is given a certain set of orthonormal
vectors, it is always possible to complete it by more vectors, up
to a complete orthonormal basis. \item[4)]If one divides the
complete orthonormal basis of space $\BH$ into two subsets of
orthonormal vectors and then one considers linear closures of each
set, one obtains two Hilbert subspaces $V$ and $V^\bot$ such that
\begin{equation}V\cup V^\bot =\BH,\quad V\cap V^\bot =0,\end{equation}
where $V\cup V^\bot$ is a linear closure of $V$ and $V^\bot$, and
$V\cap V^\bot$ their intersection.
\end{description}

Now let $V_1$ and $V_2$ be two closed subspaces of the Hilbert
space $\BH$ such that
\begin{equation}\label{plusik}
V_1\subseteq V_2,\quad V_2\cap V_1^\bot =0.
\end{equation}
We must prove that \begin{equation}V_1=V_2.\end{equation} Indeed,
$V_1$ is itself a Hilbert space. Consider its complete orthonormal
basis $\mathcal{A}$. In virtue of (\ref{plusik}), all vectors of
$\mathcal{A}$ belong also to $V_2$. Add to $\mathcal{A}$ a set
$\mathcal{B}$ such that it completes $\mathcal{A}$ in $V_2$ to a
complete orthonormal basis of the latter. This full basis is now
$\mathcal{A}\cup \mathcal{B}$. Further, add to $\mathcal{A}\cup
\mathcal{B}$ a set $\mathcal{C}$ of orthonormal vectors which
completes it to the full orthonormal basis of the Hilbert space
$\BH$. The basis in $\BH$ has the form $\mathcal{A}\cup
\mathcal{B}\cup \mathcal{C}$.

Apply Property 4 of Hilbert spaces listed above. Divide the basis
$\mathcal{A}\cup \mathcal{B}\cup \mathcal{C}$ into two sets,
namely $\mathcal{A}$ and $\mathcal{B}\cap \mathcal{C}$. Consider
their linear closures, respectively $V(\mathcal{A})$ and
$V(\mathcal{B}\cap \mathcal{C})$. Follows that \begin{equation}
V^\bot(\mathcal{A})=V(\mathcal{B}\cap \mathcal{C}), \quad
V(\mathcal{A})\cup V(\mathcal{B}\cap \mathcal{C})=\BH,\quad
V(\mathcal{A})\cap V(\mathcal{B}\cap
\mathcal{C})=0.\label{krestik}
\end{equation}
By definition $\mathcal{A}$ is a complete orthonormal basis in subspace $V_1$,
and consequently $V(\mathcal{A})=V_1$. From this and (\ref{krestik}) follows
that $V_1^\bot=V(\mathcal{B}\cap \mathcal{C})$. Also by construction
$V_2=V(\mathcal{A}\cap \mathcal{B})$, where the right-hand side means linear closure
of the vector set $\mathcal{A}\cap \mathcal{B}$. Now let
\begin{equation}V_2\cap V_1^\bot=0,\end{equation}that is
\begin{equation}V(\mathcal{A}\cap \mathcal{B})\cap V(\mathcal{B}\cap \mathcal{C})=0.
\end{equation}
The latter equation means that $\mathcal{A}\cap \mathcal{B}$ and
$\mathcal{B}\cap \mathcal{C}$ do not contain vectors in common,
i.e. that $\mathcal{B}$ is empty and
\begin{equation}\mathcal{A}\cup \mathcal{B}=\mathcal{A}.\label{dve}\end{equation} From
equations~\ref{plusik}, \ref{krestik}, and (\ref{dve}) follows
that
\begin{equation}
V_2=V(\mathcal{A})=V_1.
\end{equation}
Therefore, we obtained that, in the lattice notation, from
$V_1\leq V_2$ and $V_2\wedge V_1^\bot=0$ follows $V_1=V_2$. By
Lemma~\ref{lemorthomod} lattice $\mathcal{L}(\BH)$ is
orthomodular. Completeness of the lattice is trivial, and
atomicity follows from the fact that 1-dimensional subspaces of
the Hilbert space are atoms of the Hilbert lattice. \end{proof}

The result of Lemma~\ref{hilblatt} states that Hilbert spaces as
well as pre-Hilbert spaces are characterized among Banach spaces
by the property that the lattice of closed subspaces carries an
orthocomplementation. Further, the orthomodularity of
$\mathcal{L}$ differentiates between Hilbert spaces and
pre-Hilbert spaces. This follows from the theorems given in the
next section.

\section{From orthomodular lattices to spaces}\label{sectos}

\begin{chsum}
In this section we study whether with a complete, atomic and
orthomodular lattice can be associated a Hilbert space. The answer
is in the negative: these properties are insufficient. A different
set of requirements is then given that ensures the appearance of
the Hilbert space.
\end{chsum}

The much more interesting question than the one of the previous
section is the reverse characteristics, i.e. a set of properties
required from a lattice of closed subspaces of a vector space for
this space to be a Hilbert space. Here enters a crucial property,
which can manifest itself in different formulations but has always
something to do with requiring continuity. First, by providing a
counterexample, we explain why without requiring this additional
property one cannot obtain anything like a Hilbert space. Thus,
for the long time it has been the most important problem of
lattice theory to find out whether the properties of being
complete, atomic and orthomodular suffice for a lattice to be a
lattice of closed subspaces of a real, complex or quaternionic
Hilbert space. The result due to Keller \cite{keller} gives a
negative answer to this. To demonstrate it, assume the following
definition.

\begin{defn}
The space $(\varepsilon,f)$ is called \textbf{orthomodular space}
if $\varepsilon$ is a vector space over a field $\mathbb{K}$ with
involution $\omega$ provided with a $\omega$-product $f$:
$\varepsilon\times\varepsilon\mapsto \mathbb{K}$ such that for
$x,y\in \varepsilon$ $x\bot y$ if and only if $f(x,y)=0$, and the
projection theorem holds in $\varepsilon$:
\begin{equation}\label{projth}
\mathrm{If}\; U=U^{\bot\bot}\mathrm{\;is\; a\; subspace\; of\;
}\varepsilon\mathrm{\; then\;}\varepsilon=U+U^\bot.
\end{equation}
\end{defn}

One can construct a non-classical example of an orthomodular
space. The ordered field $\mathbb{K}$ is built in a special way of
polynomials over real numbers in the variables $x_1,x_2,\ldots$
\cite{gross}. The elements of $\varepsilon$ are the sequences
$(\xi_i)\in \mathbb{K}^{\mathbb{N}_0}$ such that
$\sum_0^\infty\xi_i^2 x_i$ converges. The form $f$ is defined by
$f((\xi_i),(\eta_i))=\sum_0^\infty\eta_i\xi_i x_i$ and $f$ gives
rise to a norm on $\varepsilon$. The space $(\varepsilon,f)$ is
complete in the norm-topology \cite[Remark 12.3]{kalmbach} and the
projection theorem (\ref{projth}) holds in $\varepsilon$
(\textit{op.~cit.}, Theorem 12.5). One then obtains that the
lattice $\mathcal{L}(\varepsilon)$ of all closed subspaces of
$\varepsilon$ is a complete orthomodular lattice
(\textit{op.~cit.}, p. 175), and it is also atomic. Meanwhile,
$\varepsilon$ has properties quite different from the properties
of Hilbert spaces. For instance, no pair of orthogonal vectors of
the same length exist in $\varepsilon$. For the probabilities on
the lattice $\mathcal{L}(\varepsilon)$ no proof of Gleason's
Theorem~\ref{gleasonth} can be expected. Therefore, one is driven
to impose more conditions on a lattice so that non-classical cases
of spaces like $\varepsilon$ be excluded.

To start with a characterization of what is sufficient to obtain a
Hilbert space, we first recall the Birkhoff-von Neumann
theorem~\cite{bvn}.
\begin{thm}[Birkhoff-von Neumann] Consider a finite-dimension\-al vector space
$V$ over a field $\mathbb{D}$ with dimension greater than 3. Let
$\mathcal{L}(V)$ be a lattice of subspaces of $V$. There exists a natural
one-to-one correspondence between orthocomplementation on $\mathcal{L}(V)$
and normed $\theta$-products $f$ on $V$, where $\theta$ is an
involutory anti-automorphism on $\mathbb{D}$.
\end{thm}

The Birkhoff-von Neumann theorem associates an involutory
anti-\-auto\-morphism with orthocomplementation on a lattice in the
finite-dimensional case only. Still, we would like to have a
general characterization, both in the finite-dimensional the
infinite-dimensional situations. Before doing this, we shall need
to specialize from the general case of any field $\mathbb{D}$ to
real or complex numbers or quaternions. This is achieved by the
following lemma.

\begin{lem}
Let $\mathbb{D}=\mathbb{R},\mathbb{C},\mathbb{H}$ and $V$ be a
vector space over $\mathbb{D}$ with $\dim V\geq 2$. Assume that
$\theta$ is an involutory anti-automorphism on $\mathbb{D}$ and
$f$ a $\theta$-product on $V$. Then
\begin{description}
  \item[(i)] if $\mathbb{D}=\mathbb{R}$ then $\theta=\mathrm{id}$.
  \item[(ii)] if $\mathbb{D}=\mathbb{C}$ then $\theta\neq\mathrm{id}$
and if $\theta$ is continuous then $\theta$ is the conjugate.
  \item[(iii)] if $\mathbb{D}=\mathbb{H}$ then $\theta$ is the conjugate.
\end{description}\label{classif}
\end{lem}
In all three cases, if we assume that $\theta$ is continuous then
it is uniquely determined.

Now we are ready embark on the search for a sufficient condition
for a lattice to give rise to a space $V$ that will be a Hilbert
space.

\begin{thm}
Let $V$ be a vector space of dimension $\geq 4$ over a field
$\mathbb{D}.$ Consider $v_1 \in V\setminus \{0\}$ and
$\mathcal{L}$ a lattice of subspaces of $V$ which satisfies the
following conditions:
\begin{enumerate}
    \item Every finite-dimensional subspace of $V$ is in $\mathcal{L}$.
    \item $U\wedge M=U+M\in \mathcal{L}$ for $M\in \mathcal{L}$ and $\dim U <\infty$.
\end{enumerate}
If $\bot$ is an orthocomplementation on $\mathcal{L}$ then there exists a
unique involutory anti-automorphism $\theta$ on $\mathbb{D}$ and a
unique $\theta$-product $f$ on $V$ such that
\begin{equation}
\left\{ \begin{array}{l}
f(v_1,v_1)=1, \\
f(v,u)=0 \Leftrightarrow v\in \Gamma
(u)^\bot,
\end{array}\right.
\label{th1cond}
\end{equation}
where $\Gamma$ is a closure operator on $V$. \label{assautomorph}
\end{thm}

\begin{proof} If $V$ is finite-dimensional then the assertion
follows from the Birkhoff-von Neumann theorem. We need to prove
(a) that $^\bot$ induces an orthocomplementation $^\prime$ on
every finite-dimensional subspace $M$ of $V$. By the Birkhoff-von
Neumann theorem there exist for $\dim M\geq 4$ an involutory
anti-automorphism $\theta_M$ of $\mathbb{D}$ and a
$\theta_M$-product on $M$ which are unique if we fix an element
$v_1\in M\setminus\{ 0\}$ with $f_M(v_1,v_1)=1$. The pair
$(\theta_m,f_m)$ satisfies (\ref{th1cond}) on $M$. Let $M$ be
fixed and
\begin{equation} f(v,u)\equiv\label{defoff}
f_N(v,u)\;\,\mathrm{for}\;\,N=M+\Gamma(u)+\Gamma(v).\end{equation}
Subsequently we need to prove (b) that $f$ is well-defined and is a
$\theta$-\-product on $V$. Finally, in (c) we show that $\theta$ and $f$
are uniquely determined.

(a) Let $M$ be a finite-dimensional subspace of $V$. Define
\begin{equation}U^\prime=U^\bot\cap M\label{defuprime}\end{equation}
for a subspace $U$ of $M$. Then
$U\subseteq W$ for a subspace $W$ of $M$ implies $W^\prime
\subseteq U^\prime$, $U\cap U^\prime=U\cap U\bot\cap M=0$ and
$U^{\prime\prime}=(U^\bot\cap M)^\bot\cap M=(U\vee M^\bot)\cap
M=(U+M^\bot)\cap M)=U$. Hence $^\prime$ is a well-defined
orthocomplementation on the lattice of subspaces of $M$.

(b) Let $M$, $W$ be finite-dimensional subspaces of $V$ and
$M\subseteq W$ such that $v_1\in M$ and $\dim M\geq 4$. If $U$ is
a subspace of $M$ then the orthocomplement $U^\prime$ defined in (\ref{defuprime})
for $M$ coincides with the intersection of $M$ with the
orthocomplement of $U$ in $W$. Hence
\begin{equation}U^\prime=\{ v\in M | f_W (v,u)=0 \:\forall u\in U\}
\end{equation}
and $f_W | _{M\times M}$ is a $\theta _W$-product on $M$ which
induces $^\prime$. By the uniqueness of such a product it
follows that $\theta _W=\theta _M$ and $f_W | _{M\times M}=f_M$.
The $\theta$-product $f$ satisfies the first of the conditions
(\ref{th1cond}) by virtue of its definition (\ref{defoff}) and
satisfies the second one since for $v\in N$ the conditions
$f(v,u)=0$, $v\in \Gamma(u)^\bot\cap N$ and $v\in\Gamma(u)^\bot$
are equivalent.

(c) Let $\omega$ be an involutory anti-automorphism of
$\mathbb{D}$ and $g$ a $\omega$-product which satisfies
(\ref{th1cond}). Choose $W$ as in (b). Then the restriction $h$ of
$g$ to $W\times W$ is a $\omega$-product on $W$ which induces
$^\prime$. The uniqueness of $\theta=\theta _W$ and $f=f _W$
implies that $\omega=\theta$ and $h=f _W$. By (\ref{defoff})
applied to $W=N$ we obtain $h(v,u)=f(v,u)$ for arbitrary vectors
$v,u\in V$. \end{proof}

\begin{thm}
Let $\BH$ be a vector space over
$\mathbb{D}=\mathbb{R},\mathbb{C}$ or $\mathbb{H}$ of dimension
$\geq 4$ and $\mathcal{L}$ a lattice of subspaces such
that\begin{description}
    \item[(i)] Every finite-dimensional subspace of $\BH$ belongs
    to $\mathcal{L}$,
    \item[(ii)] For every $U\in \mathcal{L}$ and every finite-dimensional
    subspace $V$ of $\BH$ the sum $U+V$ belongs to $\mathcal{L}$.
\end{description}
Assume that $\mathcal{L}$ carries an orthocomplementation $^\bot$.
Assume further that the associated involutory anti-automorphism
$\theta$ of Theorem~\ref{assautomorph} is continuous in case the
field $\mathbb{D}$ equals $\mathbb{C}$. Then there exists an inner
product $f$ on $\BH$ satisfying (\ref{th1cond}) which is unique up
to multiplication with a positive real constant. In particular,
$\BH$ is a pre-Hilbert space.\label{th112}
\end{thm}
\begin{proof} We shall apply Theorem~\ref{assautomorph} and
Lemma~\ref{classif}. For $v_1\in \BH\setminus\{ 0\}$ there exists
a unique involutory anti-automorphism $\theta$ on $\mathbb{D}$ and
a unique $\theta$-product $f$ on $\BH$ which satisfies
(\ref{th1cond}) and with $f(v_1,v_1)=1$. From the assumption on
$\theta$ it follows that $\theta$ is the conjugation for
$\mathbb{D}=\mathbb{H}$ or $\mathbb{C}$ and it is the identity for
$\mathbb{D}=\mathbb{R}$. Since $f$ is normed, it is an inner
product. If $g$ is an inner product on $\BH$ which satisfies
(\ref{th1cond}) then we define $a=g(v_1,v_1)$ and
$h(v,u)=g(v,u)\cdot a^{-1}$. Observe that $a>0$. Now
$h(v_1,v_1)=1$ implies $h=f$ by the uniqueness of $f$. Therefore
$g(v,u)=af(v,u)$ holds for all $v,u\in \BH$.\end{proof}

We give without proof the following two propositions about
properties of lattices of subspaces of Banach spaces
\cite{kakutanimackey}.
\begin{prop}
Let $B$ be an infinite-dimensional complex Banach space, $\mathcal{L}(B)$
the lattice of closed subspaces of $B$ and $^\bot$ an
orthocomplementation on $\mathcal{L}(B)$. Then the associated involutory
anti-automorphism $\theta$ is continuous.
\end{prop}

\begin{thm}[Kakutani-Mackey] Let $B$ be an infinite-dimensional real
or complex Banach space, $\mathcal{L}$ the lattice of closed subspaces of
$B$ and $^\bot$ an orthocomplementation on $\mathcal{L}$. Then there exists
an inner product on $B$ such that for any $U$ in $\mathcal{L}$ its
orthocomplement $U^\bot=\{v\in B |\: f(v,u)=0 \;\forall u\in U\}$.
The pair $(B,f)$ is a Hilbert space whose topology coincides with
the norm topology on $B$. The inner product $f$ is unique up to
multiplication with a real positive constant.
\end{thm}

There results are used to prove the following properties of
pre-Hilbert spaces.

\begin{prop}
Let $\BH$ be a pre-Hilbert space and $\mathcal{L}=\{U\subseteq
\BH\: |\: U=u^{\bot\bot}\}$. The following two conditions are
equivalent:
\begin{description}
\item[(i)]$\BH$ is a Hilbert space, \item[(ii)]$U+U^\bot=\BH$ for
all $U\in \mathcal{L}$.
\end{description}\label{prop115}
\end{prop}

\begin{proof} Since every Hilbert space satisfies (ii) it is
sufficient to prove that (ii) implies (i). Assume that (ii) holds.
Let $\BG$ be the completion of $\BH$ and let $x\in \BG$. One has
to show that $x\in \BH$. For this, define $z=y-x$ where $y\in \BH$
such that $x\bot(y-x)$. The sequences $(x_n)$, $(z_n)$ are chosen
for $x\bot z$ so that $x_n\bot z_m$, $x_n\bot z$, $z_n\bot x$ for
all $n,m\in \mathbb{N}$ and $\lim x_n=x$ and $\lim z_n = z$.
Further, let $U=\{ z_n\: | \: n\in \mathbb{N}\}^\bot$ and
$\mathrm{pr}: \BG\mapsto \Gamma(U)$ be the projection of $\BG$
onto the closure of $U$ in $\BG$. Then $U=U^{\bot\bot}$ implies
$U\in \mathcal{L}$ and $\BH=U+U^\bot$. The element $y\in \BH$ has
a representation $y=u+v$ with $u\in U$ and $v\in U^\bot$. We need
to prove (a) that $U^\bot\subseteq\Gamma(U)^\bot$ and (b) that
$\mathrm{pr}(y)=x$. Then $x=pr(u+v)=u\in U\subseteq \BH$ and this
shows that $x\in \BH$.

(a) Let $w\in U^\bot$. Then $g(w,u)=0$ for all $u\in U$ where $g$
is the inner product on $\BG$. Since $g$ is continuous it follows
that $g(w,u)=0$ holds for all $v\in \Gamma(U)$. Therefore $w\in
\Gamma(U)^\bot$.

(b) $x\in \Gamma(U)$ since $\lim x_n=x$ and $x_n \in U$. Let $v\in
\Gamma(U)$ and $v_n\in U$ with $\lim v_n=v$. Then $g(z_n,v_m)=0$
implies $g(z,v)=0$. Hence $z\in \Gamma(U)^\bot$ and
$\mathrm{pr}(y)=\mathrm{pr}(z)+\mathrm{pr}(x)=0+x=x$.\end{proof}

\begin{prop}
Let $\BH$ be a pre-Hilbert space and $\mathcal{L}=\{U\subseteq
\BH\: | \: U=U^{\bot\bot}\}$. The following conditions are
equivalent:
\begin{description}
\item[(i)]$\BH$ is a Hilbert space, \item[(ii)] $\mathcal{L}$ is
orthomodular.
\end{description}\label{prop116}
\end{prop}

\begin{proof} For proof that from (i) follows (ii) we refer to
section 5.1 of \cite{kalmbach83}. Let $\mathcal{L}$ be
orthomodular. We shall demonstrate that the statement (ii) of
Proposition~\ref{prop115} holds, which will be sufficient to prove
that $\BH$ is a Hilbert space. Assume there exists $U\in
\mathcal{L}$ and $z\in \BH$ such that $z\neq x+y$ holds for all
$x\in U$ and $y\in U^\bot$. Denote $B=U\wedge(U^\bot \vee
\Gamma(z))$ and $C=U^\bot\wedge(U\vee\Gamma (z))$. If $C$ if
finite-dimensional then by virtue of properties of pre-Hilbert
spaces $B+C=B\vee C$. We now show that $C$ is always
finite-dimensional. For every pre-Hilbert space, $\mathcal{L}$ is
an atomic, complete ortholattice which satisfies the exchange
axiom, i.e. if $a\geq a\wedge b$ then $a\vee b\geq b$. Since
$\mathcal{L}$ is orthomodular and $\Gamma(z)$ is an atom in
$\mathcal{L}$ with $\Gamma(z)\nsubseteq U$ one is in position to
apply Theorem 10.9 from Ref. \cite{kalmbach83} to prove that $C$
is an atom in $\mathcal{L}$. It therefore always true that
$B+C=B\vee C$.

Further, from the orthomodularity of $\mathcal{L}$ and the definition of $C$
it follows that $U\vee C=U\vee \Gamma(z)$ and \begin{equation}
B\vee C=(U\vee C)\wedge(U^\bot\vee\Gamma(z)\vee C)=(U\vee \Gamma(z))\wedge
(U^\bot\vee\Gamma(z)\vee C)\geq \Gamma(z).
\end{equation}

This has a consequence that\begin{equation}
z\in B+C=U\wedge(U^\bot \vee \Gamma(z))+U^\bot\wedge(U\vee\Gamma (z))\subseteq U+U^\bot,
\end{equation}which contradicts the initial assumption on $z$. Therefore $\BH$
is a Hilbert space.\end{proof}

\begin{cor}
Every finite-dimensional pre-Hilbert space $\BH$ is a Hilbert
space.\label{prop117}
\end{cor}
\begin{proof} Proposition~\ref{prop116} provides for the desired
outcome if we show that $\mathcal{L}=\{U\subseteq \BH\: | \:
U=U^{\bot\bot}\}$ is orthomodular. In $\BH$ holds $U\vee V=U+V$
for all (automatically finite-dimensional) subspaces $U,V\subseteq
\BH$. Let $x\in U\wedge (V\vee W)=U\wedge (V+W)$ for some $W\in
\mathcal{L}$ such that $W\subseteq U$. Then $x=x_1+x_2\in U$ for
$x_1\in V$ and $x_2\in W\subseteq U$. Hence $x_1=x-x_2\in U$ and
$x\in (U\cap V)+W= (U\wedge V)\vee V$. This proves that
$\mathcal{L}$ is modular by Definition~\ref{modular}. Since
$\mathcal{L}$ is also an ortholattice, it is
orthomodular.\end{proof}

Modularity of $\mathcal{L}$ is characteristic of
finite-dimensional Hilbert spaces. In the infi\-nite-dimensional
case $\mathcal{L}$ is always non-modular. In application of
Theorem~\ref{th112} or Corollary~\ref{prop117} we obtain the
following final lists of properties of a lattice $\mathcal{L}$
associated with the space $\BH$, which are necessary for space
$\BH$ to be a Hilbert space. Not surprisingly, these lists differ
in finite-dimensional and infinite-dimensional cases.

\begin{thm}\textbf{(finite-dimension\-al Hilbert
space characterization)}

Let $\BH$ be a finite-dimensional vector space over
$\mathbb{D}=\mathbb{R}$, $\mathbb{C}$ or $\mathbb{H}$ of dimension
$\geq 4$ and let $\mathcal{L}$ be the lattice of subspaces of
$\BH$. Assume $\mathcal{L}$ has an orthocomplementation such to
which by virtue of Theorem~\ref{assautomorph} one associates an
involutory anti-automorphism $\theta$, and for
$\mathbb{D}=\mathbb{C}$ $\theta$ is continuous. Then there exists
an inner product $f$ on $\BH$ which satisfies
\begin{equation}
U^\bot=\{ v\in \BH\;|\; f(v,u)=0\;\forall u\in U\}
\end{equation}
such that $\BH$ together with $f$ is a Hilbert space. The inner
product $f$ is unique up to multiplication by a positive real
constant. \label{fdth}\end{thm}

\begin{proof} Since conditions (i) and (ii) of
Theorem~\ref{th112} hold for $\mathcal{L}$ it follows that $\BH$
is a pre-Hilbert space. From Corollary~\ref{prop117} it follows
that $\BH$ is a Hilbert space. For $U\in \mathcal{L}$ one has
\begin{equation}
U^\bot=\bigwedge_{u\in U}\Gamma(u)^\bot=\bigcap_{u\in U} \{v\in
\BH\: |\: f(v,u)=0\},
\end{equation}
which equals $\{v\in \BH\: |\: f(v,u)=0\}$ by virtue of the
condition (\ref{th1cond}) as used in
Theorem~\ref{th112}.\end{proof}

\begin{thm}\textbf{(infinite-dimensional Hilbert space characterization)}

Let $\BH$ be an infinite-dimensional vector space over
$\mathbb{D}=\mathbb{R}$, $\mathbb{C}$ or $\mathbb{H}$ and let
$\mathcal{L}$ be a complete orthomodular lattice of subspaces of
$\BH$ which satisfies the conditions of Theorem~\ref{th112}:
\begin{description}
    \item[(i)] Every finite-dimensional subspace of $\BH$ belongs
    to $\mathcal{L}$,
    \item[(ii)] For every $U\in \mathcal{L}$ and every finite-dimensional
    subspace $V$ of $\BH$ the sum $U+V$ belongs to $\mathcal{L}$.
\end{description}
By Theorem~\ref{assautomorph} one associates an involutory
anti-automorphism $\theta$ and we assume that for
$\mathbb{D}=\mathbb{C}$ $\theta$ is continuous. Then there exists
an inner product $f$ on $\BH$ such that $\BH$ together with $f$ is
a Hilbert space with $\mathcal{L}$ as its lattice of closed
subspaces. $f$ is uniquely determined up to multiplication by a
positive real constant. \label{idth}\end{thm}

\begin{proof} By Theorem~\ref{th112} there exists an inner
product $f$ on $\BH$ which satisfies (\ref{th1cond}) and it is
unique up to multiplication by a positive real constant. $\BH$
itself is a pre-Hilbert space. Let $\mathcal{L}(\BH)=\{ U\subseteq
\BH\: | \: U=U^{\prime\prime}\}$ where $U^\prime= \{ x\in \BH\: |
\: (x,u)=0 \; \forall u\in U\}$. We need to prove that
$\mathcal{L}=\mathcal{L}(\BH)$. Then it follows by
Proposition~\ref{prop116} that $\BH$ is a Hilbert space.

Assume $U\in \mathcal{L}$. Since $\mathcal{L}$ is complete and all
1-dimensional subspaces of $\BH$ belong to $\mathcal{L}$ one
obtains\begin{eqnarray} U=\bigvee_{u\in
U}\Gamma(u)=\bigwedge_{u\in U}\{v\in \BH\: | \:
f(v,u)=0\}=\nonumber\\=\{v\in \BH\, | \, f(v,u)=0\; \forall u\in
U\}=U^\prime.
\end{eqnarray}
Therefore $U=U^{\bot\bot}=U^{\prime\prime}\in \mathcal{L}(\BH)$.

Assume $U \in \mathcal{L}(\BH)$. By (\ref{th1cond}) and
completeness of $\mathcal{L}$ one obtains\begin{equation}
U^\prime=\bigcap_{u\in U}\{v\in \BH\: | \: f(v,u)=0\}=
\bigwedge_{u\in U}\Gamma(u)^\bot\in \mathcal{L}.
\end{equation}
From the previous it follows that
$U=U^{\prime\prime}=(U^\prime)^\bot\in \mathcal{L}$. Hence
$\mathcal{L}(\BH)$ is a subset of $\mathcal{L}$. \end{proof}

\chapter{Reconstruction of the quantum mechanical formalism}\label{chaptreconstr}

\section{What do we have to reconstruct?}

Reconstruction of the quantum mechanical formalism proceeds by
building its blocks from the axioms. In this chapter we show how
to achieve this; we also complete the list of axioms, which for
the moment includes Axioms~\ref{ax1} and \ref{ax2} introduced in
Section~\ref{axioms12} and Axiom~\ref{ax3} introduced in
Section~\ref{ipobservers}. The blocks to be reconstructed are the
conventional key components of quantum theory: the Hilbert space
of observables, the Born rule with the state space, and the
unitary dynamics or evolution in time. Reconstruction of these
blocks will be undertaken in
Sections~\ref{rovellirigorous},~\ref{bornrule} and \ref{sect44}
respectively.

As a preliminary exercise, we analyze the role that each of the
above mentioned blocks plays in the quantum theory. We start with
the last block, the unitary dynamics. Conventionally, it arises
from the Schr\"{o}dinger equation in the Schr\"{o}dinger picture
(wavefunction is time-dependent, operators are time-independent)
or from the equation for the evolution operator in the Heisenberg
picture (wavefunction is time-independent, operators are
time-dependent). In quantum mechanics the time change does not
influence the synchronic algebraic structure of the theory, and
all that time evolution does is that it ``shifts'' this algebraic
structure between different time moments. It becomes clear then,
that from a mere study of the synchronic, or better say timeless,
algebraic structure of the quantum theory nothing can be inferred
about unitary time evolution. Indeed, in Section~\ref{sect44} we
see that one must add a new assumption from which the time
dynamics will follow. More will be said about the role of time in
Part~\ref{cstar} in the context of the $C^*$-algebraic approach.

The second block---the Born rule---is closely linked to
probabilities in quantum theory. In fact, our derivation in
Section~\ref{bornrule} suffices for building the state space of
quantum mechanics (density matrices) and for establishing usual
probabilistic quantum mechanical rules. We deliberately choose not
to enter into the vast domain of discussion concerning the meaning
and the philosophy of probabilities.

By means of the information-theoretic reconstruction we bring some
novelty to the discussion of the significance of the first block
of quantum theory, i.e. the Hilbert space. The Hilbert space
appeared in quantum mechanics quite \textit{ad hoc}, following the
joint work by von Neumann, Hilbert and Nordheim \cite{HvNN}. In
1926 nothing seemed to force physicists into accepting the Hilbert
space, apart from the fact that ``it was available on the market''
\cite{mittelprivate}. Also, we know that von Neumann became
greatly disillusioned in the Hilbert space quantum theory already
in a few years after he himself created it. This will be explained
and discussed in more detail in Section~\ref{vNdisill}.

Quite naturally, this leads to a question, ``Why Hilbert space?''
Or, even more surprisingly, ``What is Hilbert space?'' The
mathematical answer, as in Definition~\ref{hilbspdef}, is
well-known, and yet Chris Fuchs in a recent paper \cite{fuchshilb}
call this question ``tough.'' Why is that? The issue at stake is
to justify the use of Hilbert space in quantum theory, and the
most intriguing problem is to explain the dimensionality of the
Hilbert space. Let us quote Fuchs further:
\begin{quote}
Associated with each quantum system is a Hilbert space. In the
case of finite dimensional ones, it is commonly said that the
dimension corresponds to the number of distinguishable states a
system can ``have.'' But what are these distinguishable states?
Are they potential properties a system can possess in and of
itself, much like a cat's possessing the binary value of whether
it is alive or dead? If the Bell-Kochen-Specker theorem~\cite{BKS}
has taught us anything, it has taught us that these
distinguishable states should not be thought of in that way.
\end{quote}
From the quantum logical derivation that we propose below, the
structure of the Hilbert space will follow, but not its dimension.
However, this dimension will appear implicitly in
Equation~(\ref{exV}). The same problem of the origin of Hilbert
space dimension arises in Ref. \cite{FuchsLittleMore}, where it is
suggested that dimension is an ``irreducible element of reality.''
In Refs. \cite{FuchsPaulian,FuchsJacobs} the same author argues
that dimensionality has to do something with the ``sensitivity to
the touch, i.e. ability of the system to be modified with respect
to the external world due to the interventions of that world upon
its natural course.'' Fuchs then proposes a solution to a smaller
problem than the problem of dimension, which is the problem of
justification of quantumness of the Hilbert space. He argues that
quantumness can be viewed as a characteristics of the sensitivity
to eavesdropping. Dimension, on its part, plays a crucial role in
the possible eavesdropping strategies.

To Fuchs's ``sensitivity to the touch'' we offer an alternative
justification. Indeed, the way sensitivity to the touch is
defined, it bears a very strong ontological connotation and a
flavor of realism. The external world ``intervenes upon the
natural course'' of the quantum system. This contradicts both our
epistemological attitude and the attitude dictated by the
Kochen-Specker theorem, which calls for abandoning the assignment
of built-in properties to quantum systems and indeed is one of the
strongest arguments against realism in quantum physics. Thus,
because the realist attitude openly contradicts the philosophical
position to which we stick in this dissertation, the problem of
dimensionality must be given a different analysis devoid of
ontological commitments. This will be attempted via the
transcendental argument in Section~\ref{solersect}.

\section{Rovelli's sketch}\label{rovhilbspace}

Before we start the derivation of the Hilbert space structure from
the informa\-tion-theo\-re\-tic axioms, we present in this section
a conceptual sketch of such derivation due to Rovelli. Rovelli's
discussion of the results concerning the Hilbert space, however,
is only a sketch, i.e. it is not rigorous. He acknowledges it when
he says ``I do not claim any mathematical nor philosophical
rigor.''~\cite{RovRQM}

Let us start with the distinction between P-observer and
I-observer made in Section~\ref{ipobservers}. P-observer interacts
with the quantum system and thus provides for the physical basis
of measurement. I-observer is only ``interested'' in the
measurement result, i.e. information \textit{per se}, and he gets
information by reading it from P-observer. The act of reading or
getting information is here a common linguistic expression and not
a physical process, because I-observer and P-observer are not
physically distinct. The concept of ``being physical'' only
applies to P-observer, and by definition the physical content of
the observer is all contained in P-observer. I-observer as
informational agent is meta-theoretic, and hence the fact that its
interaction with P-observer, or the act of ``reading
information,'' is unphysical. To give a mathematical meaning to
this act, we assume that getting information is described as
yes-no questions asked by I-observer to P-observer.

The set of these yes-no questions will be denoted $W(P)=\{Q_i,
i\in I\}$. According to Axiom~\ref{ax1}, there is a finite number
$N$ that characterizes P-observer's capacity to supply I-observer
with information. The number of questions in $I$, though, can be
much larger than $N$, as some of these questions are not
independent. In particular, they may be related by implication
($Q_1\Rightarrow Q_2$), union ($Q_3=Q_1\vee Q_2$), and
intersection ($Q_3=Q_1\wedge Q_2$). One can define an always false
($Q_0$) and an always true question ($Q_\infty$), negation of a
question ($\neg Q$), and a notion of orthogonality as follows: if
$Q_1 \Rightarrow \neg Q_2$, then $Q_1$ and $Q_2$ are orthogonal
($Q_1 \bot Q_2$). Equipped with these structures, and under the
non-trivial assumption that union and intersection are defined for
every pair of questions, according to Rovelli's statement which,
as we shall see, does not hold without auxiliary assumptions,
``$W(P)$ is an orthomodular lattice.''

Rovelli proposes a few more steps to obtain the Hilbert space
structure. As follows from Axiom~\ref{ax1}, one can select in
$W(P)$ a set $c$ of $N$ questions that are independent from each
other. In the general case, there exist many such sets $c$, $d$,
etc. If I-observer asks the $N$ questions in the family $c$ then
the obtained answers form a string
\begin{equation}
s_c=[e_1,\ldots,e_N]_c. \label{sstring}\end{equation} This string
represents the information that I-observer got from P-observer as
a result of asking the questions in $c$. Note that it is, so to
say, ``raw information'' meaning that it is not yet information
about the quantum system $S$ that the I-observer ultimately wants
to have, but only a process due to functional separation between
the P-observer and the I-observer. The string $s_c$ can take
$2^N=K$ values. We denote them as
$s_c^{(1)},s_c^{(2)},\ldots,s_c^{(K)}$ so that
\begin{equation}
\begin{array}{ccc}
  s_c^{(1)} & = & [0,0,\ldots,0]_c \\
  s_c^{(2)} & = & [0,0,\ldots,1]_c \\
   & \ldots, & \\
  s_c^{(K)} & = & [1,1,\ldots,1]_c
\end{array}
\end{equation}

Now define new questions $Q_c^{(1)}\ldots Q_c^{(K)}$ such that the
yes answer to $Q_c^{(i)}$ corresponds to the string of answers
$s_c^{(i)}$:
\begin{eqnarray}
Q_c^{(1)} = [(e_1=0)\wedge (e_2=0)\wedge\ldots\wedge (e_N=0)]?
= \neg Q_1\wedge \neg Q_2\wedge\ldots\wedge \neg Q_N\nonumber\\
Q_c^{(2)} = [(e_1=0)\wedge (e_2=0)\wedge\ldots\wedge (e_N=1)]?
= \neg Q_1\wedge\neg Q_2\wedge\ldots\wedge Q_N\nonumber\\
\ldots\quad\quad\quad\quad\quad\quad\quad\quad\quad\\
Q_c^{(K)}= [(e_1=1)\wedge (e_2=1)\wedge\ldots\wedge (e_N=1)]? =
Q_1\wedge Q_2\wedge\ldots\wedge Q_N\nonumber \label{complquest}
\end{eqnarray}
To these questions we refer as to ``complete questions.''
\begin{lem}Complete questions $Q_c^{(i)}$ are mutually exclusive
\begin{equation}
Q_c^{(i)}\wedge Q_c^{(j)}=Q_0\:\forall\, i\neq j.
\end{equation}
and for them holds the distributivity law (\ref{distrib}):
\begin{equation}
Q_c^{(i)}\vee(Q_c^{(j)}\wedge Q_c^{(k)})=(Q_c^{(i)}\vee
Q_c^{(j)})\wedge (Q_c^{(i)}\vee Q_c^{(k)}).\label{qdistrib}
\end{equation}
\end{lem}
\begin{proof} Equality to the always false question of the
disjunction of any two different complete questions follows
immediately from their definition (\ref{complquest}). Because
questions $Q_1,\ldots,Q_N$ in the family $c$ are independent by
construction, distributivity holds for them and, consequently, for
the questions $Q_c^{(i)}$.\end{proof}

By taking all possible unions of sets of complete questions
$Q_c^{(i)}$ of the same family $c$ one constructs a Boolean
algebra that has $Q_c^{(i)}$ as atoms.

Alternatively, one can consider a different family $d$ of N
independent yes-no questions and obtain another Boolean algebra
with different complete questions as atoms. It follows, then, from
Axiom~\ref{ax1} that the set of questions $W(P)$ that can be asked
to P-observer is algebraically an orthomodular lattice containing
subsets that form Boolean algebras. As Rovelli says, ``This is
precisely the algebraic structure formed by the family of linear
subsets of Hilbert space.'' This concludes his sketch.

The sketch of the Hilbert space construction is not a rigorous
derivation due to two key obstacles: First, orthomodularity of the
lattice was not derived and, strictly speaking, from Rovelli's
construction one cannot derive it. Second, even if one admits that
the lattice is orthomodular, the fact that yes-no questions form
an orthomodular lattice and that it contains as subsets Boolean
algebras does not yet lead to emergence of the Hilbert space. Both
these claims will now be formalized and all the assumptions needed
on the way to rigorous proof will be made explicit.
%only commits one to the structure of union of Hilbert
%spaces and not of a single Hilbert space. Thus, this can happen to
%be the union of very low-dimensional Hilbert spaces, which allow
%for a classical and not a quantum interpretation. In the general
%case, the structure will be the one of the Hilbert space with
%superselection rules. One needs then to add a separate argument to
%solve the problem of superselection rules.

\section{Construction of the Hilbert space}\label{rovellirigorous}

This section is the highlight of the dissertation. We derive the
structure of the Hilbert space from the information-theoretic
axioms in seven steps:
\begin{enumerate}
    \item Definition of the lattice of yes-no questions.
\item Definition of orthogonal complement. \item Definition of
relevance and proof of orthomodularity. \item Introduction of the
space structure. \item Lemmas about properties of the space. \item
Definition of the numeric field. \item Construction of the Hilbert
space.\end{enumerate}

The fundamental notion of fact in the quantum logical formalism is
represented as answer to a yes-no question. Information is then
brought about by such answer, and the object that we study is the
set of yes-no questions that can be asked to the system.
Importantly, each such question that \textit{can} be asked is not
necessarily asked, and it means that one cannot state that the
information which a question may bring about is the actual
information possessed by I-observer. This possibility, but not
actuality, of bringing about information is a crucial feature of
our approach: only the information actually possessed by
I-observer is given meta-theoretically, while there is also
possible information that I-observer must take into account in
building quantum theory. As it was said in an illuminating
discussion of {Bohr's\label{bohrposs}} understanding of
complementarity~\cite{plotni}, ``\,`Possible information' is the
key phrase in Bohr's formulation, indicating a crucial distinction
between possible and actual events of measurement in quantum
mechanics.'' In this sense, we fully subscribe to Bohr's view.

Denote the set of questions that can be asked to the system as
$W(P)=\{Q_i, i\in I\}$. According to Axiom~\ref{ax1}, there is a
finite number $N\in \mathbb{N}$ that characterizes I-observer's
maximum amount of relevant information. The number of questions in
$W(P)$, though, can be much larger than $N$, as some of these
questions are not independent. Nothing stops from thinking that
index set $I$ is countably or uncountably infinite. At
\textbf{step~1} of the reconstruction, for each pair of questions
we postulate the existence of ``or'' and ``and'' logical
operations and then define the material implication.

\begin{axiom}[logical \textit{or}]\label{axiii}
$$\forall Q_1,Q_2\in I \,\exists \,Q_3\in I \,|\, Q_3=Q_1\vee Q_2,$$
where $Q_1\vee Q_2$ equals 1 if and only if any one of $Q_1$ or $Q_2$ equals 1.
\end{axiom}\addtocounter{axiomcounter}{1}

\begin{axiom}[logical \textit{and}]\label{axiv}
$$\forall \,Q_1,Q_2\in I \,\exists \,Q_3\in I \,|\,Q_3=Q_1\wedge Q_2,$$
where $Q_1\wedge Q_2$ equals 1 if and only if both $Q_1$ or $Q_2$ equal 1.
\end{axiom}\addtocounter{axiomcounter}{1}

\begin{prop}
$W(P)$ is a lattice.
\end{prop}
\begin{proof}Axioms \ref{axiii} and \ref{axiv} define infimum and supremum for
every pair of questions. The result then follows from
Definition~\ref{deflattice}.\end{proof}

As for completeness of this lattice, Definition~\ref{complete} of
complete lattice requires that lower and upper bounds be defined
for any, possible infinite, set of questions. This fact is not
entailed by any previous arguments and must be postulated
separately. As Specker notes \cite{specker1960}, it is sufficient
to enlarge the domain of propositions so that it contains
conjunctions and disjunctions of all elements. This enlargement,
however, is the subject of a separate axiom.

\begin{axiom}\label{axv}
Lattice $W(P)$ is complete.
\end{axiom}\addtocounter{axiomcounter}{1}

By disjunction of a question and its negation one defines the
always false question $Q_0=Q\wedge\neg Q$. By conjunction of a
question and of its negation one defines the always true question
$Q_\infty =Q\vee \neg Q$. Questions $Q_0$ and $Q_\infty$ serve as
lattice elements $0$ and $1$.

Lattice $W(P)$ is also atomic in virtue of being constructed of
yes-no questions. The answer to a yes-no question gives the
indivisible 1 bit of information. Then questions in $W(P)$ that
are not composed from other questions by conjunctions \textit{and}
or \textit{or} are atoms of the lattice.

As \textbf{step~2} of the reconstruction we introduce orthogonal
complementation in the lattice. It is important to distinguish the
material implication, or entailment, which is a true or false
statement about the elements of the language such as questions,
from the conditional operation often referred to as implication,
which is defined in the language itself. To be precise, ``if $A$
then $B$'' is a true or false statement and thus obeys classical
logic. On the contrary, $A\Rightarrow B$, where $\Rightarrow$
means the conditional operation, gives a third, new element of the
language. The theory of conditionals in quantum logic was
developed by Mittelstaedt \cite{mittel}. For a review we refer to
chapter 8 of Ref. \cite{redei}. In the following we shall only be
interested in the relation of material implication expressed by
the ``if - then'' phrase and we shall not enter in the discussion
of quantum logical conditionals.

\begin{defn}[material implication]
Question $Q_1$ entails question $Q_2$, transcribed as
$Q_1\rightarrow Q_2$, if in any two subsequent facts which bring
about information containing answers to $Q_1$ and $Q_2$,
respectively, it is not the case that $Q_1=1$ and $Q_2=0$, and at
least one such sequence of facts is possible:
%unless $Q_1=1$ and $Q_2=0$ in a possible measurement:
$$Q_1\rightarrow Q_2\:\Leftrightarrow\:\neg((Q_1|_M=1)\wedge(Q_2|_M=0)),$$
where $M$ denotes a fact (or a measurement). Equivalently, one can
say that I-observer never has information that $Q_1=1$ and
$Q_2=0$. The requirement that the facts be subsequent means that
no other information is allowed to emerge between these two acts
of bringing about information.
\end{defn}

\begin{defn}
Questions $Q_1$ and $Q_2$ are \textbf{orthogonal} if
\begin{equation}Q_1\rightarrow \neg Q_2.\label{deforthog}\end{equation}
Orthocomplement $Q^\bot$ is a union (conjunction) of all questions orthogonal to $Q$.
\label{oclat}\end{defn}

Note that according to the definition of implication,
orthogonality requires validity of (\ref{deforthog})
\textit{in all possible measurements}. This means that \textit{whenever} questions $Q_1$ and
$Q_2$ are asked to the system, it is not the case that $Q_1=1$ and $Q_2=1$.

\begin{lem}Definition~\ref{oclat} is in full accord with Definition~\ref{oc}.
\end{lem}
\begin{proof} Indeed, (\ref{deforthog}) by the definition of
implication is equivalent to $Q_2\rightarrow \neg Q_1$, which
insures that $(Q_1^\bot)^\bot=Q_2^\bot=Q_1$, where $Q_2=Q_1^\bot.$
Further, it is trivial to verify that $Q\wedge Q^\bot=Q_0$ and
$Q\vee Q^\bot=Q_\infty$ since $Q^\bot$ is greater or equal to
$\neg Q$. It remains to show that property (ii) of
Definition~\ref{oc} holds. Assume that $Q_1\leq Q_2$, i.e.
$Q_1\wedge Q_2=Q_1$. We need to prove that $Q_2^\bot\leq
Q_1^\bot$, i.e. $Q_2^\bot\wedge Q_1^\bot=Q_2^\bot$. The left-hand
side of this last expression denotes such questions $Q$ that
$Q_1\rightarrow\neg Q$ and $Q_2\rightarrow\neg Q$ in all possible
measurements. In its turn, these two conditions holding separately
in all measurements imply that it must not be the case that
$\left[(Q_1\vee Q_2)\wedge \neg Q\right]$. Now insert the equality
$Q_1\wedge Q_2=Q_1$. We get for the negative assumption
\begin{eqnarray}
\neg\left[ (Q_1\vee Q_2)\wedge Q\right]=(\neg Q_1\wedge \neg Q_2)\vee Q=
\left[(\neg Q_1\vee \neg Q_2)\vee \neg Q_2\right]\vee Q=\nonumber\\=\neg Q_2\vee Q.
\qquad\label{proofoc}
\end{eqnarray}
Recall that (\ref{proofoc}) must \textit{not} be the case. Then
negation of the last expression in the line entails that $\neg Q
\wedge Q_2$. Since equivalence holds everywhere in (\ref{proofoc})
and we started with $Q_2^\bot\wedge Q_1^\bot$, we conclude that
$Q_2^\bot\wedge Q_1^\bot=Q_2^\bot$, which was the needed result.
Therefore orthocomplementation as defined in $W(P)$ fulfills the
requirement for a lattice orthocomplementation.\end{proof}

The notion of orthogonality as introduced in the
Definition~\ref{oclat} is closely tied to the notion of relevance
used in Axiom~\ref{ax1}. At this \textbf{step~3} of the
reconstruction, the time is ripe to discuss the latter term.
Imagine that information obtained from a question $Q_1$ is
relevant for I-observer. We are looking for ways to make it
irrelevant. This can be achieved by asking some new question $Q_2$
that will turn $Q_1$ irrelevant. Consider $Q_2$ such that it
entails the negation of $Q_1$:
\begin{equation}
Q_2 \rightarrow \neg Q_1.\label{relev}
\end{equation}
If I-observer asks the question $Q_1$ and obtains an answer to
$Q_1$ but then asks a \textit{genuine} new question $Q_2$, it
means, by virtue of the meaning of the term ``genuine,'' that
I-observer expects either a positive or a negative answer to
$Q_2$. This, in turn, is only possible if information $Q_1$ is no
more relevant; indeed, otherwise I-observer would have been bound
to always obtain the negative answer to $Q_2$.
%The same applies to the positive answer to $Q_2$.
Consequently, we conclude that, by asking $Q_2$, I-observer makes
the question $Q_1$ irrelevant. Note further that
Equation~(\ref{relev}) fully repeats the definition of
orthogonality (\ref{deforthog}). This motivates the following
interpretative definition of the notion of relevance. Remember,
too, that relevance is meta-theoretic and must be defined in the
physical theory independently (see page \pageref{revrefm}).

\begin{figure}[htbp]
\begin{center}
\epsfysize=2.5in \epsfbox{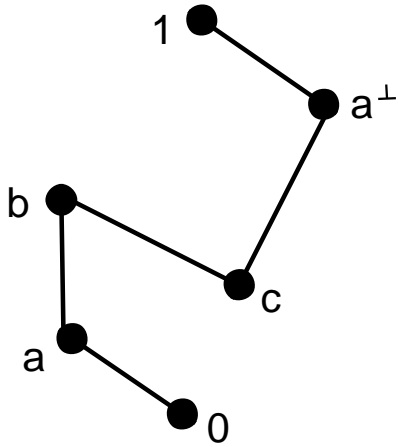}
\end{center}
\caption{\singlespacing The Notion of Relevance. Order in the
lattice is denoted by solid lines and grows from bottom to top,
i.e. $0\leq a\leq b$, etc. If there exists $c\neq 0$ such that
$c\leq b$ and $c\leq a^\bot$, then question $b$ is irrelevant with
respect to question $a$, i.e. in $b$ is contained a ``component''
of $\neg a$, and consequently, by genuinely asking $b$, one
renders the question $a$ irrelevant.} \label{relevfig}
\end{figure}

\begin{defn}
Question $Q_2$ is called \textbf{irrelevant} with respect to
question $Q_1$ if $Q_2\wedge Q_1^\bot\neq 0$. Otherwise question
$Q_2$ is called relevant with respect to question $Q_1$.
\label{revd}\end{defn}

Conceptual justification of Definition~\ref{revd} is offered on
Figure~\ref{relevfig}. Now, the amount of information mentioned in
the Axiom~\ref{ax1} is a nonnegative integer function, so $1$ is
its minimal nonzero value. We postulate that each atom in the
lattice $W(P)$ brings 1 bit of information. Let us now use
Axiom~\ref{ax1} to demonstrate orthomodularity of the lattice
$W(P)$.

\begin{prop}
$W(P)$ is an orthomodular lattice.
\end{prop}

\begin{proof}
By Axiom~\ref{ax1} there exists a finite upper bound of the amount
of relevant information. Let this be an integer $N$. Select an
arbitrary question $Q_1$ and consider a question $\tilde{Q}_1$
such that
\begin{equation}\{Q_1,\tilde{Q}_1\}\label{qqti}\end{equation} bring the maximum amount of
relevant information, i.e. $N$ bits. Notation $\{\ldots\}$ here
means a sequence of questions that are asked one after another.
Because all information here is relevant, we have by the
definition of relevance that \begin{equation}\tilde Q_1 \wedge
Q_1^\bot=0\label{qqti2}\end{equation}.

We shall now use Lemma~\ref{lemorthomod}. It is sufficient to show
that $Q_1\leq Q_2$ and $Q_1^\bot\wedge Q_2=0$ imply $Q_1=Q_2$.
Note first that the second condition means, by
Definition~\ref{revd}, that $Q_2$ is relevant with respect to
$Q_1$. Since $Q_1\leq Q_2$, we obtain that
\begin{equation}Q_2^\bot \leq Q_1^\bot.\end{equation}
Using this result and the result of Equation~\ref{qqti2}, we
derive that \begin{equation}\tilde Q_1\wedge
Q_2^\bot=0.\label{qqti3}\end{equation} By definition, it means
that question $\tilde Q_1$ is relevant with respect to $Q_2$.

Now suppose, contrary to what is needed, that $Q_2>Q_1$ and
consider the following sequence of questions:
\begin{equation}\{Q_1,\, Q_2,\, \tilde Q_1\}\end{equation}
From Equations~\ref{qqti2} and \ref{qqti3} follows that relevance
is not lost in this sequence of question, i.e. all later
information is relevant with respect to all earlier information.
However, while relevance is preserved, this sequence, in virtue of
the fact that $Q_1\neq Q_2$, brings about more information that
the sequence (\ref{qqti}). It means that we have constructed a
setting in which the amount of relevant information is strictly
greater than $N$ bits, causing a contradiction with the initial
assumption. Consequently, $Q_1=Q_2$ and the lattice $W(P)$ is
orthomodular.
\end{proof}

By now, having completed steps 1 through 3 of the reconstruction,
we obtained a complete, atomic and orthomodular lattice $W(P)$.
From Section~\ref{sectos} we know that these properties do not
suffice for emergence of the Hilbert space. Therefore, at this
\textbf{step~4} of the reconstruction, we switch from discussing
lattice $W(P)$ alone to introducing a space of which a lattice of
(certain) subspaces $\mathcal{L}$ will be isomorphic to $W(P)$.
Let us consider an arbitrary Banach space $V$ satisfying this
condition.
\begin{equation}
\mathcal{L}(V) \sim W(P) \label{exV}\end{equation} Note here that
the existence of space $V$ is a relatively moderate constraint,
for at this stage we require that space $V$ be a generic Banach
space. No assumption on the structure of the inner product is
made. Compare this assumption with what Mackey assumes in his
quantum mechanical axioms 7 and 8 \cite{mackey63}. Notation used
in Mackey's axiom 8 will be explained in detail in
Section~\ref{solersect}.
\begin{quotation}\label{mackey78}
Axiom 7. The partially ordered set of all questions in quantum mechanics
is isomorphic to the partially ordered set of all closed subspaces
of a separable, infinite dimensional Hilbert space.

Axiom 8. If $e$ is any question different from the always false question
then there exists a state $f$ in $\mathcal{S}$ such that $m_f(e)=1$.
\end{quotation}
Unlike Mackey, we neither require that the space in question be
the Hilbert space nor its infinite dimensionality. However,
similar to Mackey's axiom 8, we do require that the lattice of all
closed subspaces of $V$ be isomorphic to the lattice of questions
$W(P)$. When later we prove that $V$ has an inner product with
which it forms a Hilbert space, this requirement will be
interpreted as a requirement that to every projection operator on
a closed subspace of the Hilbert space corresponds a question, or
alternatively that cases of product spaces with
{superselection\label{super1}} rules are excluded. Indeed, had we
not chosen a single vector space $V$ ``by hand,'' we could have
considered lattices that are isomorphic to $W(P)$ but built as
direct products of several lattices $\mathcal{L}_i, i=1..n$. Such
cases are relevant in quantum field theories (for discussion see
\cite[Section 4.1]{prug}). Motivation for excluding superselection
rules comes from our search for a simpler structure;
superselection can then be reintroduced as a new meta-theoretic
restriction on the information acquired by I-observer. This
restriction will not be general in the sense of applying to
quantum theory in its most general form, but will lead to a new
information-theoretic axiom in the particular case where
superselection takes place. Note too that one cannot argue that
allowing product spaces with superselection rules could remove
quantumness by reducing the space to a product of one or
two-dimensional Hilbert spaces, in which all physics can be
described classically. The cause of quantumness is not linked with
dimension and will be presented in Section~\ref{quantsect}.

Now observe that $V$ is separable if $W(P)$ contains countably
many questions. It follows from our construction of a complete
orthogonal sequence of questions in (\ref{complquest}) and from
the existence of an isomorphism connecting $W(P)$ and a lattice of
closed subspaces of $V$. One can then consider a family of
projectors on these subspaces that will all commute and together
form a basis in $V$. Then this corresponding space will be
separable~\cite[p. 12, Theorem 2]{richtmyer}.

To summarize, at step 4 of the reconstruction we introduced the
space $V$ such that the lattice of its closed subspaces is
isomorphic to $W(P)$. We now pass to \textbf{step~5} where we
prove two lemmas concerning the space $V$.

\begin{lem}Each finite-dimensional subspace of $V$ is in $\mathcal{L}$.
\label{lem1}\end{lem} \begin{proof} For every finite-dimensional
subspace $V_0 \subseteq V$ one can choose $N$ being the smallest
integer greater than $\log _2 \dim V_0$. One can then pick no more
than $N$ questions in $W(P)$ that correspond to projections onto
one-dimensional subspaces of $V_0$. Units and intersections of any
subset of these questions are also questions and belong to $W(P)$
by Axioms \ref{axiii} and \ref{axiv}. Consequently, $V_0$, of
which all knowledge can be exhausted by such units and
intersections, belongs to $\mathcal{L}$.\end{proof}

\begin{lem}
If Q is in $W(P)$ with $Q\leftrightarrow U\in \mathcal{L}$
and $V_0$ a subspace of $V$ such that $\dim V_0<\infty$
then $U\wedge V_0\in \mathcal{L}.$\label{lem2}
\end{lem}
\begin{proof} This lemma states that to a question one can add
by operations of disjunction and conjunction any finite set of
questions and obtain yet another question. The proof is analogous
to the proof of Lemma~\ref{lem1}. Namely, choose $N$ being the
smallest integer greater than $\log _2 \dim V_0$. Then pick no
more than $N$ questions in $W(P)$ that correspond to projections
onto one-dimensional subspaces of $V_0$. Operation $\wedge$ taken
between any subset of these questions and $Q$ produces a question
which belongs to $W(P)$ in virtue of Axioms \ref{axiii} and
\ref{axiv}. By the isomorphism between $W(P)$ and $\mathcal{L}$,
this new question corresponds to a subset of $\mathcal{L}$. In
virtue of the finite number of questions concerned, we obtain that
$U\wedge V_0\in \mathcal{L}$.\end{proof}

At \textbf{step~6} of the reconstruction we study the field
$\mathbb{D}$ on which is built space $V$. According to
Theorem~\ref{assautomorph} there exists an involutory
anti-auto\-morphism $\theta$ in $\mathbb{D}$. We now first
postulate a concrete form of $\mathbb{D}$ and continuity of the
involutory anti-automorphism and then discuss the alternatives to
this postulate. Continuity will be discussed in this section,
while the concrete form of $\mathbb{D}$ will be discussed both
here and in Section~\ref{solersect}.

\begin{axiom}
The underlying field of the space $V$ is one of the numeric fields
$\mathbb{R}$, $\mathbb{C}$ or $\mathbb{H}$ and the involutory
anti-automorphism $\theta$ is continuous.\label{contaxiom}
\end{axiom}\addtocounter{axiomcounter}{1}

\begin{rem}
It is commonplace to build quantum mechanics in a Hilbert space
over the field $\mathbb{C}$. However, in one and two dimensions a
complete description in a real Hilbert space is possible. The
quaternionic Hilbert space can fully model all properties of the
complex Hilbert space, but it will also lead to novel effects that
have not been observed until now~\cite{adler}. Strictly speaking,
there is no theoretic argument in favor of one of the three fields
only; nor shall we invent an information-theoretic argument.
\end{rem}

Instead of directly postulating that one of the three fields is
involved, real numbers, complex numbers or quaternions, we could
have adopted Zieler's axiom (Co) \cite{zieler} presented below in
Section~\ref{solersect}. In full accord with the argument about
the crucial role of the continuity assumption, axiom (Co) tells
that a certain function is continuous. From this, with the help of
Pontrjagin's index theorem, Zieler deduces that the field in
question is one of the three fields named above.

Note that the continuity property assumed in this axiom is in
direct correspondence with the continuity properties which one
finds in various other proposed sets of axioms for quantum
mechanics. In section 3.7 of his book \cite{land}, Landsman
rephrases continuity into a ``two-sphere property'' which, as it
is easy to expect, requires that some algebraically built
structure be isomorphic to a topological continuous object, namely
a sphere.

Yet a different usage of the continuity axiom can be found
in Lucien Hardy's papers \cite{hardy,hardynato}. Hardy
gives five axioms from which he reconstructs quantum mechanics. They are:
\begin{quote}\begin{description}
\item[Axiom H1.] \textit{Probabilities}. Relative frequencies (measured
by taking the proportion of times a particular outcome is observed) tend
to the same value (which is called probability) for any case where a given
measurement is performed on an ensemble of $n$ systems prepared by some given
preparation in the limit as $n$ becomes infinite.
\item[Axiom H2.] \textit{Simplicity}. The number of the degrees of freedom
of a system $K$ is determined as a function of the dimension $N$ (i.e.
$K=K(N)$) where $N=1,2,\ldots$ and where, for each given $N$, $K$ takes
the minimum value consistent with the axioms.
\item[Axiom H3.] \textit{Subspaces}. A system whose state is constrained
to belong to an $M$ dimensional subspace (i.e. have support on only $M$
of a set of $N$ possible distinguishable states) behaves like a system
of dimension $M$.
\item[Axiom H4.] \textit{Composite systems}. A composite system consisting
of subsystems $A$ and $B$ satisfies $N=N_A N_B$ and $K=K_A K_B$.
\item[Axiom H5.] \textit{Continuity}. There exists a continuous reversible
transformation on a system between any two pure states of that system.
\end{description}
\end{quote}
It has been argued that one can reconstruct quantum mechanics
without Axiom H1 \cite{schack}. Still, the key role is played by
Axiom H5. It is this axiom which, in Hardy's construction,
distinguishes quantum mechanics from classical mechanics. In our
approach the latter separation will appear in
Section~\ref{quantsect} in virtue of Axiom II. This explains why
we do not need the full machinery of Hardy's H5, but only a weaker
apparatus requiring continuity of the involutory anti-automorphism
of the underlying field. Unlike this choice, in his version Hardy
postulates continuity of the transformation of states, which
requires in turn a pre-existing notion of state of the system.
Hardy's motivation that ``there are generally no discontinuities
in physics'' appears unconvincing.

With Axiom~\ref{contaxiom} and the previous results in hand, we
pass to the final \textbf{step~7} of the reconstruction of the
Hilbert space at which we formulate the main theorem of this
section.

\begin{thm}[construction of the Hilbert space]
Let $W(P)$ be an ensemble of all questions that can be asked to a
physical system and $V$ a vector space over
$\mathbb{D}=\mathbb{R},\mathbb{C},$ or $\mathbb{H}$, such that a
lattice of its subspaces $\mathcal{L}$ is isomorphic to $W(P)$.
Then there exists an inner product $f$ on $V$ such that $V$
together with $f$ form a Hilbert space.\label{constrhilb}
\end{thm}
\begin{proof} If $V$ is finite-dimensional the result follows
from Theorem~\ref{fdth} and if $V$ is infinite-dimensional it
follows from Lemmas~\ref{lem1},~\ref{lem2} and Theorem~\ref{idth}.
For application of both theorems the required continuity of
$\theta$ is assumed in Axiom~\ref{contaxiom}.\end{proof}

%The Hilbert space of Theorem~\ref{constrhilb} will be denoted
%$\BH$. Note that $\mathcal{L}$, which is a lattice of closed
%subspaces of $\BH$, is a highly reducible lattice. Indeed, if we
%take a family $c$ of $N$ questions as on page~\pageref{sstring},
%or indeed any other such family $d$, the resulting subalgebra of
%unions of sets of complete questions $Q^{(i)}_d$ is Boolean. The
%center $C(\mathcal{L})$ is therefore non-trivial. However, the
%possibility to factorize $\mathcal{L}$ into irreducible
%sublattices, in the general case, hints at, but does not
%guarantee, any such possibility of factorization for the space
%$\BH$.
Space $\BH$ is built in Theorem~\ref{constrhilb} in a manner that
does not allow
%to know if it can be presented as a product of lower dimensional Hilbert spaces or
to specify its particular elements before we know the sets of
questions in $W(P)$ that correspond to relevant information. What
is relevant is reflected in the choice of questions that are asked
by I-observer (note that in Definition~\ref{revd} relevance of a
question is defined only relatively to another question, i.e.
contextually in the meta-theoretic sense), and it comes without
surprise that the construction of tangible structure of the
Hilbert space in each particular case requires knowledge of the
questions which I-observer intends to, and can, ask.
Theorem~\ref{constrhilb} is therefore non-constructive in the
sense that it makes use of the notion of relevance which is
imposed on the quantum theory from its meta-theory, a circumstance
that underlines the importance of the loop cut of
Figure~\ref{loop01}.

\section{Quantumness and classicality}\label{quantsect}

The Hilbert space $\BH$ constructed in Theorem~\ref{constrhilb}
may happen to be decomposable into the direct product of Hilbert
spaces of smaller dimension. We avoided this possibility by saying
that to every question in $W(P)$ corresponds a closed subspace of
$\BH$ and vice versa. Indeed, were there
{superselection\label{super2}} rules present, some configurations
in the Hilbert space would be physically prohibited, for example
subspaces that intersect with many different multipliers in the
direct product. For such subspaces there would be no corresponding
question in $W(P)$, as we assumed that $W(P)$ does not contain
questions that are conventionally called ``physically
prohibited.'' This latter observation must be credited to the way
in which we have built $W(P)$: it contains all questions that can
be asked to the system, i.e. facts that can occur. If a fact is
``physically prohibited,'' it of course cannot occur. Therefore,
in the philosophy of the loop of existences that motivated the
selection of fundamental notions in Section~\ref{defmeasurement},
it makes no sense to speak of physically prohibited facts, and the
assumption of isomorphism in Equation~\ref{exV} only allows the
appearance of Hilbert spaces without superselection rules.

However, to obtain a Hilbert space without superselection rules is
not enough for building quantum theory. In 1963 Mackey
\cite{mackey63} showed that such a logical construction fits well
both the classical and the quantum cases, and one needs an
additional postulate to recover either the classical formalism or
the quantum one. Classical mechanics in the Hilbert space was
first introduced by Koopman \cite{koopman} and von Neumann
\cite{vN32a}; for a recent discussion see Ref.~\cite{bergeron}.

Mackey formulated his additional assumption which permits to
distinguish between the classical and the quantum cases as follows:
\begin{quotation}
\ldots the fundamental difference between quantum mechanics and
classical mechanics is that in quantum mechanics there are non-simultaneously
answerable questions, i.e. the set of all questions is not a Boolean
algebra.
\end{quotation}
Axiom~\ref{ax2} in our approach plays the role of Mackey's
assumption about non-simulta\-neous\-ly answerable questions. The
Hilbert space $\BH$ was solely built using the consequences of
Axiom~\ref{ax1} (and supplementary axioms), and indeed
Axiom~\ref{ax2} remained unused through the whole discussion which
preceded Theorem~\ref{constrhilb}. It is now time for this axiom
to play its role. We shall prove that Mackey's criterion of
quantumness holds, i.e. that the lattice $W(P)$ is not
distributive or, equivalently, that it is not a Boolean algebra.
This also meets Bub's requirement when he says that ``the
transition from classical to quantum mechanics involves the
transition from a Boolean to a non-Boolean structure for the
properties of a system.''~\cite{bubook}

\begin{lem}
All Boolean subalgebras of $\mathcal{L}$ are proper.\label{boolproper}
\end{lem}

\begin{proof} If I-observer asks the $N$ questions of family $c$
as on page~\pageref{sstring}, i.e. a maximum number of independent
questions, Axiom II requires that he still be able to ask a
question the answer to which is not determined by answers to
questions in the family $c$. Because with the help of $c$ one can
build Boolean subalgebras of the lattice $\mathcal{L}$, it follows
that all such subalgebras are proper and the lattice $\mathcal{L}$
itself is not Boolean. Indeed, were it not the case, one could
have asked the questions $Q_n$ of a family $d$ such as the
complete questions $Q_d^{(i)}$ corresponding to this family $d$,
as defined in (\ref{complquest}), would form a Boolean algebra
coinciding with the whole lattice $\mathcal{L}$. Answers to $Q_n$
of the family $d$ would then leave no room for a new question to
which the response would have not been determined. Since this
contradicts Axiom II, we conclude that all Boolean subalgebras of
$\mathcal{L}$ are proper.\end{proof}

\begin{cor}
The lattice of all questions $W(P)$ is not a Boolean algebra.
\end{cor}
\begin{proof} Follows from Lemma~\ref{boolproper} and
isomorphism between the lattices $\mathcal{L}$ and
$W(P)$.\end{proof}

\section{Problem of numeric field}\label{solersect}

To complete the discussion of how to obtain the Hilbert space, we
return to the problem of justification of our
Axiom~\ref{contaxiom}. In that axiom we postulated that the field
that underlies the space $V$ is one of $\mathbb{R}$, $\mathbb{C}$
or $\mathbb{H}$. Most authors also postulate this, but not all.

Let us start by looking at two attempts of justification of
Mackey's axiom 7 (see page \pageref{mackey78}), one by Zierler in
1961 \cite{zieler} and one by Holland in 1995 \cite{holland}. Both
Zierler and Holland start with the structure which follows from
Mackey's first six axioms and which is essentially the pair
$(\mathcal{L},\mathcal{S})$ of questions and states, where
$\mathcal{L}$ and $\mathcal{S}$ are described in the following
definitions.
\begin{defn}
$\mathcal{L}$ is a \textbf{countably orthocomplete orthomodular partially ordered
set} if
\begin{description}
  \item[(1)] $\mathcal{L}$ is a partially ordered set with smallest element $0$
and largest element $1$;
  \item[(2)] $\mathcal{L}$ carries a bijective map $a\mapsto a^\bot$ that satisfies
$a^{\bot\bot}=a$ and $a\leq b\Rightarrow a^\bot \geq b^\bot$ for all $a,b\in \mathcal{L}$;
  \item[(3)] for every $a \in \mathcal{L}$ the join $a\vee a^\bot=1$ and the meet
$a\wedge a^\bot=0$ both exist and have the value indicated;
\item[(4)] given any sequence $a_i$, $i=1,2,\ldots$ of elements
from $\mathcal{L}$ such that $a_i\leq a_j^\bot$ when $i\neq j$,
the join $\vee a_i$ exists in $\mathcal{L}$; \item[(5)]
$\mathcal{L}$ is orthomodular: $a\leq b\Rightarrow b=a\vee
(b\wedge a^\bot)$.
\end{description}
\end{defn}
A countably orthocomplete orthomodular partially ordered set is
different from a lattice with the same properties only in that
join and meet are not defined for each pair of questions in
$\mathcal{L}$.

\begin{defn}
$\mathcal{S}$ is a \textbf{full, strongly convex family of probability measures
on} $\mathcal{L}$ if
\begin{description}
\item[(1)] each $m\in \mathcal{S}$ is a probability measure on
$\mathcal{L}$, i.e. $m\; :\;\mathcal{L}\rightarrow \{s\; :\; 0\leq
s \leq 1\}$, $m(0)=0$, $m(1)=1$, and $m(\bigvee a_i)=\sum m(a_i)$
for any orthogonal family $\{ a_i\: :\: i=1,2,\ldots\}$ of
elements in $\mathcal{L}$; \item[(2)] $m(a)\leq m(b)$ for all
$m\in \mathcal{S}$ implies $a\leq b$ (``full''); \item[(3)]
$m_i\in \mathcal{S},\, 0<t_i\in \mathbb{R},\, i=1,2,\ldots$, and
$\sum t_i =1$ together imply $\sum t_i m_i \in \mathcal{L}$
(``strongly convex'').
\end{description}
\end{defn}

The structure $(\mathcal{L},\mathcal{S})$ is equivalent to the
structure of the set of observables, states and the probability
measure, which follows from Mackey's first six axioms~\cite[p.
68]{mackey63}. Mackey himself only states this fact and a complete
proof has been provided by Beltrametti and Casinelli \cite{bc}.

Still, Mackey's first six axioms, just as our axioms, do not
guarantee quantumness. As we said above, the latter goal is
achieved by Mackey's axiom 7. In an early attempt to justify this
axiom, Zieler proposed another list of axioms that allow one to
deduce the isomorphism postulated by Mackey (we keep Zieler's
original numbering):
\begin{quote}\begin{description}
\item[(E4), (E5), (A) and (ND)] $\mathcal{L}$ is a separable
atomic lattice, the center $\mathcal{C}(\mathcal{L})\neq
\mathcal{L}$, and element $1\in \mathcal{L}$ is not finite [see
Definition~\ref{finproj}].

\item[(M), (H)] If $a\in \mathcal{L}$ is finite, then $\mathcal{L}(0,a)$ is modular;
if $a,b$ are finite elements of the same dimension, then $\mathcal{L}(0,a)$ and
$\mathcal{L}(0,b)$ are isomorphic.

\item[(S2)] If $0\neq a\in \mathcal{L}$, then there exists $m\in \mathcal{S}$
with $m(a)=1$.

\item[(S3)] $m(a)=0$ and $m(b)=0$ together imply $m(a\vee b)=0$.

\item[(C$^\prime$), (C)] For every finite $a\in \mathcal{L}$ and
for each $i,\, 0\leq i\leq\dim a$, the set of elements $\{ x\in
\mathcal{L}\: :\: x\leq a \;\mathrm{and}\;\dim x =i\}$ is compact
in the topology provided by the metric
$$
f(x,y)=\sup \{|m(x)-m(y)|\; :\; m\in \mathcal{L} \}.
$$
For each $i=0,1,\ldots$ the set of finite elements in $\mathcal{L}$
of dimension $i$ is complete with respect to the same metric.

\item[(Co)] For some finite $b$ and real interval $I$ there exists a nonconstant
function from $I$ to $\mathcal{L}(0,b)$.
\end{description}\end{quote}

One can see that axioms (C$^\prime$), (C) and (Co) essentially
involve non-algebraic concepts, such as topology or continuity.
This comes as little surprise after we have discussed in
Axiom~\ref{contaxiom} the role of the continuity assumption.
However, Zieler's axioms appear to import too much of ``alien''
terminology, and one can do better. This is mainly due to a
beautiful theorem proved by Maria Pia Sol\`{e}r \cite{soler}.

\begin{thm}[Sol\`{e}r]\label{solth}
Let $\mathbb{D}$ be a field with involution, $V$ a left vector
space over $\mathbb{D}$, and $f$ an orthomodular form on $V$ that
has an infinite orthonormal sequence. Then
$\mathbb{D}=\mathbb{R},\mathbb{C}$ or $\mathbb{H}$, and $\{
V,\mathbb{D},f\}$ is the corresponding Hilbert space.
\end{thm}

This theorem makes use of the following definition.

\begin{defn}
An orthonormal sequence is a sequence $$\{ e_i\; :\;
i=1,2,\ldots\}$$ of nonzero vectors $e_i\in V$ such that
$f(e_i,e_j)=0$ for $i\neq j$ and $f(e_i,e_i)=1$ for all
$i$.\end{defn}

Sol\`{e}r's theorem allowed Holland to revise Zieler's postulates,
thus arriving at the following set of axioms \cite{holland}.
\begin{quote}
\begin{description}
  \item[(A1)] $\mathcal{L}$ is separable, i.e. any orthogonal family of
nonzero elements in $\mathcal{L}$ is at most countable.
  \item[(A2)] If $m(a)=m(b)=0$ for some $a,b\in \mathcal{L}$ and an $m\in \mathcal{S}$,
then there exists $c\in \mathcal{L}$, $c\geq a$ and $c\geq b$ with $m(c)=0$.
  \item[(B1)] Given any nonzero question $a\in \mathcal{L}$, there is
a pure state $m\in \mathcal{S}$ with $m(a)=1$.
\item[(B2)] If $m$ is a pure state with support $a\in \mathcal{L}$, then
$m$ is the only state, pure or not, with $m(a)=1$.
\item[(C)]Superposition principle for pure states:
\begin{description}
  \item[1.] Given two different pure states (atoms) $a$ and $b$, there is
at least one other pure state $c$, $c\neq a$ and $c\neq b$ that is a superposition
of $a$ and $b$.
  \item[2.] If the pure state $c$ is a superposition of the distinct
pure states $a$ and $b$, then $a$ is a superposition of $b$ and $c$.
\end{description}
\item[(D)] Ample unitary group: Given any two orthogonal pure states $a,b\in \mathcal{L}$,
there is a unitary operator $U$ such that $U(a)=b$.
\end{description}
\end{quote}

We note that Holland's axioms (A) and (B) appear in Ref.
\cite{bc}; (B) roughly states, in the ordinary language, that for
every question there is a state with a yes answer, and for every
pure state there is one and only question the answer to which is
yes in this state and in no other.

From Sol\`{e}r's theorem it follows that if a pair
$(\mathcal{L},\mathcal{S})$ of question space and state space
satisfies Holland's axioms A through D, then Mackey's axiom 7
follows as a consequence. The structure $\mathcal{L}$, referred to
as \textit{quantum logic}, is an orthocomplemented lattice and is
isomorphic to the orthocomplemented lattice of all closed
subspaces of a separable real, complex, or quaternionic Hilbert
space. The beauty of Sol\`{e}r's result is that it allows to
weaken our Axiom~\ref{contaxiom} by omitting the condition for the
field to be real or complex numbers or quaternions. However, in
doing so, Sol\`{e}r's theorem brings to the information-theoretic
approach a new complication.

The problem is that this theorem is only valid if the Hilbert
space is infinite-dimensional. Theorem~\ref{constrhilb} uses the
result of Theorem~\ref{fdth} which provided construction of a
finite-dimensional Hilbert space. To obtain this, we had to
postulate earlier that the underlying field is either
$\mathbb{R},\mathbb{C}$ or $\mathbb{H}$ and that its involutory
anti-automorphism is continuous. Sol\`{e}r's theorem, though
elegantly avoiding assumptions about anything but the lattice
structure, also avoids the finite-dimensional case. This is by
itself regrettable and all the more so for the science of quantum
computation: for example, to make a quantum computer work as
quantum simulator, the restriction to infinite-dimensional Hilbert
spaces is a major difficulty (see \cite{megpav2000}). It is
impossible to derive a finite-dimensional Hilbert space directly
from lattice axioms, hence to derive the version of quantum theory
needed for quantum computation. The only option left is
philosophical rather than mathematical: One must first derive the
infinite-dimensional Hilbert space and then use meta-theoretic
constraints to reduce the infinite-dimensional space to the
finite-dimensional space of qubits. In the generic situation,
information-theoretic justification of these extra meta-theoretic
constraints remains an unsolved problem.

Still, and without assuming full rigor, we propose a conceptual
argument that goes as follows: It is unclear why there may exist
any \textit{a priori} preferred dimensionality of the Hilbert
space. The symmetry between all values of dimension is preserved,
because dimensionality arises in the isomorphism between the set
of questions $W(P)$ and the lattice of closed subspaces of some
space $V$. There are no information-theoretic constraints on the
questions apart from those that enter in Axioms~\ref{ax1},
\ref{ax2} and \ref{ax3}. So we admit that all dimensions have
\textit{a priori} equal rights. Then, if we believe that the
choice of dimension must still be justified within the theory, we
are left with no particular value for the dimension and we have to
seek for a case that encompasses all the values that are possible.
Apparently, a candidate dimension that does not give preference to
any finite value is the infinity.

In the spirit of this argument one must further say, in order to
be consistent, that structure of the information-based quantum
theory allows that the dimension of the Hilbert space be infinity
or any reduction thereof, where each reduction is operationally
(\textit{a posteriori}) chosen. Like in the case with the
transcendental deduction of probabilities (see the footnote on
page \pageref{traproba}), the structure of the theory provides a
general framework but does not pick a particular value for the
dimension of the Hilbert space. Like the concrete numeric values
of probabilities, the value of dimension is chosen in the process
of application of the theory to a concrete practical situation.
Infinite-dimensional Hilbert space is then reduced to some its
finite-dimensional subspace.

If we had included Sol\`{e}r's theorem in our
information-theoretic reconstruction of the Hil\-bert space, then
it would have allowed us to weaken Axiom~\ref{contaxiom} and only
leave the requirement that the anti-automorphism associated to the
field be continuous, without making any assumption on which field
this one is. The price to pay is that we would have had to
postulate the existence of an infinite orthonormal sequence. By
the lattice isomorphism between $\mathcal{L}$ and $W(P)$, this
condition means that, in $W(P)$, there exists an infinite sequence
of orthogonal questions. Is there an information-theoretic
justification for it? The answer seems to be in the negative.
Axiom~\ref{ax2} says that one can always ask a new question; but
this fact does not guarantee that such a question will be
orthogonal to all questions that have been asked prior to this
one. The word ``new'' does not imply orthogonality. On these
grounds we believe that the assumption needed for Sol\`{e}r's
theorem is not well-justified informationally and we prefer to
postulate explicitly the form of the underlying field as this was
done in Axiom~\ref{contaxiom}.

\section{States and the Born rule}\label{bornrule}

In the choice of fundamental notions in
Section~\ref{defmeasurement} we stated that information and facts
are fundamental. This gave rise to the Hilbert space as space of
the physical theory, while subspaces of the Hilbert space
correspond to yes-no questions. Nothing has been said about the
notion of quantum state. Thus, state is a theoretical construction
that comes after the Hilbert space and that is dependent on the
Hilbert space structure. Such view is consistent with the original
Heisenberg's idea \cite{heis27} and was developed with great
persuasive power by van Fraassen \cite{vF92}. In this section we
show how the Born rule and the state space are reconstructed in
the information-theoretic approach in virtue of Axiom~\ref{ax3}.

Just like the sketch of derivation of the Hilbert space presented
in Section~\ref{rovhilbspace}, Rovelli gives a sketch for the case
of the Born rule and probabilities: From Axiom II it follows
immediately that there are questions such as answers to these
questions are not determined by $s_c$. Define, in general, as
$p(Q,Q_c^{(i)})$ the probability that a yes answer to $Q$ will
follow from the string $s_c^{(i)}$. Given two complete strings of
answers $s_c$ and $s_b$, we can then consider the probabilities
$$p^{ij}=p(Q_b^{(i)},Q_c^{(j)}).
\symbolfootnote[3]{{This introduction of probabilities does not
yet commit one to any particular view on what probabilities}
\textit{are}{. Personally, the author believes in the
trascendental deduction of the structure of probabilities
\cite{Petitot,Bitbol} and in the subjective attribution of numeric
values to probabilities \cite{savage}.}\label{traproba}}
$$ From the way it is defined, the
$2^N\times 2^N$ matrix $p^{ij}$ cannot be completely arbitrary.
First, we must have $$0\leq p^{ij}\leq 1.$$ Then, if information
$s_c^{(j)}$ is available about the system, one and only one of the
outcomes $s_b^{(i)}$ may result. Therefore $$\sum _i p^{ij}=1.$$
If we assume that $p(Q_b^{(i)},Q_c^{(j)})=p(Q_c^{(j)},Q_b^{(i)})$
then we also get $$\sum _j p^{ij}=1.$$

However, if pursued further, this introduction of probabilities
encounters some difficulties. The correct approach, as it appears
for example in the quantum logical derivation in Ref. \cite{land},
should address the question of the construction of a state space
over the Hilbert space obtained. The Hilbert space will then be
treated as space of operators acting on the state space. In this
formulation, the task of building a state space vividly reminds of
a similar problem in the theory of $C^*$-algebras, where it is
solved by the Gelfand-Naimark-Segal (GNS) construction. We shall
explore this similarity in greater detail in Part~\ref{cstar}.
Here we limit ourselves to a less structured approach; still we
avoid explicitly postulating the existence of the state space, as
done for example in Holland's axioms discussed in
Section~\ref{solersect}.

Rovelli expresses a desire to deduce the existence of the state
space and the Born rule from his third axiom, which he
unofficially formulates as follows \cite{Rovelliprivate}:
\begin{quote}
Tentative axiom 3: Different observers hold information in a
consistent way.
\end{quote}
Although this willingness is also expressed in Ref. \cite{RovRQM},
no development is proposed, and instead Rovelli postulates the
superposition principle. We do not know how to complete the
program proposed by Rovelli and we choose instead a different
approach.

In Axiom~\ref{ax3} we introduced intratheoretic
non-contextuality---this is the condition that will now allow to
obtain more of the structure of quantum theory. For
Axioms~\ref{ax1} and \ref{ax2} we have found mathematical
counterparts in the quantum logical formalism with regard to
relevance and quantumness. Now time is ripe to find such a
counterpart for Axiom~\ref{ax3}. It will be understood in terms of
probabilities as sketched by Rovelli. The axiom can then be
reformulated as a condition of independence from the physical
context which has no informational share in determining the answer
to a particular chosen question. This is to say that, if a
question corresponds to a projection operator in the Hilbert space
constructed in Theorem~\ref{constrhilb}, then probabilities can be
defined for a projector independently of the family of projectors
of which it is a member, or that in $p(Q_b^{(i)},Q_c^{(j)})$ with
fixed $Q_b^{(i)}$ probability will be the same had the fixed
question belonged not to the family $b$ but to some other family
$d$.

Non-contextuality remains a widely disputed assumption in the
literature. There exists a multitude of its versions: in
philosophy, type vs. token non-contextuality; in the foundations
of quantum theory, preparation vs. transformation vs. measurement
non-contextuality \cite{spekknctx}. We discuss the general notion
before returning to the intratheoretic non-contextuality that we
postulated in Axiom~\ref{ax3}.

Saunders is one of those who simply reject non-contextuality
because it is ``too strong to have any direct operational
meaning'' \cite{Saunders}. One should also take care to avoid the
Kochen-Specker paradox \cite{KochSpeck}, which along with
non-contextuality requires a premise of value-definiteness
\cite{stanfkoch}:
\begin{quote}
All observables defined for a quantum mechanical system have definite values
at all times.
\end{quote}
Value-definiteness obviously does not hold in
information-theoretic derivation programs like ours, but a deeper
analysis is pending.

In the usual treatment of the Kochen-Specker paradox (for example
\cite{redhead}), value-definiteness is accompanied by a rule
called the Functional Composition Principle, which states that
$[f(A)]^{| \varphi \rangle}=f([A])^{|\varphi\rangle}$. Here $A$ is
a self-adjoint operator, $[A]$ denotes the value of the
corresponding observable, and $f(A)$ denotes the observable whose
associated operator is $f(\hat{A})$. Essentially, the latter
principle states that the algebraic structure of operators should
be mirrored in the algebraic structure of the possessed values of
the observables. One then sees that, in our approach, the
Functional Composition Principle is not justified, because the
conditions of relevance imposed on a set of questions that
\textit{can} be asked do not translate into any conditions of
relevance on the values of responses to these questions.
Responses, in fact, are only given to a tiny fraction of the
questions that can be asked. Therefore, there is no reason to
think that the structure of the question lattice can be imitated
by the structure on the set of ascribed values.

Let us now return to our notion of intratheoretic
non-contextuality. This assumption is not trivial but in order to
see its force, one must first translate it into the mathematical
language of the formalism. We say that the intratheoretic context
is defined by the questions surrounding some fixed question, i.e.
by possible facts other than the given fact in which information
was brought about. In the other words, we say that information as
answer to a yes-no question is only given by the particular answer
to this particular question and not by anything else, including
other answers to other questions.

Remembering the correspondence between questions and subsets of
the Hilbert space that form a complete, atomic and orthomodular
lattice, one is now in position to prove a theorem due to Gleason
\cite{Gleason}:
\begin{thm}[Gleason]
Let $f$ be any function from 1-dimensional projections on a
Hilbert space of dimension $d>2$ to the unit interval, such that
for each resolution of the identity in projections $\{ P_k\},
k=1\ldots d$
\begin{equation}\sum _{k=1}^d P_k = I, \:\sum _{k=1}^d
f(P_k) = 1.\end{equation} Then there exists a unique density
matrix $\rho$ such that $f(P_k)=\mathrm{Tr}(\rho
P_k).$\label{gleasonth}
\end{thm}

Theorem~\ref{gleasonth} shows how the state space is built on the
Hilbert space of the Theorem~\ref{constrhilb} and how
probabilities can be evaluated on that space by means of a
trace-class operator. This justifies the Born rule. With the help
of Axiom~\ref{ax3} and Gleason's theorem we have therefore
constructed the second block of the formalism of the quantum
theory.

\section{Time and unitary dynamics}\label{sect44}

In this section we reconstruct the third and last block of the
quantum formalism after the Hilbert space and the Born rule:
unitary dynamics or evolution in time. As in the case of the Born
rule and Gleason's theorem, we use powerful theorems to minimize
the need in additional postulates. Still, additional assumptions
are unavoidable. To give a reason why it is so, observe that the
axioms introduced in the previous sections refer to the definition
of observables, states, and the Born rule. This is the Heisenberg
picture of quantum mechanics. As Rovelli says in an illuminating
discussion \cite[Section III.A]{Rovelli1}, ``In the Heisenberg
picture, the time axiom can be dropped without compromising the
other axioms or the probabilistic interpretation of the theory.''
Quantum mechanics can be represented as \textit{timeless}. If one
wishes to speak about time, then this notion has to emerge
independently.

The discussion in this section will be limited to non-relativistic
quantum mechanics. This is to say that we shall take into account
time dynamics postulated along with the notion of \textit{fact} in
Section~\ref{defmeasurement}. If one treats only facts, and not
time, as fundamental, thus not willing to assume that time is
introduced axiomatically, then one has to show how time arises
from the interplay of the three fundamental notions. This requires
a general algebraic approach and will be further discussed in
Section~\ref{nfrole}.

Following Rovelli's approach, every yes-no question can be
labelled by the time variable $t$ indicating the time at which it
is asked. Denote as $t\rightarrow Q(t)$ the one-parameter family
of questions defined by the same procedure performed at different
times. Then recall that, by Theorem~\ref{constrhilb}, the set
$W(P)$ has the structure of a set of linear subspaces in the
Hilbert space. Assume that time evolution is a symmetry of the
theory under the shift of the real variable $t$. From this
assumption immediately follows that the set of all questions asked
by I-observer to P-observer at time $t_2$ is isomorphic to the set
of all questions at time $t_1$. The isomorphism has some specific
properties, namely it does not intermingle with the relevance of
information. Because relevance is defined in connection with
orthogonal complementation in the lattice, we require from the
isomorphism that it commutes with orthocomplementation, thus
ensuring that the relations between questions which existed at
time $t_1$ are fully transferred onto relations between the
respective images of these questions at time $t_2$. In other
words, there exists a transformation $U(t)$ such that the inner
product $f$ is preserved
\begin{equation}
f\left(U(t_2-t_1) Q_1(t_1),U(t_2-t_1)
Q_2(t_1)\right)=f\left(Q_1(t_1), Q_2(t_1)\right),
\end{equation}
where $f$ is applied to the elements of the Hilbert space of the
Theorem~\ref{constrhilb}, which isomorphically correspond to
questions. We can now apply Wigner's theorem~\cite{wigner}. By its
virtue transformation $U$ is either unitary or antiunitary, with a
possible phase factor which can be included in the norm $f$.
Antiunitary case is excluded by considering the limit
$t_2\rightarrow t_1$ and requiring that in this limit $U$ becomes
an identity map. Consequently, $U$ is unitary.

Unitary matrices $U(t_2-t_1)$ form an Abelian group. One can write
the composition law
\begin{equation}
U(t_1+t_2)=U(t_1)U(t_2).
\end{equation}
We require that $t\rightarrow U(t)$ be weakly continuous and then
by Stone's theorem \cite[Theorem 6.1]{prug} obtain that
\begin{equation}
U(t_2-t_1)=\exp {[-\mathrm{i}(t_2-t_1)H]},\end{equation} where $H$
is a self-adjoint operator in the Hilbert space, the Hamiltonian.

Recall the distinction between I-observer and P-observer in
Section~\ref{ipobservers}. P-observer as a physical system
interacts with another physical system $S$, and the questions are
being asked by I-observer to P-observer. In order to include the
system $S$ in the theory, we need to make one more step, namely we
need to connect the dynamics of the interaction between physical
systems with the what theory says with regard to the dynamics of
information acquisition by I-observer.

Interaction between P-observer and the quantum system should be
viewed as physical interaction between just any two physical
systems. Still, because I-observer then reads information from
P-observer and because we aren't interested in what happens
between P-observer and $S$ after the act of reading information by
I-observer from P-observer, we can treat P-observer as an
ancillary system in course of its interaction with $S$. After the
reading by I-observer the ancillary system ``decouples.'' Thus,
such an ancillary system would have interacted with $S$ and then
would be subject to a standard measurement described
mathematically on its Hilbert space via a set of ``yes-no''
orthogonal projection operators.

So far, for P-observer we have the Hilbert space and the standard
Born rule. The fact that P-observer is treated as ancilla allows
us to transfer some of this structure on the quantum system $S$. A
new non-trivial assumption has to be made, that the time dynamics
that has previously arisen in the context of I-observer and
P-observer alone, also applies to the P-observer and $S$. In other
words, there is only one time in the system. Time of I-observer is
the one in which one can grasp the meaning of the words ``past''
and ``future'': only what happened between P-observer and $S$ in
the past of the act of reading counts, and the future of that act
has no informational impact. The unique time is thus the time in
which are defined a ``before the act of bringing out information''
and an ``after the act of bringing about information.'' The
hypothesis of unique time is useful for the purposes of this
section and will be invalidated by the discussion in
Section~\ref{nfrole}.

Assume now, as we proposed in Ref.~\cite{grinbijqi, grinbvaxjo},
that both the physical interaction of P-observer with $S$ and the
process of asking questions by I-observer to P-observer take place
in one and the same time. Since (a) until I-observer asks the
question that he chooses to ask, sets of questions at different
times are isomorphic and the evolution is unitary, and (b) time at
which I-observer asks the question only depends on I-observer, one
concludes that the interaction between the quantum system and
P-observer must respect the unitary character all until the
decoupling of the ancilla. Now write,
\begin{equation}\rho_{SP}\rightarrow U\rho_{SP}U^\dag.\end{equation}
After asking a question corresponding to a projector $P _b$,
probability of the yes answer will be given by \begin{equation}
p(b)=\mathrm{Tr}\left(U(\rho _S\otimes\rho _P)U^\dag(I\otimes P
_b)\right).\end{equation} Because the systems decouple, trace can
be decomposed into
\begin{equation}p(b)=\mathrm{Tr} _S (\rho_S
E_b),\end{equation} where all presence of the ancilla is hidden in
the operator \begin{equation}E_b=\mathrm{Tr} _P\left(
(I\otimes\rho _P)U(I\otimes P
_b)U^\dag\right),\label{eoper}\end{equation} which acts on the
quantum system $S$ alone. This operator is
positive-semi\-definite, and a family of such operators form
resolution of identity. They are not, however, mutually
orthogonal. Such operators form positive operator-valued measures
(POVM) \cite{Peres}.

What we have achieved must be now described as follows: by
neglecting the physical component of measurement via factoring out
P-observer and treating measurement as purely informational, we
made the move, from the description of measurement as yes-no
questions asked by I-observer to P-observer, to the description of
measurement as POVM. Information-theoretic derivation of quantum
theory therefore leads to a natural introduction of POVM in virtue
of the selected information-theoretic axioms and fundamental
notions. Importance of this fact must not be underestimated:
POVMs, we remind, are the essential tool in the science of quantum
computation, and the use of this tool can now be justified based
on information-theoretic principles.

\section{Summary of axioms}

We now bring together all axioms used in the derivation of the
formalism of quantum theory. The key information-theoretic axioms
are:

\setcounter{axiomcounter}{1}
\begin{axiom}There is a maximum amount of relevant information that can be
extracted from a system.\end{axiom}
\addtocounter{axiomcounter}{1}
\begin{axiom}It is always possible to acquire new information about a system.
\end{axiom}
\addtocounter{axiomcounter}{1}
\begin{axiom}If information $I$ about a system has been brought about, then it
happened independently of information $J$ about the fact of bringing
about information $I$.\end{axiom}
\addtocounter{axiomcounter}{1}

Auxiliary axioms to which no information-theoretic meaning was
given are:

\begin{axiom}
For any two yes-no questions there exists a yes-no question
to which the answer is positive if and only if the answer to at least
one of the initial question is positive.
\end{axiom}\addtocounter{axiomcounter}{1}
\begin{axiom}
For any two yes-no questions there exists a yes-no question
to which the answer is positive if and only if the answer to both
initial questions is positive.
\end{axiom}\addtocounter{axiomcounter}{1}
\begin{axiom}
The lattice of questions is complete.
\end{axiom}\addtocounter{axiomcounter}{1}
\begin{axiom}
The underlying field of the space of the theory is one of the
numeric fields $\mathbb{R}, \mathbb{C}$ or $\mathbb{H}$ and the
involutory anti-automorphism $\theta$ in this field is continuous.
\end{axiom}
From the full set of axioms it follows that (1) the theory is
described by a Hilbert space which is quantum and not classical;
(2) over this Hilbert space one constructs the state space and
derives the Born rule.

By way of the additional assumption of an isomorphism between the
sets of questions corresponding to different time moments, unitary
dynamics is introduced in the conventional form of Hamiltonian
evolution.

The conceptual framework in which meta-theory is consistently
separated from the theory requires that the observer be
functionally separated into observer as physical system
(P-observer) and observer as meta-theoretic entity or
informational agent (I-observer). This, in turn, leads to a
reinterpretation of the notion of measurement so that the
interaction between I-observer and the physical system is formally
described via a positive operator-valued measure. Such a
description meets the needs of the approach used by the science
quantum information and computation.

We conclude by reiterating that, taken together, the above results
allow one to reconstruct the three main blocks of the formalism of
quantum theory.
% as needed for the science of quantum information.
%We therefore conclude that an information-theoretic approach is self-consistent:
%it gives conceptual foundations for quantum theory, which in turn leads to
%practical methods giving rise to quantum information techniques which had initially
%motivated our selection of the foundational principles.
%Thus, information-theoretic derivation of quantum theory has been
%achieved.

\part{Conceptual foundations of the $C^{*}$-algebraic approach}\label{cstar}

\chapter{$C^*$-algebraic formalism}\label{cstarf}

In Part~\ref{part2}, with the help of quantum logic, we derived
the formalism of quantum theory. In Part~\ref{cstar} we consider a
different approach, the one of the theory of $C^*$-algebras. The
derivation program here will be reduced to a problem of
information-theoretic \textit{interpretation} of the algebraic
approach. When such an interpretation will be given, theorems of
the $C^*$-algebra theory will then permit to recover the formalism
of quantum theory. Thus we change our attitude from the one of
mathematical derivation in Part~\ref{part2} to the attitude of
conceptual justification and philosophical analysis in
Part~\ref{cstar}. Although this change of attitude seems to lead
to more modest results, discussion in Chapter~\ref{itvalapp} will
be largely innovative: to the best of our knowledge, very little
has been said in the literature concerning conceptual aspects of
the Tomita theory of modular automorphisms and the Connes-Rovelli
thermodynamic time hypothesis. To start the exposition, in
Chapter~\ref{cstarf} we present basic elements of the
$C^*$-algebraic formalism.

\section{Basics of the algebraic approach}\label{vnprp}

Content of the algebraic quantum theoretic formalism will be
exposed here following
Refs.~\cite{ConnesBook,ConnesRovelli,haaglocal,redei}.

\begin{defn}
In the linear space $\CB(\BH)$ of bounded operators on a Hilbert
space $\BH$ consider a system of $\varepsilon$-neighbourhoods of
operator $A$ defined by $||A-B||<\varepsilon$. The topology
defined by this system of neighbourhoods is called the
\textbf{norm} or the \textbf{uniform topology} in $\CB(\BH)$.
\end{defn}

In quantum mechanics, a density matrix is a positive linear
operator $\omega$ with unit trace on the Hilbert space $\BH$ and
it defines a normalized positive linear functional over
$\mathcal{A}$ via
\begin{equation}\label{eq1}
  \omega(A)=\mathrm{Tr}\:(A\omega)
\end{equation}
for every $A\in\mathcal{A}$. If one takes an arbitrary selection
of $\omega$ for a fixed $A$, this will define a system of
neighbourhoods of $A$.

\begin{defn}
Topology provided by the system of seminorms
$|\:\mathrm{Tr}\:(A\omega)\:|$ is called the ultraweak or weak
*-topology on $\CB(\BH)$ induced by the set of states $\omega$.
\end{defn}

In particular, if $\omega$ is a projection operator on a pure
state $\Psi\in\BH$, namely if
\begin{equation}\label{eq2}
  \omega=|\Psi\rangle \langle\Psi|,
\end{equation}
then Equation~\ref{eq1} can be rewritten as the quantum mechanical
expectation value relation
\begin{equation}\label{eq3}
  \omega(A)=\langle\Psi | A |\Psi\rangle.
\end{equation}

With the uniform and weak *-topologies one defines two classes of
algebra.

\begin{defn}A \textbf{concrete }$C^*$\textbf{-algebra} is a subspace $\mathcal{A}$
of $\CB(\BH)$ closed under multiplication, adjoint conjugation
(denoted as $^*$), and closed in the norm topology.
\end{defn}\begin{defn}A \textbf{concrete von Neumann algebra }is a
$C^*$-algebra closed in the weak *-topology.
\end{defn}

From these concrete notions that have their roots in quantum
mechanics one imports the intuition for definition of the
following abstract algebraic notions.

\begin{defn}An \textbf{abstract }$C^*$\textbf{-algebra} and an
\textbf{abstract von Neumann algebra} (or a $W^*$-algebra) are
given by a set on which addition, multiplication, adjoint
conjugation, and a norm are defined, satisfying the same algebraic
relation as their concrete counterparts. Namely, a $C^*$-algebra
is closed in the norm topology and a von Neumann algebra is also
closed in the weak *-topology. \label{algdef}\end{defn}
\begin{defn}
A \textbf{state} $\omega$ over an abstract $C^*$-algebra
$\mathcal{A}$ is a normalized positive linear functional over
$\mathcal{A}$.\label{defstate}\end{defn}

\begin{defn}A state $\omega$ is called \textbf{faithful} if, for
$A\in\mathcal{A}$, $\omega(A)=0$ implies
$A=0$.\end{defn}\begin{defn} A vector $x$ belonging to the Hilbert
space $\BH$ on which acts a $C^*$-algebra $\BA$ is called
\textbf{separating} if $Ax=0$ only if $A=0$ for all $A\in\BA$.
\end{defn}

Given a state $\omega$ over an abstract $C^*$-algebra
$\mathcal{A}$, the Gelfand-Naimark-Segal (GNS) construction
provides us with a Hilbert space $\BH$ with a preferred state
$|\Psi _0\rangle$ and a representation $\pi$ of $\mathcal{A}$ as a
concrete $C^*$-algebra of operators on $\BH$, such that
\begin{equation}\label{eq4}
  \omega(A)=\langle\Psi _0 | \pi(A) |\Psi _0 \rangle.
\end{equation}

In the following $\pi(A)$ will be denoted as simply $A$.

\begin{defn}
Given a state $\omega$ on $\BA$ and the corresponding GNS
representation of $\mathcal{A}$ in $\BH$, a \textbf{folium}
determined by $\omega$ is a set of all states $\rho$ over
$\mathcal{A}$ that can be represented as
\begin{equation}\label{eq5}
  \rho(A)=\mathrm{Tr}\:[A\hat{\rho}],
\end{equation}
where $\hat{\rho}$ is a positive trace-class operator in
$\BH$.\label{deffolium}\end{defn}

\begin{rem}Consider an abstract $C^*$-algebra $\mathcal{A}$ and a
preferred state $\omega$. Via the GNS construction (\ref{eq4}) one
obtains a representation of $\BA$ in a Hilbert space $\BH$.
Definition~\ref{deffolium} then introduces a folium of $\omega$,
which determines a weak topology on $\BA$. By closing $\BA$ under
this weak topology we obtain a von Neumann algebra
$\BM$.\label{howvN}\end{rem}

To continue the mathematical presentation, von Neumann factors can
be classified into three types \cite{murvN36}. Assume the
following series of definitions and results.

\begin{defn}
Commutant of a arbitrary subset $\BM\subseteq\CB(\BH)$ is a subset
$\BM^\prime\subseteq\CB(\BH)$ such that
\begin{equation}
B\in\BM^\prime\;\Leftrightarrow\;\forall A\in\BM\quad [B,A]=0.
\end{equation}
\end{defn}

\begin{thm}[von Neumann's double commutant theorem]
Let $\BM$ be a self-adjoint subset of $\CB(\BH)$ that contains
$I$. Then:
\begin{description}
    \item[(i)] $\BM^\prime$ is a von Neumann algebra.
\item[(ii)] $\BM^{\prime\prime}$ is the smallest von Neumann
algebra containing $\BM$.
\item[(iii)]$\BM^{\prime\prime\prime}=\BM.$
\end{description}\label{vNdoubc}
\end{thm}

\begin{defn}
A von Neumann algebra $\BM$ is called a factor if its center
$\BM\cap\BM^\prime$ is trivial, i.e. it consists only of the
multiples of identity.
\end{defn}

\begin{thm}[{\cite[Proposition 6.3]{redei}}]
The lattice of projections (self-adjoint, idempotent operators)
$P(\BM)$ of a von Neumann algebra is a complete orthomodular
lattice. Furthermore, this lattice generates $\BM$ in the sense
that $P(\BM)^{\prime\prime}=\BM$. \label{th63}\end{thm}

Theorem~\ref{th63} is of central importance for classification of
von Neumann algebras. It shows that a classification can be
achieved by investigating the lattice structure.

\begin{defn}
Two projections $A$ and $B$ in $\BM$ are called equivalent if
there is an operator in $\BM$ (``partial isometry'') that takes
vectors in $A^\bot$ to zero and is an isometry between the image
subspaces of $A$ and $B$. \label{defiso}\end{defn}

Definition~\ref{defiso} establishes an equivalence relation in
$P(\BM)$ and it allows to introduce a partial ordering of
projections. Intuitively, $A\preceq B$ means that the dimension of
the image subspace of $A$ is smaller or equal to the dimension of
the image subspace of $B$. The order $\preceq$ is in fact a total
order on $P(\BM)$ and, as a consequence, two von Neumann factors
cannot be isomorphic if the orderings of the corresponding
factorized projection lattices are different. To determine the
order type, the following concept is crucial.

\begin{defn}
Projection $A$ is called finite if from $A\sim B\preceq A$ follows
that $A=B$, i.e. if it is not equivalent to any proper
subprojection of itself. \label{finproj}\end{defn}

\begin{thm}[classification of von Neumann factors]
If $\BM$ is a von Neumann factor then there exists a map $d$
(unique up to multiplication by a constant) defined on $P(\BM)$
and taking its values in the closed interval $[0,\infty]$ which
has the following properties:
\begin{description}
\item[(i)] $d(A)=0$ if and only if $A=0$ \item[(ii)] If $A\bot B$,
then $d(A+B)=d(A)+d(B)$ \item[(iii)] $d(A)\leq d(B)$ if and only
if $A\preceq B$ \item[(iv)] $d(A)<\infty$ if and only if $A$ is a
finite projection \item[(v)] $d(A)=d(B)$ if and only if $A\sim B$
\item[(vi)] $d(A)+d(B)=d(A\wedge B)+d(A\vee B)$
\end{description}
\label{classth}\end{thm}

Types of von Neumann factors, well-defined in virtue of
Theorem~\ref{classth}, are listed in Table~\ref{classtable}.
\begin{table}\caption{Classification of von Neumann
factors}\begin{center}\begin{tabular}{|l|l|l|}
\hline  \textbf{Range of} $\mathbf{d}$ & \textbf{Type of factor} $\mathbf{\BM}$& \textbf{Lattice} $\mathbf{P(\BM)}$ \\
\hline
  $\{0,1,2,\ldots n\}$ & $I_n$ & modular, atomic, \\ &&non-distributive if $n>2$
\\\hline
  $\{0,1,2,\ldots \infty\}$ & $I_\infty$ & orthomodular,
non-modular,\\&& atomic\\\hline
  $[0,1]$ & $II_1$ & modular, non-atomic \\\hline
  $[0,\infty]$ & $II_\infty$ & non-modular, non-atomic \\\hline
  $\{ 0,\infty \}$ & $III$ & non-modular, non-atomic \\ \hline
\end{tabular}\end{center}\label{classtable}\end{table}

\section{Modular automorphisms of $C^*$-algebras}\label{modusect}

Consider now an abstract $C^*$-algebra $\mathcal{A}$ and an
arbitrary faithful state $\omega$ over it. The state $\omega$
defines a representation of $\mathcal{A}$ on the Hilbert space
$\BH$ via the GNS construction with a cyclic and separating vector
$|\Psi\rangle\in\BH$. This, in turn, defines a von Neumann algebra
$\BM$ with a preferred state. We are now concerned with
1-parameter groups of automorphisms of $\BM$. They will be denoted
$\alpha ^\omega _t : \BM\rightarrow\BM$, with $t$ real.

Consider the operator $S$ defined by
\begin{equation}\label{eq6}
SA|\Psi\rangle=A^*|\Psi\rangle.
\end{equation}
One can show that $S$ admits a polar decomposition
\begin{equation}\label{eq7}
S=J\Delta^{1/2}_\omega,
\end{equation}
where $J$ is antiunitary and $\Delta_\omega$ is a self-adjoint,
positive operator. The Tomita-Takesaki
theorem~\cite{TomitaTakesaki} states that the map $\alpha _t
^\omega : \BM\rightarrow\BM$ such as
\begin{equation}\label{eq8}
\alpha _t^\omega A=\Delta _\omega^{-it}A\Delta _\omega ^{it}
\end{equation}
defines a 1-parameter group of automorphisms of the algebra $\BM$.
This group is called the group of modular automorphisms, or the
modular group, of the state $\omega$ over the algebra $\BM$.

\begin{defn}An automorphism $\alpha_{inner}$ of the algebra $\BM$ is
called an \textbf{inner automorphism} if there is a unitary
element $U$ in $\BM$ such that
\begin{equation}\label{eq9}
\alpha _{inner} A=U^* AU.
\end{equation}\end{defn}
Not all automorphisms are inner. We therefore consider the
following equivalence relation in the family of all automorphisms
of $\BM$: two automorphisms are equivalent when they are related
by an inner automorphism $\alpha_{inner}$, namely
$\alpha^{\prime\prime}=\alpha_{inner}\alpha^\prime$ or
\begin{equation}\label{eq10}
\alpha ^\prime (A)U=U\alpha^{\prime\prime}(A),
\end{equation}
for every $A$ and some unitary $U$ in $\BM$. The resulting classes
of automorphisms will be denoted as outer automorphisms, and their
space as $\mathrm{Out}\:\BM$. In general, the modular group
(\ref{eq8}) is not a group of inner automorphisms. It follows that
$\alpha _t$ projects down to a non-trivial 1-parameter group in
$\mathrm{Out}\:\mathcal{M}$, which we denote as $\tilde{\alpha}
_t$. The Cocycle Radon-Nikodym theorem~\cite{ConnesBook} states
that two modular automorphisms defined by two states of the von
Neumann algebra are inner-equivalent. All states of the von
Neumann algebra $\BM$, or of the folium of the $C^*$-algebra $\BA$
that has defined $\BM$, thus lead to the same 1-parameter group in
$\mathrm{Out}\:\mathcal{M}$, or in other words $\tilde{\alpha} _t$
does not depend on the normal state $\omega$. This means that the
von Neumann algebra possesses a canonical 1-parameter group of
outer automorphisms, for which an information-theoretic
interpretation will be suggested in Section~\ref{nfrole}.

From the Cocycle Radon-Nikodym theorem follows the intertwining
property
\begin{equation}
(D\omega _1\;:\;D\omega _2)(t)\;(\alpha _t ^{\omega _2})=(\alpha
_t ^{\omega _1})\;(D\omega _1\;:\;D\omega
_2)(t),\label{intertwcoc}\end{equation} where $(D\omega
_1\;:\;D\omega _2)(t)$ is the Radon-Nikodym cocycle \cite[Section
V.2.3]{haaglocal}. If, for a particular value of $t$, the modular
automorphism $\alpha _t ^\omega$ is inner, then, as a consequence
of Equation~\ref{intertwcoc}, it is inner for any other normal
state $\omega^\prime$. Therefore the set of $t$-values
\begin{equation}\mathcal{T}\;=\;\{t\: : \: \alpha _t
^\omega\;\mathrm{is}\;\mathrm{inner}\}\end{equation}is a property
of the algebra $\mathcal{M}$ independent of the choice of
$\omega$. If $\mathcal{M}$ is not a factor then $\mathcal{T}$ is
the intersection of the sets $\mathcal{T}_k$ corresponding to
factors $\mathcal{M}_k$ occurring in the central decomposition of
$\mathcal{M}$. In case $\mathcal{M}$ is a factor, we notice that
$0\in\mathcal{T}$ and, if $t_1,t_2\in\mathcal{T}$, then $t_1\pm
t_2\in\mathcal{T}$. So $\mathcal{T}$ is a subgroup of
$\mathbb{R}$, i.e. subgroup of the group of real numbers with
addition as the group operation.

Connes \cite{connesclass} showed that $\mathcal{T}$ is related to
the spectrum of the modular operators $\Delta _\omega$ that appear
in Equation~\ref{eq7}. He defined the spectral invariant
\begin{equation}
S(\mathcal{M})=\bigcap _\omega \mathrm{Spect}\:\Delta _\omega,
\end{equation}
where $\omega$ ranges over all normal states of $\mathcal{M}$, and
the set
\begin{equation}
\Gamma(\mathcal{M})=\{ \lambda\in \mathbb{R}\: :\: e^{i\lambda t
}=1\quad\forall \: t\in \mathcal{T}\}.
\end{equation}
Connes's result is that
\begin{equation}
\Gamma(\mathcal{M})\supset \ln (S(\mathcal{M})\setminus 0)
\end{equation}
and that $\ln (S(\mathcal{M})\setminus 0)$ is a closed subgroup of
the multiplicative group $\mathbb{R}_+$. Type $III$ von Neumann
algebras are classified according to the value of $S(\mathcal{M})$
as shown in Table~\ref{classtable2}.

\begin{table}\caption{Connes's classification of von Neumann
factors}\begin{center}\begin{tabular}{|l|l|}
\hline  \textbf{Range of} $S(\mathcal{M})$ & \textbf{Type of factor} $\mathcal{M}$\\
\hline
  $\{1\}$ & $I$ and $II$ \\\hline
  $\{ 0 \cup \lambda ^n,\; n \in \mathbb{Z} \}$ & $III_\lambda\quad (0< \lambda <1)$ \\\hline
  $\mathbb{R}_+$ & $III _1$ \\\hline
  $\{ 0,1 \}$ & $III _0$ \\ \hline
\end{tabular}\end{center}\label{classtable2}\end{table}

The last notion of the von Neumann algebra theory that we
introduce here is the notion of hyperfinite algebra.

\begin{defn}
A von Neumann algebra $\BM$ is called \textbf{hyperfinite} if it
is the ultraweak closure of an ascending sequence of finite
dimensional von Neumann algebras.
\end{defn}

Clearly, a type $I_\infty$ von Neumann algebra is hyperfinite,
because it is the limit of the matrix type $I_n$ algebras of
finite dimensional subspaces. Two important results can be proved
about two other types of von Neumann algebras:

\begin{prop}[{Murray and von Neumann \cite{murvn4}}]There is only one hyperfinite factor
of type $II_1$ up to isomorphism.
\end{prop}

\begin{prop}[{Haagerup \cite{haagerup} based on Connes \cite{connesclass2}}]There is only one hyperfinite factor
of type $III_1$ up to isomorphism.\label{haaco}
\end{prop}

In Ref. \cite[Section V.6]{haaglocal} proof is provided using the
tools of local algebraic quantum theory for the claim that
algebra $\mathcal{M}(K)$ of a diamond is isomorphic to the
hyperfinite type $III_1$ von Neumann factor. A diamond $K$ is a
spatiotemporal region defined as
\begin{equation}\label{diamondef}
K_r=\{x\: :\: |x^0|+|\mathbf{x}|<r\}\end{equation} and it is
characteristic of it that modular automorphisms act on a diamond
geometrically (Hislop and Longo theorem~\cite{hisl}).
Hyperfiniteness of $\mathcal{M}(K)$ follows from the possibility
to insert a type $I$ von Neumann factor $\BN$ between the algebras
of two concentric diamonds with radii $r_2>r_1$ (``split
property''):
\begin{equation}
\mathcal{M}(K_{r_1})\subset \BN \subset \mathcal{M} (K_{r_2}).
\end{equation}
This, in turn, was shown in Ref. \cite{buchholz} to be a
consequence of the Buchholz-Wichmann nuclearity assumption
\cite{buchwich}, which is necessary and sufficient to ensure
``normal thermodynamic properties,'' namely the existence of
KMS-states for all positive $\beta$ for the infinite system and
for finitely extended parts (equivalent to absence of the Hagedorn
temperature \cite{hagedorn}). Thus, the chain of logical relations
is as follows:
\begin{eqnarray}
\nonumber\mathrm{KMS\;states\;at\;all\;}\beta\Leftrightarrow
\mathrm{nuclearity}\Rightarrow
\mathrm{split\;property}\Rightarrow\\\nonumber
\Rightarrow\mathrm{hyperfinite\;type\;}III_1 \mathrm{\;factor.}
\end{eqnarray}
We now explain what the KMS states are and what role they play.

\section{KMS condition}\label{skms}

Let $\mathcal{A}$ be a $C^*$-algebra. Consider the
1-parameter family of automorphisms of operators $A \in \mathcal{A}$
given by
\begin{equation}\label{eq20}
\gamma _t A=e^{it/H}Ae^{-it/H}.
\end{equation}
In the following we shall use the conventional language and say
that the automorphisms are defined by the time evolution $t$ and
that $H$ is the hamiltonian. However, equation (\ref{eq20}) can be
viewed purely formally, as the definition of a group of
automorphisms, without giving any physical meaning to symbols $t$
and $H$. We now look at the system from the thermodynamical point
of view.

\begin{defn}A state $\omega$ over $\mathcal{A}$ is called a
Kubo-Mar\-tin-Schwinger (or KMS) state at inverse temperature
$\beta=1/k_b T$ ($k_b$ being the Boltzmann constant and $T$ the
absolute temperature), with respect to $\gamma _t$, if, for all
$A,B\in\BA$, the function
\begin{equation}\label{eq21}
f(t)=\omega (B(\gamma _t A))
\end{equation}
is analytic in the strip
\begin{equation}\label{eq22}
0<\mathrm{Im}\;t<\beta
\end{equation}
and
\begin{equation}\label{eq23}
\omega((\gamma _t A)B)=\omega(B(\gamma _{t+i\beta} A)).
\end{equation}
\end{defn}
The most important element of this definition is that, in the
right-hand side of Equation~\ref{eq23}, to the parameter $t$ with
a conventional meaning of time variable is added the product of
the imaginary unit $i$ by the inverse temperature $\beta$. One can
therefore view the KMS condition as a generalized Wick rotation,
imposing a certain relation between dynamical and thermodynamical
quantities. Justification given to the particular form
(\ref{eq23}) of the KMS condition is always \textit{a posteriori}:
it so happens that, with this specific choice of the relation
between statistics and dynamics, one obtains correct predictions,
including such ones as for example the Unruh effect. The working
success of the prediction-making procedure justifies the form of
the equation. It remains an open problem in the foundations of
physics to uncover the principles that give rise to the fact that
a certain mathematical relation between physical quantities on the
complex plane (multiplication by $i$) receives clearly
preferential treatment over all other possible relations. As it is
the case with the Wick rotation in quantum field theory, KMS
condition at the imaginary time can be seen as a consequence of
locality and of the spin-statistics connection. Conversely, more
fundamentally and undoubtedly more interestingly for philosophers,
one can view the spin-statistics connection and locality as
consequences of the KMS condition.

In the case of systems with a finite number of the degrees of
freedom, KMS condition reduces to Gibbs condition \cite[Section
V.1.2]{haaglocal}
\begin{equation}\label{eq24}
\omega=N e^{-\beta H}.
\end{equation}
Following Ref. \cite{HHW}, one can postulate that the KMS
condition represents a correct physical extension of the Gibbs
postulate (\ref{eq24}) to infinite dimensional systems. It is
interesting to note that the authors who introduced the KMS
condition in quantum statistical mechanics were led to this
condition by the way starting from the Gibbs postulate. We refer
to the review paper \cite{BF} for a description of this point of
view. However, we shall see that, for the information-theoretic
justification of the algebraic approach, the fact that the KMS
condition is a generalized form of the Wick rotation is more
significant than the fact that it is a generalization of the Gibbs
postulate. The two lines of development can be brought together in
speaking of the \textit{twofold meaning} of the KMS condition.

The following link between the KMS condition and the
Tomita-Takesaki theorem (\ref{eq8}) was established in Ref.
\cite{TomitaTakesaki}. It is arguably one of the most important
and profound theorems in all physics of the second half of the
XXth century.\begin{thm}Any faithful state is a KMS state at the
inverse temperature $\beta=1$ with respect to the modular
automorphism $\gamma_t$ it itself
generates.\label{kmsconnect}\end{thm}

Thus, exactly as it is in the context of classical mechanics, an
equilibrium state contains all information on the dynamics which
is defined by the hamiltonian, apart from the constant $\beta$.
This means that the information about dynamics can be fully
replaced by the information about the thermal state. Indeed,
imagine that the statistical state $\rho$ is known. Then,
remembering that $\beta=1$, take the quantity $H=-\ln \rho$, treat
it as the hamiltonian, and take its one-parameter flow
\cite[Sect.~3.4]{rovellibook}. This will supply full information
about dynamics, where $t$ is none but the parameter of the
hamiltonian flow.

We close this section by discussing the role of thermodynamics in
the information-theoretic approach rooted in the philosophy of the
loop of existences. As we have seen, quantum theory based on a
$C^*$-algebra and a state over it contains all information that is
needed for the theory, including dynamics; what it does not
contain is the possibility to modify $\beta$, i.e. to modify the
temperature. When at the end of Section~\ref{modusect} we required
the existence of KMS states at all $\beta$, it was implicitly
assumed that modification of the value of $\beta$ does not have
its origin inside the theory and must be motivated somehow else.
Recall now the distinction between theory and meta-theory made by
cutting the loop on Figure~\ref{loop01}. One obtains that the
theory describing modification of temperature, which we call
thermodynamics, does not belong to this loop cut, as the loop cut
with its information-theoretic view of quantum theory provides
only for a fixed value of $\beta$. Therefore, thermodynamics,
insofar as it describes the change in temperature, belongs to
meta-theory of the information-based quantum theory. Is such a
position surprising?

The answer is that the place of thermodynamics in the loop cut of
Figure~\ref{loop1} is to be expected. This is due to the
conceptual link between such terms as information and entropy, and
also the link between entropy and temperature that is described by
thermodynamics. Because information is a meta-theoretic concept in
the information-based quantum theory, any theory having
information for its object of study falls necessarily into the
domain of meta-theory. The conceptual link between information and
entropy consists in the definition of information in statistical
physics as relative entropy. In the physical theory, facts, seen
as acts of bringing-about information, are measurement results.
Szilard \cite{szilard} argued that the measurement procedure is
fundamentally associated with the production of entropy, and
Landauer \cite{landauer61} and Bennett \cite{bennett82}, refuting
Szilard's argument, showed that entropy increase comes from the
erasure of information, say, in the preparation of the system. To
erase information means to render it irrelevant in the sense of
Axiom~\ref{ax1}. We discussed the concept of relevant information
in Definition~\ref{revd} and explained on page~\pageref{revrefm}
that any such definition must originate in meta-theory; it can now
be seen that the concept of relevance is tied to thermodynamics.

The Szilard-Landauer-Bennett debate still continues \cite{en1,
en2, bubmax} and we do not take a particular side in it in this
dissertation. Another debate into which we do not enter is the one
concerning applicability of Shannon's vs. von Neumann's entropy
\cite{BZentropy,timpson}. But the very existence of these two
debates shows that thermodynamics has its say in the
information-theoretic approach, which is instantiated, at least,
in the definition of relevant information and in the temporality
of facts. To justify this last claim, we shall return to questions
connected with thermodynamics and the KMS formalism in the
discussion of time in Section~\ref{partchoi}.

\chapter{Information-theoretic view on the
$C^*$-algebraic approach}\label{itvalapp}

\section{Justification of the fundamentals}\label{itjsect}

In this section we show how the algebraic approach arises in the
context of fundamental notions of system, information, and fact
introduced in Chapter~\ref{sect3}. But before doing that, we pay
homage to an early attempt to justify the algebraic approach to
quantum mechanics that was made in the seminal book by G\'{e}rard
Emch \cite{Emch}.

The \textit{raison d'\^{e}tre} of the algebraic approach, for
Emch, is that, besides the standard quantum effects, it
successfully describes phase transitions and nonperturbative
phenomena which the Hilbert space formalism fails to incorporate.
Needless to say, this is very far from our information-theoretic
point of view.

Emch gives a set of ten axioms that provide for the whole of
quantum mechanics. He postulates that a physical system is given
by the set of observables and proposes the first five axioms that
structure this set of observables. Axiom 6 then aims at
establishing that this set is a Jordan-Banach algebra, a direct
generalization of the notion of $C^*$-algebra. Axioms 7 and 8
install a topology on the set of observables, axiom 9 introduces
the GNS construction, and axiom 10 provides for the uncertainty
principle. At no place in the whole axiomatic construction,
however, is anything said about time or about the dynamic aspect
of the theory. But Emch's quantum theory is not timeless: time
evolution is further defined as a group of automorphisms \cite[pp.
163, 300]{Emch} connected with the KMS condition \cite[p.
205]{Emch}. This last suggestion, together with the view that a
quantum system if a set of operators, are the only elements that
we shall borrow from Emch.

Emch's axioms 1 through 5 establish the structure of the set of
observables. Note that at this stage there is no space nor time
assumed, so one cannot use the geometric intuition in determining
the structure of what one observes. Instead, one can only employ
the abstract intuition about the algebraic structure of
observables. It is in these circumstances that Emch postulates
that observables form a vector space and possess certain other
non-trivial properties. We must add to this that it remains to be
seen how a selection of axioms that installs a great deal of
\textit{a priori} mathematical structure on the set of observables
could be justified. What is needed is an \textit{interpretation}
of the algebraic approach. Our interpretation will be given along
the lines of the information-theoretic approach, and we now start
laying it out. As it was argued in Section~\ref{defmeasurement},
the first step always consists in giving a translation into the
mathematical language of each of the fundamental notions of the
information-theoretic approach.

A $C^*$-algebra is interpreted as a mathematical counterpart of
the fundamental notion of system. We have said that, in the
quantum logical approach, system is represented as physical
system, to which refers information obtained in elementary
measurements in the form of answers to yes-no questions. Imagine
for a moment the inverse optics: one could postulate that a large
family of elementary propositions defines what a physical system
\textit{is}. We employ the inverse optics here only in the formal
sense: instead of saying that the mathematical counterpart of the
notion of system is the physical system of the quantum logical
approach, we now formally represent the system as a $C^*$-algebra.

Further, as stated in Section~\ref{defmeasurement}, facts are acts
of bringing about information and, in the physical theory, they
are represented as measurement results. Usually we characterize a
system not separately, but together with the information about it.
Indeed, the system is mathematically described by a family of
operators that form a $C^*$-algebra. These operators have the
potential to frame an act of bringing-about information and,
consequently, to give rise to a fact. One observes that operations
such as to characterize a system by a family of operators and to
be given some information about the system come closely connected,
both conceptually and formally.

Therefore, let us now consider a system and a fact. The fact is an
act of bringing-about information, so there is some information
available about the system. While the system is mathematically
represented as a $C^*$-algebra of observables, we postulate that
the information that was brought about in the chosen fact is
represented as a state over this $C^*$-algebra in the sense of
Definition~\ref{defstate}. The notion of state as a positive
linear functional is a translation of the concept of information
into mathematical terms. This definition also falls in line with a
recent observation by Duvenhage that ``we can define information
as being the state on the observable algebra'' \cite{duven}.

Let us look at how our terminological translation corresponds to
the conventional one, where information is correlation between
measurement results. In the conventional quantum mechanics,
measurement results receive theoretical treatment due to
introduction in the theory of the concept of \textit{preparation}.
In almost any textbook on quantum mechanics one will find a
phrase, ``The system is prepared in a such-and-such state.'' Now,
when we prepare a system, we make a catalogue of \textit{all our
knowledge} about this system. Indeed, to prepare a system means to
set it up in accordance with our requirements to the system. These
requirements are nothing but information about the system or our
current knowledge thereof. Quantum mechanical preparation thus
means that we make a list of, or exhibit, all knowledge about the
system. Once the list has been compiled, the system has been
prepared in a state corresponding to information on this list. An
important element here is to accept that it is \textit{all} our
knowledge. Indeed, if an observer \textit{genuinely} wants to
learn something, it means that at present, as of the time before
learning a new fact, the observer does not know it and does not
possess information contained in that fact. What is going to be
measured in a specially prepared setting is yet completely unknown
at the preparation stage, and the catalogue of information that
corresponds to preparation bears no trace of the particular
information that is yet to be brought about. The argument here can
be regarded as an equivalent of the condition of intratheoretic
non-contextuality discussed in Section~\ref{ipobservers}.

Recall now that the ``what is to be measured'' is just a
collection of operators in a $C^*$-algebra according to our
definition of system. ``Completely unknown'' with respect to these
operators means that the genuine state over the algebra, in the
sense of information state, corresponds to no \textit{a priori}
information or no \textit{a priori} knowledge. To say the same
phrase in the language of thermodynamics amounts to requiring that
the prepared state over the algebra of observables correspond to
infinite temperature or, in the terminology of the KMS formalism,
to $\beta=0$.

It so happened historically that von Neumann's original idea about
how to derive quantum mechanics was related to the conclusion that
the prepared state over the algebra of observables corresponds to
infinite temperature. To illustrate the analogy, we open a
parenthesis where we give a sketch of von Neumann's derivation.

\section{Von Neumann's derivation of quantum mech\-anics}\label{vNdisill}

\begin{chsum}
This historic section falls out of the main development of the
dissertation. It offers a perspective on how were born the key
ideas of quantum theory, like the use of the Hilbert space or the
algebraic approach, and a well-informed reader may skip it.
\end{chsum}

Bub \cite{bubvN} and R\'{e}dei \cite{redei} give a concise
exposition of von Neumann's attempt to derive the probabilistic
structure of quantum mechanics. In a 1927 paper on the
mathematical foundations of quantum mechanics \cite{vN165}, the
heart of the whole theory is the ``statistical Ansatz.'' It states
that the relative probability that the values of the pairwise
commuting quantities $S_i$ lie in the intervals $I_i$ if the
values of the pairwise commuting quantities $R_j$ lie in the
intervals $J_j$ is given by
\begin{equation}\label{statansatz}
\mathrm{Tr}\left[E_1(I_1)E_2(I_2)\ldots E_n(I_n)F_1(J_1)F_2(J_2)\ldots F_m(J_m)\right],
\end{equation}
where $E_i(I_i)$ and $F_j(J_j)$ are the spectral projections of the corresponding
operators $S_i$ and $R_j$ belonging to the respective intervals. Note that we are using
here not the von Neumann's original notation, but R\'{e}dei's account of it coined
out in the modern terms.

In Ref.~\cite{vN166} von Neumann made an attempt to ``work out
inductively,'' a phrase that meant, for von Neumann, a requirement
that the statistical Ansatz (\ref{statansatz}) be derived from the
basic principles of the theory. The starting point of the
derivation is the assumption of an \textit{elementary unordered
ensemble} (``elementar ungeordnete Gesamtheit''). Von Neumann also
calls this ensemble a fundamental ensemble in Ref. \cite{vN167}
and in the same paper appears a characterization ``ensemble
corresponding to `infinite temperature'\,''. For von Neumann this
is an \textit{a priori} ensemble $E$ of which one does not have
any specific knowledge. Every system of which one knows more is
obtained from this ensemble by selection: one checks the presence
of a certain property $P$, e.g. that quantity $S$ has its value in
the set $I$, and one collects into a new ensemble those elements
of the \textit{a priori} ensemble that have the property $P$. This
new ensemble $E^\prime$ is therefore derived from $E$. On
$E^\prime$ one can compute the relative probability defined in the
Ansatz (\ref{statansatz}). Relative here means relative to the
condition $P$. Computation of the probability is done via checking
again the presence or absence of a certain property and collecting
those elements that have this property. Because von Neumann was a
partisan of the von Mises frequency interpretation of
probabilities \cite{vonMises}, he believed that one must simply
calculate the frequency of occurrence of the selected elements in
ensemble $E$. Identifying ensembles with expectation value
assignments and assuming the formalism of quantum mechanics, von
Neumann then showed that each ensemble can be described by a
positive operator $U$, such that the description in question is
given by
\begin{equation}\label{uq}\mathrm{Tr}(UQ).\end{equation} Statistical operator $U$
of the \textit{a priori} ensemble $E$ is the identity operator $I$.

Importance of the \textit{a priori} ensemble can be seen as
follows. The formula $\mathrm{Tr}(UQ)$ is not yet what von Neumann
wants to achieve, for the goal is to obtain the statistical Ansatz
(\ref{statansatz}). Suppose that we only know of the system $S$
that the values of the pairwise commuting quantities $R_j$ lie in
the intervals $J_j$. ``What statistical operator for this ensemble
should be inferred from this knowledge?'' asks von Neumann.
Assuming that it was the \textit{a priori} ensemble on which we
checked that the quantities $R_j$ lie in the intervals $J_j$, and
that we have collected those members of $E$ on which this property
was found present into a new ensemble $E^\prime$, von Neumann
proved that the statistical operator is indeed
$F_1(J_1)F_2(J_2)\ldots F_m(J_m)$ needed for Equation
\ref{statansatz}.

In this derivation the \textit{a priori} ensemble plays a
distinguished role. Its statistical operator is the identity $I$,
so it can be viewed as completely unselected, primary ensemble
from which all other ensembles, carrying particular properties,
are obtained. In our discussion in Section~\ref{itjsect}, this
corresponds to saying that at the preparation stage one creates a
catalogue of all knowledge, the genuine state is a state at
infinite temperature or at $\beta=0$, which has the significance
of not yet knowing the information that will be brought about by
the new facts. The quantum mechanical theory, so to say, starts at
the point of the observer not knowing anything, at the price of
collecting all his previous knowledge in the definitions of
algebra and a state on it.

An expected but telling analogy arises from the fact that von
Neumann himself used thermodynamical language and thermodynamical
considerations to speak about the \textit{a priori} ensemble,
which immediately brings to mind the thermodynamical origin of the
KMS condition. In the sequel of his work, von Neumann, who had to
stick to the frequency interpretation of probability, was forced
to remove some important assumptions about the \textit{a priori}
ensemble. Thus, already in Ref. \cite{vN32} he drops a phrase
which in Ref. \cite{vN166} reads,
\begin{quote}
The basis of a statistical investigation is always that one has an
``elementary unordered ensemble'' $\{ S_1,S_2,\ldots\}$, in which
``all conceivable states of the system $S$ occur with equal
relative frequency;'' one must associate the distribution of
values on this ensemble with those systems $S$, on the states of
which one does not have any specific knowledge.
\end{quote}
As R\'{e}dei argues, von Neumann was moved to reject this language
because of its inconsistency with his view on probabilities as
relative frequencies (in the theory appear infinite probabilities
that cannot be interpreted as frequencies). Meanwhile, nothing
precludes from safeguarding the original reasoning if one chooses
some other philosophy of probability, e.g. subjective
probabilities \cite{savage}.

To clarify the parallel, let us now give the main consequence of
the existence of the \textit{a priori} ensemble in von Neumann's
derivation of the statistical Ansatz. Facing the clash between the
necessary but infinite \textit{a priori} probability and the
frequency interpretation, von Neumann was left with one option
only, which was to consider the appearance of infinite,
non-normalizable \textit{a priori} probabilities as a pathology of
the Hilbert space quantum mechanics and to try to work out a
well-behaving non-commutative probability theory, one in which
there exists normalized \textit{a priori} probability or, as says
von Neumann, ``a priori thermodynamic weight of states.'' This
program was successfully completed by classification of factors
and the discovery of the type $II_1$ factor. Indeed, on the
lattice of a type $II_1$ factor the needed probability exists and
is given by the trace.

How deeply von Neumann became disillusioned in the Hilbert space
quantum mechanics is especially clear from his 1935 letter to
Birkhoff \cite[p. 112]{redei}:
\begin{quote}
I would like to make a confession which may seem immoral: I do not
believe absolutely in Hilbert space any more. After all Hilbert
space (as far as quantum mechanical things are concerned) was
obtained by generalizing Euclidean space, footing on the principle
of ``conserving the validity of all formal rules''$\ldots$ Now we
begin to believe that it is not the \textit{vectors} which matter,
but the lattice of all linear (closed) subspaces. Because: 1) The
vectors ought to represent the physical \textit{states}, but they
do it redundantly, up to a complex factor only, 2) and besides,
the states are merely a derived notion, the primitive
(phenomenologically given) notion being the qualities which
correspond to the \textit{linear closed subspaces}. But if we wish
to generalize the lattice of all linear closed subspaces from a
Euclidean space to infinitely many dimensions, then one does not
obtain Hilbert space, but that configuration which Murray and I
called ``case $II_1$.'' (The lattice of all linear closed
subspaces of Hilbert space is our ``case $I_\infty$.'')
\end{quote}
Von Neumann's repetitive reference to the ``a priori thermodynamic
weight of states'' now gets a clear meaning: the usual trace on an
infinite dimensional Hilbert space gives a thermodynamic weight
via the \textit{a priori} unordered ensemble, but this trace does
not exist as a finite quantity. To have a finite \textit{a priori}
thermodynamic weight of states, von Neumann proposes to switch
from the type $I_\infty$ factor algebras, which are just
collections of all closed linear subspaces of an infinite
dimensional Hilbert space, to type $II_1$ factor algebras. Note
that, as R\'{e}dei notices, ``a priori'' in the context of type
$II_1$ factors acquires a new meaning: it reflects the symmetry of
the system. Indeed, Equation~\ref{uq} arises from the fact that
the trace is a unique positive linear normalized functional on a
type $II_1$ factor that is invariant with respect to all unitary
transformations. The meaning of ``a priori'' as reflecting
symmetries of the system immediately reminds of the transcendental
view of quantum physics \cite{Bitbol,Petitot}.

Unfortunately, having made the first step right, von Neumann made
a wrong second step: type $II_1$ algebras do not make things
easier in quantum theory. We now explain the modern alternative
von Neumann's views.

\section{An interpretation of the local algebra theory}\label{partchoi}

Development of the algebraic quantum theory that followed the
early work by von Neumann showed that quantum theory as type
$II_1$ von Neumann algebra is not a viable solution. Algebras in
the quantum theory of infinite systems, i.e. quantum field theory,
involve factors of type $III$ and, further, of subtype $III_1$
(see Table~\ref{classtable2}); an extended argument for this was
given by Haag \cite{haaglocal}. For our approach this means that
some of the assumptions that have led, following von Neumann's
path, to favoring type $II_1$ factors must be rejected as biased.
It is now time to change the attitude: in this section we assume
the formal results of the local algebra theory briefly presented
on page~\pageref{diamondef} and we give them an
information-theoretic interpretation. Such an interpretation will
then allow to treat these results as theorems deriving the
formalism of quantum theory in the context of the
information-theoretic approach. To state clearly the goal of this
section, it is to discuss the theory of local algebras and to give
to the algebraic approach a novel justification, but without
presenting any novel mathematical results.

The most natural critique of the chain of assumptions that have
led to von Neumann's erroneous preference for type $II$ algebras
is of course to say that, while the selection of a $C^*$-algebra
with a state over it as formal counterparts of the notions of
system and information was perhaps justified, the point about no
\textit{a priori} knowledge is questionable. This is indeed
R\'{e}dei's position. We now show that the former selection itself
contains no fewer built-in assumptions than the latter one.

When one starts to build a theory by choosing a $C^*$-algebra and
by saying that a linear positive functional on it corresponds to
the notion of state, one commits himself to a great deal of
presupposed structure. This is manifest in the fact that, with the
help of the GNS construction, a $C^*$-algebra and a faithful state
on it give rise to the representation in a Hilbert space. To
compare, the whole quantum logical enterprise of
Section~\ref{rovellirigorous} aimed at obtaining the Hilbert
space. In the $C^*$-algebraic approach, as a consequence of the
postulated linearity and positivity, it is given for free.

What are the essential inputs that one adheres to in choosing a
$C^*$-algebra and a state over it? The first such input is the
structure of the $C^*$-algebra itself. This can be weakened to
Jordan-Banach or to Segal algebras, which then leads to loosing
much of the deductive power of the theory. The second input is
more peculiar and often overlooked. As hinted above, it lies in
saying that physical states are states over the algebra, while
states are defined as linear positive functionals. Both these
properties of states: linearity and positivity, are to be
justified from the general information-theoretic principles. It
appears that there are no arguments coming from within the theory
that could be used to this purpose. Furthermore, in the spirit of
Section~\ref{rovsect}, one would like to justify why no such
arguments are available. States, as argued in that section, are
relative states and require a reference to the observing system.
In Schr\"{o}dinger's language \cite{schro35}, the quantum state is
the most compact representative of expectation catalogues that
give lists of results the observer may obtain for the specific
observable he may choose to measure. We say, using our
terminology, that it is just a catalogue of all relevant
information available to I-observer. Consequently, linearity or
any other property of states can only arise from the consideration
of particular properties of the I-observer. The theory of
I-observer belongs to meta-theory of the information-based
physical theory, and therefore one needs a different loop cut
(Figure~\ref{loop1}) to justify linearity or positivity of states.
%On the other hand, the observing system is a physical system as
%any, and we have forbidden since Section~\ref{rovsect} to a system
%to have unique, individualizing features. From an ordinary
%physical system one cannot obtain linearity of the physical state
%space. Now, the only plausible way to justify these qualities
%becomes the one of the second cut of the loop of existences on
%Figure~\ref{loop1}.
If one looks at information as being based on some physical
support, then one will possibly deduce the necessary properties of
information states; but such a point of view is complementary to
the one that had been chosen throughout all of the previous
discussion, i.e. to treating physics as based on information.

As argued above, linearity and positivity of states cannot be
justified in the loop cut of Figure~\ref{loop01}. In the quantum
logical approach there was only one notion that could not be so
justified: relevance of information. Algebraic approach, by
treating states on the algebra as information states, uses at
least two properties that remain unjustified from within the
theory. In this sense, quantum logical approach goes somewhat
deeper into the structure of quantum theory, because it assumes
less: it aims at explaining, not only why the theory on the
Hilbert space is quantum rather than classical, but also why the
Hilbert space itself emerges based on only one meta-theoretic
definition of relevance. In the algebraic approach, if one
postulates linearity and positivity, one then immediately obtains
the Hilbert space in virtue of the GNS construction \ref{eq4}.

Let us now return to the information-theoretic justification of
the theory of local algebras. We have seen how the fundamental
notions of system, information and fact receive their respective
mathematical meanings. It is now time to ask how one can make
sense of the information-theoretic Axioms~\ref{ax1} and~\ref{ax2}
of Section~\ref{axioms12} and of Axiom~\ref{ax3} of
Section~\ref{ipobservers}.

To start the discussion, before going to the first axiom, we
observe that our interpretation of the fundamental notions already
justifies the passage from a $C^*$-algebra to a von Neumann
algebra in case I-observer has some (or none) information about
the system. Information is represented as a state over the
algebra, and via the GNS construction one obtains a representation
of $\BA$ in a Hilbert space $\BH$. Definition~\ref{deffolium} then
introduces a folium of $\omega$, which determines a weak topology
on $\BA$. By closing $\BA$ under this weak topology, as explained
in Remark~\ref{howvN}, we obtain a von Neumann algebra $\BM$.
Therefore, with each state over a $C^*$-algebra one associates a
von Neumann algebra. In the theory of local algebras the algebra
in question is normally a von Neumann, and not a $C^*$-, algebra,
and we wish to remove the state-dependence of the definition of a
von Neumann algebra by a $C^*$-algebra. This can be achieved, for
example, by considering the universal enveloping von Neumann
algebra of a $C^*$-algebra~\cite[p.~120]{takesaki}. However,
although we were able to give information-theoretic justification
of the passage from a $C^*$-algebra to the state-dependent von
Neumann algebra, we do not know whether such a justification
exists for replacing $C^*$-algebra with a von Neumann algebra with
regard to representation of the notion of system; and, on the
other hand, this is exactly what is required if one considers a
von Neumann algebra in a manner independent of the state. All we
can say at this stage is that, in the same \textit{fiat} way in
which we postulated that the fundamental notion of system is
formally represented by a $C^*$-algebra, one may postulate that it
is represented by a von Neumann algebra.

As a consequence of the above discussion, where necessary we shall
take the algebra to be a von Neumann algebra. Let us now give
sense in the algebraic formalism to Axiom~\ref{ax1}. We have the
freedom to choose an algebraic meaning for the phrase ``amount of
relevant information is finite.'' If one recalls that information
is associated with states on a $C^*$-algebra, an immediate
suggestion would be to treat the amount of information as some
measure on the state space and to require that this measure be
finite. Note that such a proposal ignores the presence of the
adjective ``relevant'' before the term ``information.'' Now, if
one follows the named path, then a seemingly natural candidate is
the function $d$ used in Theorem~\ref{classth} for classification
of von Neumann factors. However, this function is defined on
projections, and in our current framework information and facts
correspond not to a particular kind of operators within the
$C^*$-algebra, but to the states on the algebra. Also, to require
that $d$ take finite values would mean a restriction to type $I_n$
or type $II_1$ algebras and would exclude quantum field theories,
as it was previously the case with von Neumann's derivation of
quantum mechanics. We need something else.

Our choice of translation of Axiom~\ref{ax1} into the algebraic
terms is to require that the von Neumann algebra representing the
system be hyperfinite. Fell \cite{fell} showed that a folium of
the faithful representation $\pi _\omega$ of a $C^*$-algebra $\BA$
is weakly dense in the set of all states over $\BA$. Therefore, in
the context of the $C^*$-algebraic approach, with only a finite
amount of relevant information, we can never find out if the state
belongs to the given folium. This, in turn, means that the theory,
generically, cannot tell us which information states are the
possible states, once a particular von Neumann algebra had been
chosen. However, we want to preserve this capacity of the theory
as it is an essential component of its predictive power. To do so,
we extend the theory beyond finite amounts of information and
consider ``infinite amounts'' of information, the quotation marks
meaning that some of this information will necessarily be
irrelevant for I-observer. Let us reiterate that it is crucial to
be in position to respond to the above discussed question, i.e. to
determine if the state belongs to the folium of another state.
This is because it is only by comparing the previously possessed
with the incoming information that one can decide if
representation of the system as a given von Neumann algebra holds
or if the folium on the $C^*$-algebra has changed and the
corresponding weak closure, giving a von Neumann algebra, has
changed too. To compare information means to compare the states,
and one is then forced not limit the $C^*$-algebraic approach to
only one equivalence class of representations.

Now, once we have decided to take into consideration the full
variety of the representations of $\BA$, we must make sure that,
by the acts of bringing about more information, we shall be able
to approach this theoretic idealization with a sufficiently high
precision; or otherwise the theory would contain a surplus that
could be removed from it without damaging its
information-theoretic content. Compare this idea with the
requirement of absence of superselection rules in the quantum
logical approach (see pages~\pageref{super1} and
\pageref{super2}). Absence of superselection rules was postulated,
in order to guarantee that to every projector on a closed subspace
of the Hilbert space corresponds a question in $W(P)$ and that
there are no such subspaces about which information can never be
brought about. In other words, only such elements are considered
that fall in the domain of \textit{possible information}, in the
spirit of the quotation from Bohr given on
page~\pageref{bohrposs}. Similarly with the algebraic formalism:
only that now the surplus to be avoided are those states which
cannot be approached with a finite amount of information. We
require that, in the weak *-topology, the precision of state
detection shall tend to infinitely high in the limit of the
infinite number of acts of bringing-about information. This, in
turn, means that we require that $\BA$ be a limit of finite
dimensional algebras, i.e. a hyperfinite algebra. If one only
considers type $III$ algebras, as dictated by the local algebras'
theory, one can say that the algebra must be \textit{the}
hyperfinite algebra, in virtue of Theorem~\ref{haaco}.

At the same time, the requirement of hyperfiniteness will
guarantee that we have fully observed Axiom~\ref{ax2}. To satisfy
the constraint of this axiom, and because information is
mathematically represented as a state over the algebra, we ought
to make sure that, by the acts of bringing about information, one
can always change folium and thus switch to a representation of
the $C^*$-algebra that is not equivalent to the previous one.
Hyperfiniteness supplies precisely what is needed: the algebra is
sufficiently rich so that one can always change folium and bring
in novel information, but at the same time, because there is only
one hyperfinite algebra of each of the types $II$ and $III_1$, the
algebra will remain the same, and, in accordance with
Axiom~\ref{ax1}, one will be able to come infinitely close to it
by pursuing a chain of finite dimensional algebras. Thus
hyperfiniteness is a unique balance between two constraints: that
there be non-equivalent representations defining different folia
and that one could get information with any degree of precision
from a finite sequence of facts.

To move now to the discussion of Axiom~\ref{ax3}, its meaning is
not significantly different from what we have had in the quantum
logical reconstruction. In virtue of presence of
$\sigma$-additivity in von Neumann algebras, Gleason's
Theorem~\ref{gleasonth} is applicable so as to justify the
probabilistic interpretation and the Born rule. In the same sense
as in the quantum logical formalism, Gleason's theorem gives rise
to the state space with the Born rule.

Now that the choice of the hyperfinite von Neumann algebra in the
theory of local algebras has been given an information-theoretic
interpretation, we explore in the next section the question that
was studied in Section~\ref{quantsect} in the context of the
quantum logical approach; namely, the problem of quantumness of
the algebra. For this, we analyze the only existing, as of today,
attempt at information-theoretic \textit{derivation} of quantum
theory by means of the algebraic formalism.

\section{CBH derivation program}\label{cbhsect}

Clifton, Bub and Halvorson (CBH) \cite{Bub} and Halvorson
\cite{halvor} proved a series of results, gathered under the title
``CBH theorem,'' showing equivalence between certain
information-theoretic constraints and the algebraic properties
possessed by quantum $C^*$-algebras. CBH show, for a composite
system, $\BA+\BB$, consisting of two component subsystems, $\BA$
and $\BB$, that \textrm{(i)} the requirement of `no superluminal
information transfer via measurement'\symbolfootnote[2]{We use
single quotes instead of double quotes as elsewhere in the text to
preserve the original choice by the authors of the CBH article,
for whom this phrase clearly has more of a literal, i.e.
empirical, and not simply a metaphoric, sense.} entails that the
$C^*$-algebras $\BA$ and $\BB$ whose self-adjoint elements are the
observables $A$ and $B$, commute with each other (i.e. all
$A\in\BA$ and $B\in\BB$ commute; this is also called the condition
of kinematic independence), and \textrm{(ii)} the condition of `no
broadcasting' of a quantum state entails that $\BA$ and $\BB$
separately are noncommutative. Then, adding an independence
condition for the algebras, they show the existence of nonlocal
entangled states on the $C^*$-algebra $\BA\vee\BB$ that $\BA$ and
$\BB$ jointly generate. This guarantees the presence of nonlocal
entangled states in the mathematical formalism used in the theory,
but does not yet guarantee that these states, a resource available
mathematically, are actually instantiated. In his second paper
Halvorson shows that the third information-theoretic constraint,
`no bit commitment', delivers this missing component, thus
completing the proof of the CBH theorem.

We first discuss the significance of information-theoretic
constraints used in the CBH theorem. The sense of the `no
superluminal information transfer' constraint, the term being
chosen by CBH, is that when Alice and Bob (conventional names for
physical systems) perform local measurements, Alice's measurements
can have no influence on the statistics for the outcomes of Bob's
measurements, and vice versa. CBH go on to say that ``otherwise
this would mean instantaneous information transfer between Alice
and Bob'' and ``the mere performance of a local measurement (in
the nonselective sense) cannot, in and of itself, transfer
information to a physically distinct system.'' Upon reading these
statements, one has a feeling that for CBH \textit{distinct} and
\textit{distant} are synonyms, and it is this very issue that we
shall explore. CBH explain to their reader that the
$C^*$-algebraic framework includes not only the conventional
quantum mechanics, but also quantum field theories; we add that it
also includes generally covariant settings, i.e. theory on a
manifold. In all of these, one has to deal with $C^*$-algebras.
However, neither in quantum mechanics or quantum field theory
formulated as timeless theories \cite{rovellibook}, nor in the
generally covariant formalism, there exist space and time that
play any special role. If one wishes to give information-theoretic
axioms from which to derive the quantum $C^*$-algebraic framework,
one must not assume the spatiotemporal structure; indeed, only in
some particular cases of hand-picked $C^*$-algebras will one be
able to single out the preferred notion of time.

We shall offer several critical points concerning the CBH theorem.
For this, let us have a closer look at how the authors' language
is reflected in their mathematical formalism. They give the
following definition:

\begin{defn}[{\cite[Section 3.2]{Bub}}] Operation $T$ on algebra $\BA\vee\BB$ conveys
no information to Bob if
\begin{equation}(T^{*}\rho) |_{\BB}=\rho |_{\BB}\mathrm{\;for\;all\;states\;}
\rho\mathrm{\;of\;}\BB.\label{defnoinf}\end{equation}\label{defnoinff}\end{defn}
An operation here is understood as a completely positive linear
map on algebra $\BA$ and $T^{*}\rho$ is a state over the algebra
defined for every state $\rho$ on the \textit{same} algebra $\BA$
as
\begin{equation}
(T^{*}\rho)(A)=\frac{\rho(T(A))}{\rho(T(I))}\label{eq84}
\end{equation}
at the condition that $\rho(T(I))\neq 0$. Nonselective
measurements $T$ are the ones that have $T(I)=I$, and then
$\rho(T(I))=\rho(I)=||\rho ||=1$. CBH explain that, in their view,
Definition~\ref{defnoinff} entails
\begin{equation}T(B)=B\mathrm{\;for\;all\;}B\in\BB.\label{cninf}\end{equation}

CBH then assert that if the condition (\ref{cninf}) holds for all
self-adjoint $B\in\BB$ and for all $T$ of the form
\begin{equation}T=T_E(A)=E^{1/2}AE^{1/2}+(I-E)^{1/2}A(I-E)^{1/2},\label{teform}\end{equation}
where $A\in \BA\vee\BB$ and $E$ is a positive operator in $\BA$,
then algebras $\BA$ and $\BB$ are kinematically independent
\cite[Theorem~1]{Bub}. CBH seek for kinematic independence of
algebras in order to show that the algebras of two distinct
systems commute, and this is derived from the assumption of
$C^*$-independence and from the condition (\ref{defnoinf}), where
$C^*$-independence is brought into the discussion to grasp the
meaning of the fact that systems $\BA$ and $\BB$ are distinct.
Mathematically, $C^*$-independence means that for any state
$\rho_1$ over $\BA$ and for any state $\rho_2$ over $\BB$ there is
a state $\rho$ over $\BA\vee\BB$ such that $\rho | _{\BA}=\rho_1$
and $\rho | _{\BB}=\rho_2$. $C^*$-independence does not follow
from and does not entail kinematic independence. In the CBH paper,
Definition~\ref{defnoinff} is equated with the `no superluminal
information transfer by measurement' constraint. The term
``superluminal'' is an evident spatiotemporal concept designating
velocities that exceed the speed of light. In the discussion of
this constraint, however, no light quanta or any other carriers
that actually transfer information are considered and indeed no
space-time at all is necessary: the mathematics involved is purely
algebraic. Then, the question is whether one could give a
different meaning to this condition, without bringing in
spatiotemporal concepts that do not naturally belong to the
language of the algebraic approach. Before suggesting an answer to
this question, we stop to present two critical points concerning
Definition~\ref{defnoinff} and its discussion in the CBH paper.

Our first critique is connected with the phrasing of
Definition~\ref{defnoinff} itself. If, following the CBH authors,
in this definition $\rho$ is to be taken as a state over $\BB$,
then the definition does not make sense: operation $T$ is defined
on $\BA\vee\BB$ and consequently, in accordance with (\ref{eq84}),
$T^{*}\rho$ is defined for the states $\rho$ over $\BA\vee\BB$. If
one follows the CBH definition with a state $\rho$ over $\BB$,
then there would be no need to write $\rho |_{\BB}$ as CBH do, for
a simple reason that $\rho |_{\BB} = \rho$. To suggest a remedy,
we extend the reasoning behind this definition and reformulate it
in three alternative ways.
\begin{itemize}\item The first one is to require that in
Definition~\ref{defnoinff} the state $\rho$ be a state over the
algebra $\BA\vee\BB$. \item The second alternative is to consider
states $\rho$ on $\BB$ but to require a different formula, namely
that $(T|_{\BB})^{*}\rho=\rho$ as states over $\BB$. \item
Finally, the third alternative proceeds as follows: Take arbitrary
states $\rho _1$ over $\BA$ and $\rho _2$ over $\BB$ and, in
virtue of $C^*$-independence, consider the state $\rho$ over $\BA
\vee \BB$ such that its marginal states are $\rho _1$ and $\rho
_2$ respectively. Then $T^{*}\rho$ is also a state over
$\BA\vee\BB$. If its restriction $(T^{*}\rho) |_{\BB}$ is equal to
$\rho _2$, then $T$ is said to convey no information to
Bob.\end{itemize}

With the original formulation of Definition~\ref{defnoinff}, proof
of Equation~\ref{cninf} is problematic. We show how to prove this
equation with each of the three alternative definitions. First
observe the following remark.
\begin{rem}
Each $C^*$-algebra has sufficient states to discriminate between
any two observables (i.e., if $\rho(A)=\rho(B)$ for all states
$\rho$, then $A=B$). \label{reminff}\end{rem} \noindent To justify
(\ref{cninf}), the CBH authors then say:
\begin{quote}
$(T^*\rho)|_{\BB}=\rho |_{\BB}$ if and only if
$\rho(T(B))=\rho(B)$ for all $B\in\BB$ and for all states $\rho$
on $\BA\vee\BB$. Since all states of $\BB$ are restrictions of
states on $\BA\vee\BB$, it follows that $(T^*\rho)|_{\BB}=\rho
|_{\BB}$ if and only if $\omega(T(B))=\omega(B)$ for all states
$\omega$ of $\BB$, i.e., if and only if $T(B)=B$ for all $B\in
\BB$.
\end{quote}
Let us examine this derivation under each of the three alternative
definitions of conveying no information. By the definition of
$T^*$, we have $(T^*\rho)(B)=\rho(T(B))$ for all states $\rho$
over $\BA\vee\BB$. To obtain from this that $\rho(T(B))=\rho(B)$,
one must show that $(T^*\rho)(B)=\rho(B)$, and this is equivalent
to saying that $(T^*\rho)|_{\BB}=\rho |_{\BB}$ for all states
$\rho$ over $\BA\vee\BB$. Now, according to CBH, one would need to
show that $\rho(T(B))=\rho(B)$ if and only if
$\omega(T(B))=\omega(B)$ with states $\rho$ over $\BA\vee\BB$ and
$\omega$ over $\BB$. The latter formula, however, is not
well-defined: operator $T(B)$, generally speaking, is not in
$\BB$. Fortunately, we are salvaged by the first alternative
reformulation of Definition~\ref{defnoinff}: because
$\rho(T(B))=\rho(B)$ is true for all states $\rho$ over
$\BA\vee\BB$, we obtain directly that $T(B)=B$ in virtue of
Remark~\ref{reminff}.

The second alternative definition of conveying no information
makes use of an object such as $(T|_{\BB})^{*}\rho$. To give it a
meaning in the algebra $\BB$, one needs to impose a closure
condition on the action of $T$ on operators $B\in \BB$: namely,
that $T$ must not take operators out of $\BB$. The problem here is
the same as the one we encountered in the discussion of the
previous alternative, and it is only by assuming the closure
condition that one is able to obtain that $T(B)=B$.

In the third alternative, for the state $\rho$ over $\BA\vee\BB$,
write from the definition of $T^*$ that $(T^*\rho)(B)=\rho(T(B))$.
The result $(T^*\rho)(B)$ is the same as $(T^*\rho)|_{\BB}(B)$,
and this is equal to $\rho _2 (B)$. Consequently, $\rho(T(B))=\rho
_2(B)=\rho (B)$. Can we now say that this holds for all states
$\rho$ over $\BA\vee\BB\;$? The answer is obviously yes, and this
is because each state over $\BA\vee\BB$ can be seen as an
extension of its own restriction to $\BB$. Therefore, one has to
modify Definition~\ref{defnoinff} for it to be formally correct,
and this entails a modification in the proof of
Equation~\ref{cninf}.

The second critique of the CBH program has to do with postulating
$C^*$-indepen\-dence. Notions of independence of algebras are a
legion \cite{florig}; why, then, take $C^*$-independence as a
mathematical representation of the distinction between the
systems? For this we must look back at the origins of the notion
of $C^*$-independence. It was first introduced in Ref.~\cite{hk}
under the name of statistical independence; this was due to the
fact that Haag and Kastler wanted to give a mathematical meaning
to the ability to prepare any states on two algebras by the same
preparation procedure. As Florig and Summers importantly note, if
one has an entangled pair, then it generates $C^*$-independent
algebras that are not kinematically independent. Now read again
the phrase from the CBH article that we have already quoted: The
sense of the `no superluminal information transfer' constraint is
that ``when Alice and Bob perform local measurements, Alice's
measurements can have no influence on the statistics for the
outcomes of Bob's measurements.'' So which is the statistical
independence: $C^*$-independence or the `no superluminal
information transfer' constraint? This is where we have to look at
the meaning of the mysterious term ``superluminal'' that in the
CBH case has nothing to do with faster-than-light transfer of
information. In fact, conveying no information as defined in
\ref{defnoinff} does not prohibit only superluminal communication;
it prohibits all information transfer whatsoever. The real meaning
of the CBH condition is thus that nonselective POV measurements
can convey no information to Bob at all. As for selective
measurements, the authors themselves grant that they ``trivially
change the statistics of observables measured at a distance,
simply in virtue of the fact that the ensemble relative to which
one computes the statistics has changed.''

Now, if the operation $T$ is nonselective, the most important
thing that does not change is that the identity operator remains
in the image of $T$. Presence of the identity is a \textit{sine
qua non} for all algebras in the CBH paper. However, if the
identity is present in the algebra, the latter becomes quite
special; for instance, according to Theorem~\ref{vNdoubc},
requiring that the algebra be unital is a first step on the way to
von Neumann algebras. More seriously, which operators are included
in $\BB$ determines Bob's observational capacities. Consider, for
example, Alice and Bob as two entangled particles; then the
identity will generally not be a part of their algebras. In an
example from Ref.~\cite{florig}, the following operators on the
6-dimensional complex Hilbert space are considered:
\begin{equation}
E=\left(\begin{tabular}{cccccc}
  1 & 0 & 0 & 0 & 0 & 0 \\
  0 & 0 & 0 & 0 & 0 & 0 \\
  0 & 0 & 1 & 0 & 0 & 0 \\
  0 & 0 & 0 & 0 & 0 & 0 \\
  0 & 0 & 0 & 0 & 1 & 0 \\
  0 & 0 & 0 & 0 & 0 & 0 \\
\end{tabular}\right),
F=\left(\begin{tabular}{cccccc}
  1 & 0 & 0 & 0 & 0 & 0 \\
  0 & 1 & 0 & 0 & 0 & 0 \\
  0 & 0 & 0 & 0 & 0 & 0 \\
  0 & 0 & 0 & 0 & 0 & 0 \\
  0 & 0 & 0 & 0 & 1/2 & 1/2 \\
  0 & 0 & 0 & 0 & 1/2 & 1/2 \\
\end{tabular}\right).\end{equation}
Each of these operators generates a $C^*$-algebra. These algebras
$\BE$ and $\BF$ are $C^*$-independent but evidently do not
commute. They also do not contain the identity. According to the
CBH view, the entangled systems $\BE$ and $\BF$ are distinct, but
the transfer of information by measurement is possible between
them. The general form of operation $T$ acting on operators from
$\BE$ and $\BF$ is to leave the diagonal elements untouched and to
nullify all others, so it does not preserve the form of $B$. One
can now see that the notion of system in the CBH understanding is
quite peculiar: by requiring kinematic independence, they for
example contradict Rovelli's requirement (see
Section~\ref{rovsect}) that everything be equally treated as
physical system. They indeed see a $C^*$-algebra as a collection
of operators ``sitting'' in some place, that includes the identity
as the operator that corresponds to \textit{doing nothing} on the
part of the observer. In other words, to be $C^*$-independent is
not enough for being distinct: there has to be a supplementary
intuitive assumption of the local identity of systems made along
the way. In Rovelli's sense, state on an algebra and the
information that it reflects are observer-dependent concepts; then
the point of the first CBH constraint is to say that the
information obtained in measurement can either be possessed
exclusively by Alice or exclusively by Bob, i.e. the observer who
performs the measurement in question and who obtains the new fact
in which information is brought about.

In an attempt to escape from the above identified intuitive
assumption, let us reformulate the CBH mathematical results, which
we fully endorse, in a different language. As a possible
additional assumption to $C^*$-independence, one can directly
\textit{postulate} that to be physically distinct means to be
kinematically independent. Then, to derive kinematic independence
would amount to explaining what it means to be physically
distinct, based on the statistical independence; and this will be
the meaning of Definition~\ref{defnoinff}. A methodological
argument for this latter choice goes as follows:
$C^*$-independence is a notion that relies on the notion of state.
In the conceptual framework of Section~\ref{itjsect}, the notion
of state represents information that I-observer has about the
system, while the notion of operator, which is an element of a
$C^*$-algebra, contributes to the definition of the system as
such. As we have seen, for the CBH, too, the choice of operators
that are included in the $C^*$-algebra is crucial for
comprehension of the concept of observer. It is then natural to
require that the fact that two systems are distinct be expressed,
first of all, in the same language as used to define what a system
is; i.e. in the language of the $C^*$-algebraic constituent
operators and not the one of the states.

Only after one had postulated what it means for two physical
systems represented as $C^*$-algebras to be distinct, it comes
without surprise that in order to establish this difference
between the two systems practically, one will appeal to
constraints on how information about one system relates to
information about the other. Further, because the notion of
information has so reemerged and because information is
represented by states on the algebra, one expects a definition in
terms of states; and indeed Definition~\ref{defnoinff} speaks the
language of states.

Let us now clarify what we formally mean by \textit{distinct
physical systems}.

\begin{defn}
Two systems represented as $C^*$-algebras $\BA$ and $\BB$ are
distinct if $\forall A\in\BA, B\in\BB$ $[A,B]=0$. In the standard
terminology, we say that, by definition, systems are physically
distinct if they are kinematically independent.
\label{distdef}\end{defn}

The meaning of the notion of distinct physical systems here
becomes operational. This is due to the following theorem which
rephrases the first theorem by CBH:

\begin{thm}[information-theoretic criterion for two systems to be physically distinct]
If all POV measurements on system $\BA$ provide no information on
system $\BB$ (in the sense of Definition~\ref{defnoinff}), then
systems $\BA$ and $\BB$ are physically distinct.
\label{distth}\end{thm}

With the reformulations \ref{distdef} and \ref{distth} of the CBH
result, we have liberated the discussion from the spatiotemporal
language that appeared in the usage of terms like ``superluminal''
or ``local'' and that does not belong to the natural language of
algebra. The term ``locality'' was introduced in the theory of
algebraic independence conditions by Kraus \cite{kraus2,Kraus},
who formulated the condition of strict locality for $W^*$-algebras
that we do not present here to avoid heaping too many definitions.
Under the assumption of kinematic independence, strict locality is
equivalent to $C^*$-independence \cite[Proposition~9]{florig}. In
our language, this means that if two systems are distinct, then
strict locality would be equivalent to statistical independence: a
strange condition that links together words belonging to different
vocabularies. Indeed, algebra is the mathematical science of
structure, and that ``$A$ is distinct from $B$'' is a perfectly
structural claim that need not refer to spacetime concepts like
locality. One then sees that the strangeness arises from the use
of the term ``locality,'' and it is this use that must be
questioned.

The second CBH information-theoretic constraint is the `no
broadcasting' condition whose aim is to establish that algebras
$\BA$ and $\BB$, taken separately, are non-Abelian. Broadcasting
is defined as follows:

\begin{defn}[{\cite[Section 3.3]{Bub}}]
Given two isomorphic, kinematically independent $C^*$-algebras
$\BA$ and $\BB$, a pair $\{\rho _1,\rho _2\}$ of states over $\BA$
can be broadcast in case there is a standard state $\sigma$ over
$\BB$ and a dynamical evolution represented by an operation $T$ on
$\BA\vee\BB$ such that $T^*(\rho _i\otimes\sigma) |_{\BA}=T^*(\rho
_i\otimes\sigma) |_{\BB}=\rho _i$, for $i=0,1$. A pair $\{\rho
_1,\rho _2\}$ of states over $\BA$ can be cloned just in case
$T^*(\rho _i\otimes\sigma)=\rho _i\otimes\rho _i$ ($i=0,1$).
\end{defn}

Equivalence between the `no broadcasting' condition and
non-Abelianness of the $C^*$-algebra is then derived from the
following theorem:

\begin{thm}Let $\BA$ and $\BB$ be two kinematically independent $C^*$-algebras.
Then:\begin{description}\item[(i)] If $\BA$ and $\BB$ are Abelian
then there is an operation $T$ on $\BA\vee\BB$ that broadcasts all
states over $\BA$.\item[(ii)] If for each pair $\{\rho _1,\rho
_2\}$ of states over $\BA$, there is an operation $T$ on
$\BA\vee\BB$ that broadcasts $\{\rho _1,\rho _2\}$, then $\BA$ is
Abelian.
\end{description}\label{nobroadth}\end{thm}

It is an interesting fact that in the section where broadcasting
is discussed, although it, too, is a term with explicit
spatiotemporal connotations, the authors never refer to
broadcasting as actually transferring information in space. Such
is not the case with the two other information-theoretic
constraints. It is perhaps due to the fact that initial intention
was to use the `no cloning' condition, with the word ``cloning''
being free of spatial connotations. However, one fact deserves
closer attention: that non-Abelianness of the algebras $\BA$ and
$\BB$, taken one by one, is proved by assuming that they are
kinematically independent. It means that quantumness, of which
non-Abelianness is a necessary ingredient, is not a property of
any given system taken separately, as if it were the only physical
system in the Universe, but in order to derive the quantum
behaviour, one must consider the system in the context of at least
one other system that is physically distinct from the first one.
As a consequence, for example, this forbids the possibility of
treating the whole Universe as a quantum system, echoing our
remark on page~\pageref{unino}. For the remainder of the
discussion of the second constraint we agree with the conclusions
made by the CBH authors.

The third, `no bit commitment' constraint is discussed in Section
3.4 of Ref. \cite{Bub}. The section opens with the following
claim:
\begin{quote}
We show that the impossibility of unconditionally secure bit
commitment between systems $\BA$ and $\BB$, in the presence of
kinematic independence and noncommutativity of their algebras of
observables, entails nonlocality: spacelike separated systems must
at least sometimes occupy entangled states. Specifically, we show
that if Alice and Bob have spacelike separated quantum systems,
but cannot prepare any entangled state, then Alice and Bob can
devise an unconditionally secure bit commitment protocol.
\end{quote}

This citation essentially involves spatiotemporal terms. One is
then tempted to analyze the CBH proof so as to enlist the
occurrences of formal space-time considerations in it. The
derivation starts by showing that quantum systems are
characterized by the existence of non-uniquely decomposable mixed
states: a $C^*$-algebra $\BA$ is non-Abelian if and only if there
are distinct pure states $\omega _{1,2}$ and $\omega _{\pm}$ over
$\BA$ such that $\frac{1}{2}(\omega _1+\omega
_2)=\frac{1}{2}(\omega _+ +\omega _-)$. This result is used to
prove a theorem showing that a certain proposed bit commitment
protocol is secure if Alice and Bob have access only to
classically correlated states (i.e. convex combinations of product
states).

\begin{thm}[the CBH `no bit commitment' theorem]
If $\BA$ and $\BB$ are non-Abelian then there is a pair $\{\rho
_0,\rho _1\}$ of states over $\BA\vee\BB$ such that:
\begin{enumerate}
    \item $\rho _0 |_{\BB}=\rho _0 |_{\BB}$.
    \item There is no classically correlated state $\sigma$ over
    $\BA\vee\BB$ and operations $T_0$ and $T_1$ performable by
    Alice such that $T^*_0\sigma=\rho _0$ and $T^*_1\sigma=\rho
    _1$.
\end{enumerate}
\label{nobitcommth}\end{thm}

From this theorem the authors deduce that the impossibility of
unconditionally secure bit commitment entails that ``if each of
the pair of \textit{separated}\symbolfootnote[2]{Our emphasis.}
physical systems $\BA$ and $\BB$ has a non-uniquely decomposable
mixed state, so that $\BA\vee\BB$ has a pair $\{\rho _0, \rho
_1\}$ of distinct classically correlated states whose marginals
relative to $\BA$ and $\BB$ are identical, then $\BA$ and $\BB$
must be able to occupy an entangled state that can be transformed
to $\rho _0$ or $\rho _1$ at will by a local operation.'' The term
``separated'' is essential and, nevertheless, its precise meaning
is not defined in the CBH article. In Theorem~\ref{nobitcommth}
one requires that algebras $\BA$ and $\BB$ be non-Abelian. This
latter fact is taken as a consequence of Theorem~\ref{nobroadth},
which, in turn, requires that algebras $\BA$ and $\BB$ be
kinematically independent. So the meaning of ``separated'' must be
no more than to say that the systems are distinct in the sense of
the Definition~\ref{distdef}. There are no mathematical reasons to
claim, as the authors do in the above cited passage, that they
have taken into account the case when Alice and Bob have
``\textit{spacelike} separated systems.''
Theorem~\ref{nobitcommth} means that if systems $\BA$ and $\BB$
are distinct and unconditionally secure bit commitment is
impossible, then these systems can actually be in an entangled
state. To be in an entangled state here means that information
about systems $\BA$ and $\BB$ is such that any act of bringing it
about will necessarily provide one with the information about the
system $\BA$ and, \textit{logically} linked to it, with the
information about the system $\BB$. At no place here enters any
spatiotemporal language. Note the importance of the word
``actually'': in fact, presence of entangled states in the
mathematical formalism has long been guaranteed by non-Abelianness
and the kinematic and the $C^*$-independencies of algebras
\cite{sumwer}. The CBH authors devise the whole argument in order
to demonstrate that the entangled states, mathematically allowed,
are \textit{actually}---or shall we say
\textit{necessarily}---non-locally instantiated.

The authors of the CBH article then discuss a result converse to
Theorem~\ref{nobitcommth} which is arguably more interesting:
namely, in their terminology, that nonlocality---``the fact that
spacelike separated systems occupy entangled states''---entails
the impossibility of unconditionally secure bit commitment. We
have already seen that the term ``nonlocality'' is superfluous in
the algebraic context, although for this converse result it is not
an issue of first importance. The derivation relies on the
availability of the Hughston-Jozsa-Wootters (HJW)
theorem~\cite{HJW} for arbitrary $C^*$-algebras. The most general
proof up-to-date was given by Halvorson \cite{halvor}; it covers
the cases of type I von Neumann factors, type I von Neumann
algebras with Abelian superselection rules and the case of a
$C^*$-algebra whose commutant is a hyperfinite von Neumann
algebra. Let us stress the term \textit{hyperfinite}. Halvorson
claims that it remains an open question whether an analogue of the
HJW theorem holds for general $C^*$-algebras that are not
necessarily nuclear. Recall that nuclearity, mentioned in
Section~\ref{vnprp}, is the cause of hyperfiniteness of the type
$III_1$ von Neumann factors, and it is equivalent to the
requirement for the system to have normal thermodynamic
properties. Halvorson's desire to establish the analogue of the
HJW theorem in absence of nuclearity may therefore be prevented
from realization by the theory itself. The phrase ``normal
thermodynamic properties'' means that KMS states exist for all
positive $\beta$ for the system and its finitely extended parts,
and this is intimately linked to information-theoretic
interpretation of the formalism of local algebras. There may exist
no information-theoretic approach as such beyond the limits of
applicability of the KMS condition.

We have given in Section~\ref{partchoi} an
infor\-ma\-tion-theoretic interpretation in which hyperfiniteness
is justified based on Axioms~\ref{ax1} and \ref{ax2}. In this
section we offered critique of the extensive use of spatiotemporal
notions in the CBH articles. We must now explain how space and
time, instead of being postulated, can arise in the algebraic
information-theoretic framework. This, in turn, will involve the
KMS formalism, and hyperfiniteness as the condition of
well-definedness of the KMS states will be required.

\section{Non-fundamental role of spacetime}\label{nfrole}

\epigraph{\ldots the concepts of space and time by their very
nature acquire a meaning only because of the possibility of
neglecting the interactions with the means of measurement.}{Bohr
\cite[p.~99]{bohr1934}}

At many occasions in the history of quantum theory it has been
noticed that time and the ordering of wavefunction collapses are
unrelated, of which we cite two: First was the point emphasized by
Dirac \cite{dirac} and later discussed by Hartle \cite{hartleD}
and Rovelli \cite{Rovelli1,Rovelli2}. The argument here is very
general: The formalism of quantum mechanics allows a sequence of
measurements not ordered in the time in which the system evolves.
Thus, we can measure $B(t)$ and \textit{later} measure
$A(t^\prime)$, with $t^\prime <t$. In the standard Copenhagen
interpretation we then say that the wavefunction is projected
twice: \textit{first} on the eigenstate of $B(t)$ and
\textit{then} on the eigenstate of $A(t^\prime)$. This sequence of
projections describes the conditional probability of finding at
$A(t^\prime)$ the system that will have been detected at $B(t)$.
Such a probability can be understood either as subjective or as
objective in terms of frequencies: none of this changes the
inverse order of detection events with respect to the time in
which the system evolves.  In an illuminating passage following
this example, Rovelli writes:
\begin{quote}
The example suggests that the ordering of the collapses is not
determined by $t$. Rather, the ordering depends on the
\textit{question} that we want to formulate. The ordering is
usually related to $t$ only because we are more interested in
calculating the future than the past.
\end{quote}

The idea that the ordering depends on the question that we want to
formulate is in full accord with the conceptual approach that we
have chosen in Chapter~\ref{sect3}, where questions correspond to
facts as acts of bringing about information. Facts, in turn,
belong to fundamental notions on which rests the physical theory.
Thus time ordering is secondary, and it comes without surprise
that quantum theory can be formulated as \textit{timeless} quantum
theory \cite[Chapter 5]{rovellibook}.

The same idea is echoed in the thought of Peres who studies the
second occasion when scientists realized how little the
conventional linear time means to a quantum system. Discussing
quantum teleportation, Peres writes:

\begin{quote}Alice and Bob are not real people. They are inanimate
objects. They know nothing. What is teleported instantaneously
from one system (Alice) to another system (Bob) is the
applicability of the preparer's knowledge to the state of a
particular qubit in these systems. \cite{peresibm}\end{quote}
Applicability of the preparer's knowledge is the same thing as
Rovelli's ``question that we want to formulate.'' In our approach,
it corresponds to the concept of relevance of information for
I-observer. Indeed, by saying that ``they know nothing'' Peres
places Alice and Bob in the domain of purely physical, i.e.
intratheoretic, and the metatheoretic function of informational
agent, or I-observer, is transferred to an external ``preparer.''
If one now returns to the fundamental view in which the von
Neumann cut is put to position zero, and all systems are treated
on equal grounds, then the metatheoretic function of I-observer
can as well belong to Alice or to Bob, but this will not change
Peres's argument: what is ``teleported'' is relevant information.
Quotation marks mean that no information is actually
instantaneously transferred, because information states, as we
have emphasized, are relational, and information in question is
always possessed by one I-observer only, i.e. exclusively Alice or
exclusively Bob. Communication of information from Alice to Bob
via a classical channel falls out of the field of interest of the
information-based quantum theory with a given observer, as any
other theory of communication of information between distinct
informational agents requires a loop cut of Figure~\ref{loop1}.

\begin{figure}[htbp]
\begin{center}
\epsfysize=2.5in \epsfbox{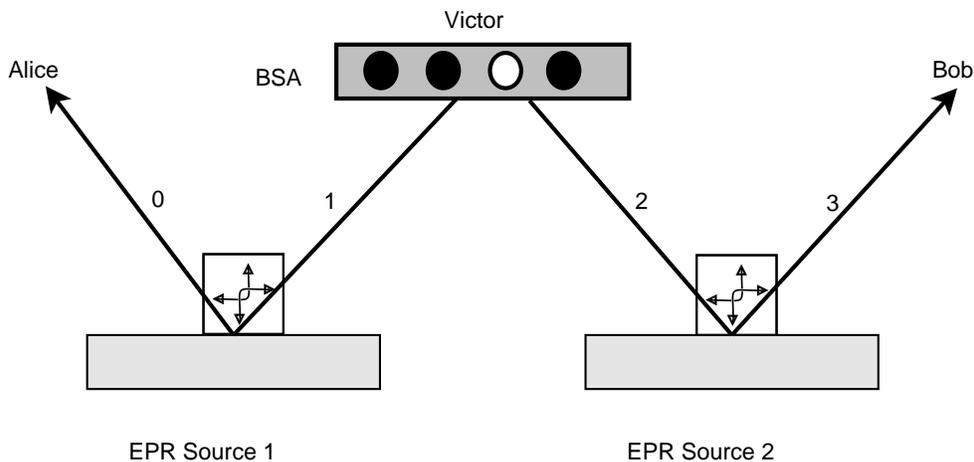}
\end{center}
\caption{\singlespacing Scheme of entanglement swapping, as
adopted from \cite{baz}. Two pairs of entangled particles 0-1 and
2-3 are produced by two Einstein-Podolsky-Rosen (EPR) sources. One
particle from each of the pairs is sent to two different
observers, say particle 0 is sent to Alice and particle 3 to Bob.
The other particles 1 and 2 from each pair are sent to Victor who
subjects them to a Bell-state analyzer (BSA), by which particles 0
and 3 become entangled although they may have never interacted in
the past.} \label{entswap}
\end{figure}

The above mentioned second occasion has to do with the
long-lasting debate that was originally started by Einstein and
Bohr who discussed the double-slit experiment \cite{EPR,bohr35},
later continued by Wheeler in the form of the ``delayed-choice''
experiment \cite{wheel78}, and that we present here in the version
having to do with quantum information, which is called
``entanglement swapping'' \cite{jwhz1,ryff,jwhz2}
(Figure~\ref{entswap}). 

Contrary to the CBH paper discussed in Section~\ref{cbhsect}, here
the authors, who also employ the quantum computational language of
Alice and Bob, state very clearly that their usage of terms like
``locality'' has nothing to do with spacetime separation. The only
important factor is that Alice, Bob and Victor be distinct
physical systems. Irrelevance of the temporal ordering may even
give rise to seemingly paradoxical situations, like in the
following passage:

\begin{quote}
It is now important to analyze what we mean by ``prediction.'' As
the relative time ordering of Alice's and Bob's events is
irrelevant, ``prediction'' cannot refer to the time order of the
measurements. It is helpful to remember that the quantum state is
just an expectation catalogue. Its purpose is to make predictions
about possible measurement results a specific observer does not
know yet. Thus which state is to be used depends on which
information Alice and Bob have, and ``prediction'' means
prediction about measurement results they will learn in the future
independent of whether these measurements have already been
performed by someone or not\ldots It is irrelevant whether Alice
performs her measurement earlier in any reference frame than Bob's
or later or even if they are spacelike separated when the
seemingly paradoxical situation arises that different observers
are spacelike separated. \cite{jwhz2}
\end{quote}

It is clear from the discussion of the entanglement swapping and
from Dirac's argument given above that the concept of two
\textit{distinct} physical systems (e.g. observers) in the
information-based quantum theory has very little to do with the
spacetime separation between the systems. What role do then space
and time play? In our program of the foundation of quantum theory,
there is no place for space and time among the fundamental notions
of the theory. They are, consequently, non-fundamental and need to
be derived from the fundamental notions and the axioms. We propose
a way to achieve this for the notion of time. As for space, we can
only say that the allegedly very important role of the spatial
notion of locality has been overestimated, as we intended to show
in Section~\ref{cbhsect}. In the information-theoretic approach,
locality as the criterion of distinction between systems can be
replaced by a different, properly information-theoretic criterion.
Perhaps, a consistent mathematical approach to reconstructing
space in the context of the information-theoretic approach will
proceed by the methods of {loop\label{spaceprob}} quantum
gravity~\cite{rovellibook}.

To return to the problem of time, the intuition here is to use the
ideas from thermodynamics. Indeed, if quantum mechanics can be
formulated as timeless theory, then one has to look elsewhere for
reasons why time is so special a parameter. An interesting
possibility \cite[p.~100]{rovellibook} is that it is the
statistical mechanics, and therefore thermodynamics, that singles
out $t$ and gives it special properties.

In the algebraic approach, we have a $C^*$-algebra with a
preferred state, giving rise to the Hilbert space representation.
One then defines a von Neumann algebra as explained in
Remark~\ref{howvN}, and, in a von Neumann algebra, Gleason's
Theorem~\ref{gleasonth} is applicable so as to justify the
probabilistic interpretation and the Born rule. This construction
allows to build all elements of the quantum theory except unitary
dynamics. As discussed in Section~\ref{sect44}, in a non-generally
covariant setting it is impossible to derive spacetime without
introducing additional assumptions. We also know from
Equation~\ref{eq20} and Proposition~\ref{kmsconnect} that, in a
non-generally covariant theory, an equilibrium state is the one
whose modular group is the time translation group.

Now consider generally covariant theories. The theory is given by
the hyperfinite $C^*$-algebra $\mathcal{A}$ of generally covariant
physical operators, states $\omega$ over $\mathcal{A}$ and no
additional information about dynamics. Each state $\omega$ that
represents information about the system is generically impure, for
it cannot but approach---recall that the amount of information is
finite---the large number of the degrees of freedom allowed in a
hyperfinite $C^*$-algebra. The hypothesis in Ref.
\cite{ConnesRovelli} (see also \cite{Heller}) is that to define
time in such a case, one must look at the thermodynamics of the
system. In one phrase, ``time is a side effect of our ignorance of
the microstate''~\cite{rovelliprivate2004}; we should like to
shorten this assertion even further: time is ignorance; or yet in
a third way: time is not knowing. When I-observer chooses to throw
away some previously available information as irrelevant, it gives
rise to time. To translate this idea into formal terms, we say
that time is a state-dependent notion and is given by the modular
group $\alpha ^\omega_t$ of $\omega$ as defined in
Equation~(\ref{eq8}). This time flow will be denoted as
\textit{thermal time}. Connes's and Rovelli's thermal time
hypothesis reads:
\begin{quote}
In nature, there is no preferred physical time variable $t$. There
are no equilibrium states $\rho _0$ preferred \textit{a priori}.
Rather, all variables are equivalent; we can find the system in an
arbitrary state $\rho$; if the system is in state $\rho$, then a
preferred variable is singled out by the state of the system. This
variable is what we call time.~\cite[p.~101]{rovellibook}
\end{quote}

The fact that time is determined by the KMS state, and therefore
the system is always in thermodynamic equilibrium with respect to
the thermal time flow, does not imply that its evolution is
frozen. In a quantum system with an infinite number of the degrees
of freedom, what we generally measure is the effect of small
perturbations around a thermal state. In other words, facts bring
about new information and thereby define new states, but on the
scale of the $C^*$-algebra of the system, each new state does not
drastically differ from the old state. In a generally covariant
setting, given the algebra of observables $\mathcal{A}$ and a
state $\omega$, the modular group gives a time flow $\alpha
^\omega _t$. Then, the theory describes physical evolution in the
state-dependent thermal time in terms of amplitudes of the form
\begin{equation}\label{eq26}
F_{A,B}(t)=\omega(\alpha _t(B)A),
\end{equation}
where $A$ and $B$ are operators in $\mathcal{A}$. The quantity
$F_{A,B}(t)$ is related to the probability amplitude for obtaining
information pertaining to $B$ in a fact that will be established
after ``waiting'' for time $t$ following a preparation $\M$, i.e.
departing from a state $\omega_A$ that describes information about
the complete knowledge of $\M$. Time $t$ here is the thermal time
determined by the state $\omega_A$ of the system. In a generally
covariant setting the thermal time is the only definition of time
available. The essence of the definition is then that the
quotation marks around the word \textit{waiting} must be removed.

In a theory in which a geometrical definition of time is assumed
independently from the thermal time (as in Section~\ref{sect44}),
arises a problem of relating the two times. From the study of the
non-relativistic limit of generally covariant theories with
thermal time one obtains that the latter is proportional to
geometrical time, and the temperature can be interpreted as a
ratio between the two. Connes and Rovelli \cite{ConnesRovelli}
study the non-relativistic limit, where modular time is preserved
but the conventional time also becomes meaningful, and show that
the modular group of Equation~(\ref{eq8}) and the time evolution
group in the non-relativistic limit introduced in
Equation~(\ref{eq20}) are linked:
\begin{equation}
\alpha^\omega_t=\gamma_{\beta t}.\end{equation}

In the spirit of Bohr's quotation put in the epigraph to this
section we must now show how from the state-dependent notion of
time one can, by way of neglecting certain information, make sense
of the state-independent notion of time. It is the time of the
state-independent notion of time that indexes acts of bringing
about information and turns them into facts, an assumption that we
made for the non-generally covariant theory in
Section~\ref{defmeasurement}. Note that Bohr's words are also
closely tied with our discussion in Section~\ref{ipobservers} of
the necessity to distinguish between I-observer and P-observer. If
one places himself in a world-picture in which there is no cut
(Figure~\ref{loop0}), then one would have to accept simultaneously
that time (and space) can be derived within a physical theory, but
it also determines the possibility of meta-theory of that physical
theory. Both aspects of the concept of time cannot be described in
a single theory, for otherwise that would render it logically
circular. What Bohr says one must neglect for space and time to
arise is that measurement is physical, i.e. the existence of
P-observer. It corresponds to cutting the loop
(Figure~\ref{loop01}) and neglecting the fact the information is
physical, i.e. that it has some physical support like, for
instance, a human body, and thereby one will render the concept of
time a topic open for a theoretic justification. We base the
theory on information and we are thus uninterested, as it was the
case with factoring out P-observer, in the loop cut of
Figure~\ref{loop1}. However, we must justify why, by neglecting
information, I-observer, or the informational agent, acquires the
possibility to observe a single state-independent flow of time
instead of the variety of different state-dependent notions of
time.

In the covariant setting, in general, the modular flow is not an
inner automorphism of the algebra, namely, there is no hamiltonian
in $\BM$ that generates it. However, as shown in
Section~\ref{vnprp}, the difference between two modular flows is
always an inner automorphism and, therefore, any modular flow
projects on the same 1-parameter group of elements in
$\mathrm{Out}\:\BM$. Consequently, the flow $\tilde{\alpha} _t$
defined after Equation~(\ref{eq10}) is canonical: it depends only
on the algebra itself. To factorize the states into classes of
states of which modular automorphisms are inner-equivalent means
to neglect information: only that information is kept which is
characteristic of the class, and information that distinguishes
states within the class is lost. The passage from the
state-dependent modular time flow to the flow $\tilde{\alpha} _t$
is therefore achieved via neglecting information, in full accord
with Bohr's idea.

As follows from Table~\ref{classtable2}, in type $I$ and type $II$
von Neumann algebras the canonical modular flow is frozen at
modular time $t=0$: indeed, evolution is unitary and no
information can be brought about by the no-collapse
Schr\"{o}dinger dynamics. In type $III_1$ von Neumann algebra,
which corresponds to the theory of local algebras which we
interpreted information-theoretically in Section~\ref{partchoi},
the modular time flow covers all $\mathbb{R}_+$, thus coinciding
with the intuition of infinite linear time; but it is now the
algebra that determines the ``intuitive'' time flow. Therefore, a
von Neumann algebra contains an intrinsic dynamics, and the time
needs no more to be externally postulated: indeed, it can be
derived intratheoretically in the context of the
information-theoretic approach, with the conceptual help of
thermodynamics that belongs to meta-theory of this approach, but
without any interference of thermodynamics in the actual formalism
of the theory.

To conclude, let us briefly summarize the key ideas of this
section. In an infor\-ma\-tion-theoretic framework we start with
the fundamental notions of system, information and fact. In the
algebraic formalism a system is interpreted as a C*-algebra and
information is interpreted as state over this algebra. There is no
space and no time yet, for we have not postulated anything like
space or time. Via the KMS formalism every state gets its flow, so
each information state has its own flow; we call it
state-dependent time. What are the consequences?
\begin{itemize}
\item Time is a state-dependent concept. Unless the state is
changed time does not change. A change in the state means a change
in information. A change in information can be brought about in a
new fact. At each fact state-dependent time ``restarts.'' We see
that the temporality of facts (variable $t$ that indexes facts)
has nothing to do with the state-dependent notion of time.

\item Thermodynamics has not played any role so far. To view a
state as a KMS state at $\beta=1$ and to define the flow, we need
not say that a state over $C^*$-algebra is a thermodynamical
concept. Therefore, this allows to separate thermodynamics as
meta-theory in the information-theoretic approach. To achieve
this, take the modular time of the state, perform the Wick
rotation, and call the result temperature. If we now change the
temperature independently of the modular time, we shall thus have
added a new degree of freedom with respect to the
information-theoretic approach. Evidently, this degree of freedom
may not come from within the approach; so it must be
meta-theoretic and related to the notions that were merely
postulated in the information-theoretic approach. Such notions are
information and fact, but also relevance. This is how, at least at
the conceptual level, one explains the origin of the link between
information and thermodynamics.

\item Assume the information-theoretic interpretation of the local
algebra theory in which Axioms~\ref{ax1} and \ref{ax2} justify why
the $C^*$-algebra of the system is hyperfinite. Then, if no new
information is brought about, and if the algebra is a type $III_1$
factor, the spectrum of $t$ is from $0$ to $+\infty$. It is a
satisfactory result that the internal, state-dependent time
behaves as one would think the time must behave: it is a real
positive one-dimensional parameter.
\end{itemize}

Time is a state-dependent notion but one would wish to have also a
state-in\-de\-pen\-dent time. Why would one wish that? Because we
are accustomed to the linear time that does not depend on the
information state. The word ``accustomed'' translates as a
requirement to obtain Newtonian time in the limit. Now, to obtain
this state-independent notion we factorize by inner automorphisms
and pick up the whole class of these that will correspond to one
outer automorphism. What have we done in information-theoretic
terms? To each modular automorphism corresponds a state that
defines it; by factorizing over modular automorphisms we neglect
the difference between these states and therefore neglect the
differences in information that we have in these different states.
Thus state-independent time becomes an issue of rendering some
information irrelevant.

We have said that time is ignorance. In fact, the word
``ignorance'' is perhaps not the best pick; the problem is that
ignorance has a strong flavor of being able to, but not knowing
something. In fact, there is no ``being able to.'' The state as
information state is given from meta-theory, and there is nothing
inside the theory that tells one how to pass from one state to
another (i.e. the measurement problem is not solved, but
\textit{dissolved}, see Section~\ref{dissom}). So if we ``were
able'' to know more, that would have defined another state over
the algebra and another state-dependent time, which is not the
case.

To formulate the main idea even shorter, let us come back to
Bohr's words in the epigraph: ``The concepts of space and time by
their very nature acquire a meaning only because of the
possibility of neglecting the interactions with the means of
measurement.'' We explained that if we functionally separate the
observer into meta-theoretical informational agent I and physical
system P, we are then able to define facts as answers to yes-no
questions posed by I to P and, in the course of interaction of P
with a physical system S, by chasing P out of the formalism, these
yes-no questions translate into POV measurement of I on S.
P-observer is the ancilla. So we see that POV measurements emerge
as an act of neglecting that the observer is a physical system. By
themselves, POV measurements are just positive operators that span
a $C^*$-algebra, and, as we said, a $C^*$-algebra corresponds to
the notion of system. Consequently, to determine the system, i.e.
a $C^*$-algebra, one must ``put oneself'' on the metalevel with
respect to that system by leaving the informational agent and
factoring out P-observer.

Now, each von Neumann algebra has a unique state-independent time.
Put the two together: by ``neglecting the interactions with the
means of measurement'' (Bohr) and therefore getting rid of
P-observer in the formalism, we define the algebra and its
state-independent time. This is how time acquires a meaning
exactly as Bohr wanted it.

As Einstein said, ``time and space are modes by which we think and
not conditions in which we live'' \cite{einfor}. Let us rephrase
Einstein and reconciliate him with Bohr: time and space are the
modes by which information is operated with and are not the
unjustified postulates in the information-based physical theory.

\part{Conclusion}\label{part4}

\chapter{Summary of information-theoretic approach}

\section{Results}

John von Neumann was a great, and the only, scientist of the first
70 years of the XXth century who made major contributions to both
quantum theory and the theory of information, and in quantum
theory he contributed to both quantum logic and algebraic quantum
theory. Although von Neumann's interest dates back to late 1920s,
it was in 1940s that he and his collaborators, taking inspiration
from physical sciences, taught their colleagues in biology,
psychology, and social science to speak the language of
information. The new language proved so successful that over time
it became possible to take it back to physics and to teach physics
itself a new language. Furthermore, time has been ripe since 1970s
for the world-picture as a whole, i.e. the philosophy of the human
theoretical inquiry into nature, to be built around the notion of
information.

The new world-picture is not akin to many its predecessors. The
attempts proved futile to reduce the full enterprise of
theoretical inquiry to relying upon information as the first
notion. Such a reductionist point of view cannot be defended
because of its circularity. Here, the futility and the circularity
are due to the fact that information, too, can be taken as object
of study, but this in a separate theory, which, obviously, will no
more be able to have information as the first notion. The
theories, then, are mutually connected by what they choose as
their basis and as their object of study, and there exists no set
of primary concepts common to each and every theory. Such a
situation amounts to a picture of the theoretical inquiry as a
loop of existences. Consistent exposition of the epistemological
attitude of the loop of existences, with its consequences for
distinguishing theory from meta-theory, is the \textbf{first
highlight} of this dissertation.

Theories have flourished since 1940s studying information by the
means and tools of physics. To give just one result, computers are
the greatest achievement of this current of human thought. Areas
like artificial intelligence strive to demystify operations with
information, its storage and communication, and cognitive science
aims at giving a theory of mind. On the other part of the loop,
information itself has been put in the very foundation of physics,
and so since the appearance of the science of quantum information
in 1980s. Questions have been raised: Can physics be derived from
information-theoretic postulates? What are these postulates? What
other assumptions must be added to them? As the \textbf{second
highlight} of this dissertation, we have given one possible answer
for a part of physics which is the quantum theory.

Two key axioms: that the amount of relevant information is finite
and that it is always possible to acquire new information, suffice
to grasp the essence of the quantum-theoretic structure.
Mathematically, they need to be formulated in one of the
formalisms of quantum theory and properly adjusted to the needs of
this formalism; thus, being supplied with additional assumptions,
they give rise to the conventional quantum theory. By means of the
quantum logical formalism, we have shown how to achieve the goal
of derivation of the Hilbert space and other blocks of which
consists the formalism of quantum theory. Also, all along the
derivation we have studied the role that play the additional
assumptions and have compared our system of axioms with the
existing alternatives.

Reconstruction by means of the quantum logical formalism has not
met the need for an information-theoretic justification of the
notions of space and time. To give such a justification along the
lines of the algebraic formalism, we have first interpreted this
formalism in information-theoretic terms. As the \textbf{third
highlight} of the dissertation, this interpretation together with
the argument for non-fundamental role of time belong to a field
seldom ploughed of the conceptual analysis of the $C^*$-algebraic
formalism in the theory of local algebras.

The importance of the information-theoretic approach to quantum
theory must not be underestimated. Apart from being an integral
part of the world-picture that implies the loop of existences,
this approach allows to view quantum theory as \textit{a theory of
knowledge}, i.e. a particular epistemology. From the general
epistemology it differs in imposing two axiomatic constraints on
the kind of knowledge that will be studied: that the amount of
information must be finite and that it must always be possible to
acquire new information. While the first constraint appears
plausible even for the most general theory of knowledge, the
second one clearly distinguishes quantum theory as theory of
knowledge from, say, classical physics as theory of knowledge, for
which no such axiom can be formulated. Indeed, the significance of
Axiom~\ref{ax2} lies in non-Abelianness of the structure of
observables such as lattice or $C^*$-algebra. Let us repeat once
again: quantum theory is a theory of knowledge; it is not a theory
of micro-objects nor of the physical reality. Its two key axioms,
perhaps with a different set of supplementary axioms than that of
Chapter~\ref{chaptreconstr}, will allow to apply the essentially
quantum theoretic approach to areas of human theoretical inquiry
other than the theory of micro-objects. As one of the areas of
potential interest we cite the application of the quantum
mechanical ideas to cognitive psychology and
economics~\cite{zwirn}.

The importance of the information-theoretic approach to quantum
theory must not be overestimated. This approach responds to the
need of giving a sound foundation to quantum physics, but it does
not bring any added value to the way in which quantum theory is
applied in the daily work of an ordinary physicist.
Information-theoretic approach to the foundations of physics
belongs to the area of theory, as opposed to application, and even
to the philosophy of science, although its development was
inspired by the purportedly practical field of quantum
information. Thus the information-theoretic approach cannot, for
instance, help to make the world economy grow faster or poor
people live a happier life, at least in the short run. Like poetry
in W.H.~Auden's words, it makes nothing happen; but it creates a
new language for science and by doing so imposes on the human
thought a novel pattern.

\section{Open questions}

Many questions that are raised in the context of the
information-theoretic approach to reconstructing quantum theory
were left open in this dissertation. These questions are listed
below, and despite our effort the list is most probably
incomplete.
\begin{enumerate}
\item Although they install the structure of a complete lattice,
Axioms~\ref{axiii}, \ref{axiv} and \ref{axv} have not been given
an information-theoretic justification. One such justification
could be based on the capacities offered to human beings by their
language: namely, in the language any two questions can be
concatenated or united in a longer question by a conjunction. But
to reason so would mean to assume that I-observer is a human agent
possessing a language, something that we have tried to avoid in
Section~\ref{rovsect}. Even if to carry on with this assumption,
it will still be necessary to decide whether human language has
the complexification capacity \textit{de facto} or only \textit{in
abstracto}, especially when applied to very large or countably
infinite sets of questions, as requires Axiom~\ref{axv}.
Information-theoretic approach, in the choice of Axioms~\ref{ax1}
and \ref{ax2}, aims explicitly at eliminating all abstract
structure never to be exemplified. It would be a pity if the
justification of Axioms~\ref{axiii}, \ref{axiv} and \ref{axv} had
to be at odds with this aim.

\item Information-theoretic meaning of Axiom~\ref{contaxiom} is
unclear and so is the one of its replacement offered by the
Sol\`{e}r theorem~\ref{solth}. We discussed this question in
Section~\ref{solersect}.

\item The appeal to Gleason's theorem~\ref{gleasonth} is not
completely justified by Axiom~\ref{ax3} of intra-theoretic
non-contextuality. The condition of Gleason's theorem involves a
function $f$ but nothing is said about the origin and meaning of
this function. It is easy to see that to justify the appearance of
$f$ amounts to explaining the origin of probabilities in quantum
theory. Although the Born rule fulfils in part this task,
information-theoretic meaning of the function $f$ remains to be
uncovered.

\item A series of assumptions about time evolution were made in
Section~\ref{sect44}. Although we have said that these assumptions
cannot be properly justified on the information-theoretic grounds
without exploring the other cut of the loop, it remains to be seen
how, in this other cut of the loop (Figure~\ref{loop1}), emerge
these very assumptions. Partially this task has been carried out
by the demonstration of classical limit of the modular time
hypothesis by Connes and Rovelli.

\item We deliberately postulated the absence of superselection
rules in the Hilbert space and gave an argument for this choice of
ours (see pages~\pageref{super1} and \pageref{super2}). We are
however ready to acknowledge a decisive weakness of this argument:
in Hilbert spaces of the quantum theory as it is conventionally
used, superselection rules are usually present. One needs to find
a way out of this dilemma.

\item Section~\ref{nfrole} treats of the problem of time in
algebraic quantum theory, but only a few lines are consecrated to
the problem of space. More research is needed that will perhaps go
in the direction described on page~\pageref{spaceprob}.

\item Reaching out both to the conceptual foundations of the
information-theoretic approach laid in Part~\ref{part1} and to the
concrete mathematical problems described in Part~\ref{cstar}, the
question of justification of the link between thermodynamics and
quantum theory (or equivalently, of the Wick rotation) remains
unanswered. Indeed, it would be too ambitious to pretend to have
found an answer to this question. What is clear, though, is that
the answer may only come from a meta-theoretic analysis in which
the two theories concerned will be somehow intertwined in one
context. To close the chapter, we suggest as a joke that a
mathematical formalization of the loop of existences may play the
role of such context: indeed, the imaginary unit $i$ is encoded in
the equation of a circle, and, as we argued, thermodynamics and
quantum theory lie in different cuts of the circle which is the
loop of existences. So to connect them would mean to pass from one
part of the circle to another, i.e. make a rotation, and this
requires a reference to $i$. We are of course fully aware of the
non-scientific (as of today) character of this proposal but we end
with a proverb which goes, ``In every joke there is a grain of
truth.''
\end{enumerate}

\chapter{Other research directions}

\section{Physics and information in cognitive
science}\label{cognsc}

In this closing chapter of the Conclusion, we discuss questions
pertaining to other research directions that arise in the context
of the ideas explored in the dissertation. The first such question
concerns the theory that emerges if the loop of
Section~\ref{loopsect} is cut as on Figure~\ref{loop1}; this is to
say that we analyze a theory which is based on physics as datum
and has information for the object of inquiry, thus aiming at
giving a theoretic account of how to operate with, store,
represent, and communicate information. These areas fall into the
large domain of cognitive science, i.e. the scientific study of
mind. The Oxford English Dictionary defines the word
\textit{cognitive} as ``pertaining to the action or process of
knowing.'' In a science of information that is based on physics,
the concept of information is to be viewed as the means by which
biological or even social questions from the study of mind could
be reduced to problems of physics. This was Norbert Wiener's view
\cite[p.~114]{Dupuy}, and we start by explaining the philosophy
that underlies it.

\begin{figure}[htbp]
\begin{center}
\epsfysize=2.5in \epsfbox{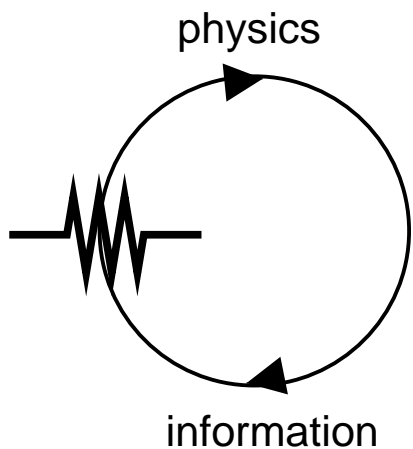} \caption{Connectionism: With
its roots in the first cybernetics, connectionism asserts that
objects have no symbolic value. Meaning and mind arise from
matter, and in the theory there is no intermediate level of
concepts between physics and
information.}\label{loop4}\end{center}
\end{figure}
Two main currents of thought in cognitive science are
connectionism and cognitivism. Connectionism (Figure~\ref{loop4}),
with its roots in the first cybernetics of Macy conferences,
asserts that meaning and mind are associated with matter because
they \textit{arise} from it. The matter in question is a neuronal
network in the brain, and thinking is an algorithm operating on
the neuronal machine. Meaning then has no essence, or rather its
essence is just its appearance. Neural network is a complex
system, and the mind is ``perfectly susceptible to a physicalist
approach provided that we rely upon the qualitative macrophysics
of complex systems and no longer upon the microphysics of
elementary systems''~\cite{natpheno}. No argument is however given
that would allow one to reject a particular physical theory, and
indeed in 1986 Roger Penrose, coming from a domain initially very
remote from cognitive science, that of quantum gravity,
proposed~\cite{penrose86} that consciousness, which is one of the
main objects of study in cognitive science, be seen as linked to
the deep microphysics, and this without abandoning complexity. The
contradistinction in views leaves open the question of which
physical theory in the physicalist doctrine must be taken as the
basis on which relies the theory of mind.

\begin{figure}[htbp]
\begin{center}
\epsfysize=2.5in \epsfbox{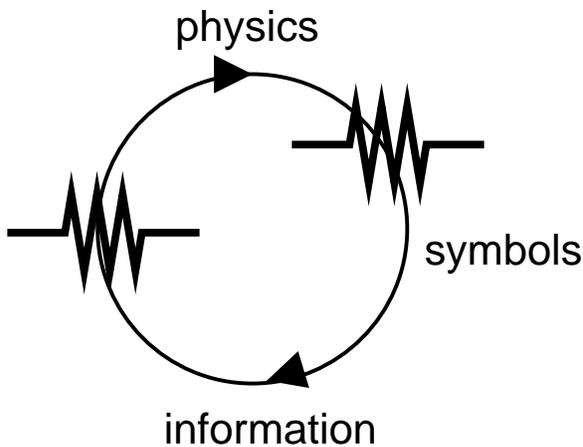} \caption{Cognitivism: What is
essential for the emergence of mind is not a concrete causal
structure but an abstract symbolic organization, which remains
invariant when one passes from one physical system to
another.}\label{loop5}\end{center}
\end{figure}

In our world-picture of Section~\ref{loopsect} connectionism and
its physicalist paradigm correspond to the loop cut so that the
theory of information is based on physics as datum. However,
besides the two configurations of Figures~\ref{loop01}
and~\ref{loop1} that only use one cut in the whole loop, one can
think of theories that arise in two or more loop cuts. One such
theory, and indeed a major current of thought in the philosophy of
cognitive science, is known under the name of \textit{cognitivism}
(Figure~\ref{loop5}).

\begin{figure}[htbp]
\begin{center}
\epsfysize=2.5in \epsfbox{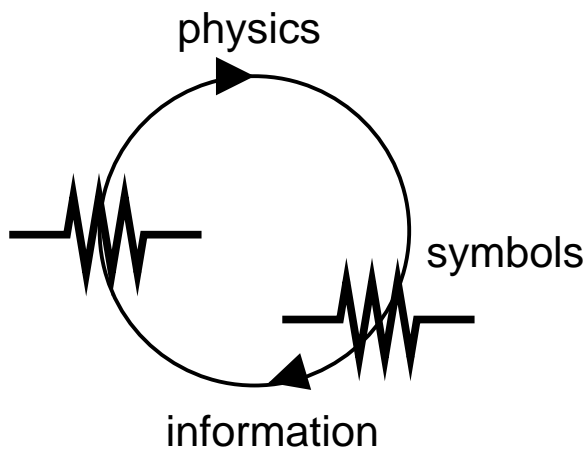} \caption{A cognitivist needs
a theory of how the symbolic level arises from
physics.}\label{loop6}\end{center}
\end{figure}

Cognitivism asserts that if the mind arises as a result of
implementing a certain algorithm, or a program, in the physical
world, then any implementation of the same program on a different
hardware, no matter what it may be, would produce a mind endowed
with the same properties. Therefore, what is essential for
emergence of the mind is not the concrete physical causal
organization of the material system possessing a mind; what is
essential is the abstract organization, which remains invariant
under the change of the material system. This abstract
organization is symbolic, meaning that the level on which it
operates is the level of symbols. On the cognitivist view, symbols
have three aspects: physical, syntactic and semantic. Syntactic
computations are rooted in the physical processes, but ``syntax by
itself is neither constitutive of nor sufficient for
semantics''~\cite{searle}. Thus a cognitivist theory of mind is
directly grounded in the symbolic and only indirectly in the
physical, in virtue of the fact that a theory of symbols,
physicalist in itself, requires a different loop cut
(Figure~\ref{loop6}) than the cognitivist cognitive science of
Figure~\ref{loop5}.

A theory that is urged on the cognitivist approach by the
necessity to consider, not only the loop cut of
Figure~\ref{loop5}, but also the one of Figure~\ref{loop6}, is a
\textit{grand oubli} of the proponents of cognitivism. They tend
to forget the second of the loop cuts altogether and focus their
research on the symbolic level as if it were the only fundamental
level; those cognitivists who call themselves physicalists are in
fact no more than scientists whose reflection went deep enough to
recognize the necessity of the second theory, but without ever
achieving practical results. The physics of cognitivists is a
physics of philosophers that is unconnected with the actual
physics of physicists. When a scientist seriously addresses the
need for a theory of which the schema is drawn on
Figure~\ref{loop6}, he is at once inclined to pass in the camp of
connectionists and to remove the second loop cut thereby obtaining
a theory of Figure~\ref{loop4}.

Let us now return to the choice of physical theory on which a
theory of mind may rely. We are going to give an argument showing
that if one adopts the connectionist view of Figure~\ref{loop4},
then the theory of consciousness cannot rely on classical physics,
although it still can rely on quantum physics. Two assumptions
that we make are as follows: \begin{itemize} \item Consciousness
is an object of theoretical inquiry, i.e. there exists a
\textit{theory} of consciousness. \item Assumption of strong
physicalism, i.e. every proposition of the theory of
con\-scious\-ness can be translated into a proposition of physical
theory, even though this latter proposition may be quite
complex.\end{itemize}

Both these assumptions are far from being consensual among
cognitive scientists and philosophers. Concerning the first one,
we deliberately abstain from discussing whether consciousness is a
phenomenon~\cite{nedblock,metz} and if it has a place in the loop
of existences. Perhaps it does not, and then consciousness is
purely epiphenomenal. For example, such is nowadays the case with
the notion of life, although some 150 years ago a rare scientist
would call life epiphenomenal. We simply assume that consciousness
is a legitimate object of theoretical inquiry.

Regarding the second assumption, its proponents are a few but
include such philosophers as John Searle, who asserts that all
mental phenomena must be reduced, at the last instance, to the
level of physical fields and fundamental
interactions~\cite{searle95,searle2001}. Although we do not
endorse Searle's \textit{ontological} physicalism and instead
propose the loop \textit{epistemology}, both lead to the
assumption of strong physicalism that we make in the sequel.

In order to find out which physical theory can serve as foundation
for the theory of consciousness, we follow a filtering procedure.
This procedure consists in taking a particular property of
consciousness that must be explained by the theory of
consciousness and checking which physical theories are capable of
giving an account of that property. In fact, we shall only be
concerned with one such property: self-referentiality. The
requirement of taking into account self-referentiality of
consciousness will lead to a situation when only some, and not
other, physical theories, which can be a foundation for the theory
of consciousness, will survive filtering. Filtering criteria,
including the one of self-referentiality, are non-constructive in
the sense that they allow to eliminate candidate theories but they
do not tell one how the theory of consciousness can be built using
physical theories that will have survived filtering.

We start by treating observation as a semantic concept. Generic
statement of a physical theory has the form, ``The state of the
system has such and such properties.'' Irrespectively of the
meaning of the term \textit{state} which, as we argued in
Section~\ref{rovsect}, must be relational, this generic form of
the physical statement permits, instead of speaking about the
validity statements of the theory, to speak about sets of states:
to every statement corresponds a set of states in which the
statement is valid. To verify a statement about the system means
to make an observation of the system and to check if the observed
state falls into the expected set of states. In this sense
observations contribute to set up semantics of the theory.

Largely avoiding some crucial philosophical aspects of the
discussion in Chalmers's illuminating book~\cite{chalm}, we assume
that ``I am aware that'' is a predicate of the theory of
consciousness. In light of the semantic role of observations, ``I
am aware that'' is at the same time an observation in the theory
of consciousness and a semantic statement belonging to the theory
of consciousness. For the reason of simplicity, in the following
argument we take the theory of consciousness to contain only the
predicate ``I am aware that.''

Let us now give several definitions. A theory is semantically
complete if and only if objects and processes that are necessary
for testing and interpreting the theory are themselves included
among the phenomena described by the theory~\cite[p.~4]{mittel2}.
Meta-theory of a given theory is a theory that contains predicates
about the predicates of the theory. Follows that if a theory is
semantically complete, then its meta-theory is a subset of the
theory.

A theoretical statement is self-referential if it refers to the
states of the system which, in their turn, refer to this very
statement (i.e. the set of states)~\cite{tarski,breuer}. In every
semantically complete theory one necessarily finds
self-referential statements. The converse does not hold: presence
of a self-referential statement in a theory does not make the
theory semantically complete.

\begin{figure}[htbp]
\begin{center}
\epsfysize=2.5in \epsfbox{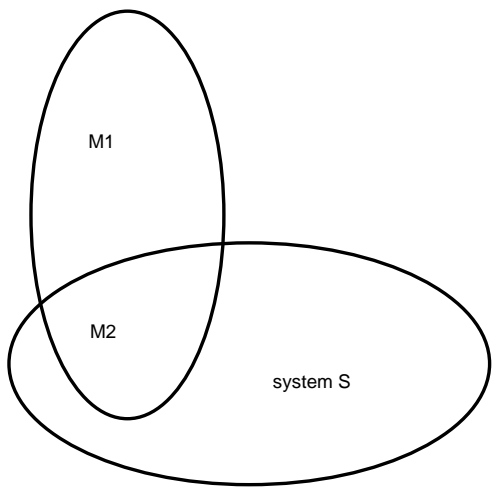}
\end{center}
\caption{Self-referential consistency: Observation of $S$ by
$M=M_1+M_2$ provides information about the state of $S$, including
certain information about $M_2$. This information must be
compatible with the fact that $M_2$ is a part of the measuring
apparatus.}\label{consi}
\end{figure}

The concept of self-reference leads to introducing the concept of
self-referential inconsistency (Figure~\ref{consi}). In
self-referential statements observation of the system (which is a
semantic proposition) is made from inside the system, and this
observation provides information not only about the system as
such, but also about the measuring apparatus which is a part of
the system. The latter information must be consistent with the
fact that this measurement apparatus is indeed a
\textit{measurement} apparatus: for instance, the information
obtained must not preclude the apparatus from existing.
Self-referential consistency is a necessary requirement for any
self-referential theory, because self-referential inconsistency
leads to logical paradoxes. From this we learn an important
lesson: If in a theory there are self-referential propositions
then one must impose the condition of self-referential
consistency.

Petersen writes, ``To define the phenomenon of consciousness, Bohr
used a phrase somewhat like this: a behaviour so complex that an
adequate account would require references to the organism's
self-awareness.''~\cite{aagebohr} Somewhat in the spirit of Bohr's
idea, we now show that self-referentiality of consciousness
implies self-referentiality of the theory of consciousness, which
in turn implies self-referentiality of the physical theory on
which relies the theory of consciousness.

``I am aware that I am aware'': this statement, viewed as a
linguistic statement about the state of the system, reports a
valid observation and thus belongs to meta-theory of the theory of
consciousness. On the other hand, ``I am aware that I am aware''
is a statement of the type ``I am aware that'' and is itself a
state of consciousness, so it belongs to the theory of
consciousness. Every act of observation in the theory of
consciousness, which we agreed to limit to ``I am aware that''
statements, is therefore self-referential.

\begin{figure}[htbp]
\begin{center}
\epsfysize=2.5in \epsfbox{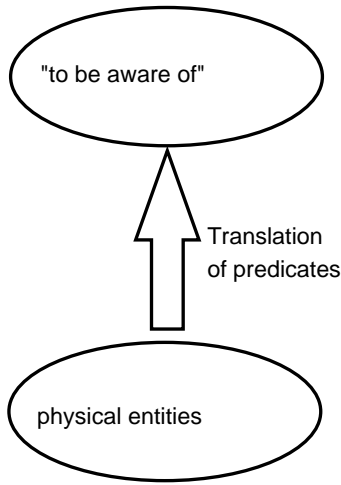}
\end{center}
\caption{Translation of theoretic predicates in virtue of the
assumption of strong physicalism.}\label{translpred}
\end{figure}

Let us now show that self-referentiality of the theory of
consciousness implies self-referentiality of the physical theory
that serves as a foundation to the theory of consciousness.
According to the assumption of strong physicalism, every predicate
of the theory of consciousness can be translated into a predicate
of the physical theory (Figure~\ref{translpred}). Consider the
predicate ``I am aware that I am aware.'' Put in the place of each
of the two clauses ``I am aware'' its physical counterpart. We
obtain a predicate of the physical theory which at the same time
belongs to the theory and to meta-theory. This proof works if
translation of the predicate ``I am aware'' into the language of
physics does not depend on the content of the referring part of
the predicate: evidently, referents of the two clauses ``I am
aware'' are different, and their translations may therefore
differ.

Consider now the opposite: namely, predicate translation depends
on the referent. Translation is possible for any referent, so let
us take as referent an arbitrary semantic statement of the form
``such and such properties are true,'' which belongs to the
meta-theory of the physical theory. Add to it ``I am aware that;''
appears a statement that belongs to the theory of consciousness.
Now translate this statement into the language of physical theory
in virtue of the assumption of strong physicalism. Starting with a
meta-theoretical physical statement, we have thus obtained a
statement of the physical theory itself. This confirms that
physical theory on which relies the theory of consciousness is
self-referential.

In short, what this procedure allows to achieve can be called ``a
new g\"{o}delization'' fully analogous to the original idea of
G\"{o}del's: ``The language of the formal system used by G\"{o}del
\ldots does not contain any expressions referring explicitly to
meta-theoretical concepts. But after assigning numbers to the
propositions, these numbers can be interpreted as expressions of
the language referring to its own propositions.''~\cite{breuer}
Instead of assigning to every proposition a number, as did
G\"{o}del, we add to it a clause ``I am aware that'' that allows
to put in correspondence with each semantic statement over the
physical theory (i.e. observation) a physical state.

Having established that the physical theory must be
self-referential, we would like to use this result to complete the
filtering procedure. For this, we return to the notion of
self-referential inconsistency and show that classical physics
viewed as self-referential theory is inconsistent.

Key intuition comes from Einstein's words that measuring
instruments which we use to interpret theoretical expressions must
be really existing physical objects. Skip the word ``really'' and
focus on the word ``existing'': this will lead to the check by
self-referential consistency. In a theory of consciousness,
measuring instrument is the human brain. If the theory runs into a
contradiction when the brain elements are considered as measuring
instruments, then the theory is inconsistent. One sort of such
brain elements are hydrogen atoms. Consider a human observer $O$
who observes hydrogen atoms in his own brain and assume that the
theory of consciousness relies on classical physics. Result of
this observation can be represented as ``I am aware that hydrogen
atoms in my brain have property $P$ predicted by classical
physics.'' This observation, according to the new g\"{o}delization
procedure, is itself a predicate of classical physics. Now,
because predictions of classical physics about hydrogen atoms do
not allow the existence of hydrogen atoms, being projected within
the domain of classical physical on $M_2$ of Figure~\ref{consi},
they prevent the very existence of observer $O$. Consequently,
classical physics is self-referentially inconsistent. It cannot
serve as a foundation for the theory of consciousness.

As for quantum theory as basis of the theory of consciousness, it
passes filtering by the criterion of self-referentiality:
Mittelstaedt~\cite{mittel2} in the discussion of the
objectification postulate gives a classification of situations
where quantum theory might appear to be self-referentially
inconsistent and then, based on Breuer's result~\cite{breuer},
proves the impossibility of such situations. This, however, does
not guarantee that there exist no other reasons why the theory of
consciousness may not rely on quantum physics as an underlying
physical theory. So if for classical physics this question is
settled in the negative, for quantum physics it remains open to
future investigation.

\section{Two temporalities in decision theory}

We have seen in Section~\ref{nfrole} that in the algebraic quantum
theory interpreted in information-theoretic terms there arise two
temporalities: \begin{description}\item[(a)] a state-dependent
notion of time which is characterized by the I-observer's
information state, and \item[(b)]a state-independent notion of
time which is obtained by neglecting certain information and
therefore factoring over whole classes of state-dependent
temporalities.\end{description} It is the second,
state-independent time that indexes facts as acts of bringing
about information. If for the first, state-dependent time one can
say that its range of values, in the hyperfinite type $III_1$ von
Neumann algebra, covers all positive real numbers, nothing at this
level of precision can be said about the state-independent time.
So there is no obvious reason why one would think that the
state-independent time is ``linear'' in the usual sense and covers
all $\mathbb{R}_+$. Still, it is this very time that the
informational agent perceives as indexing facts in which
information is brought about.

\begin{figure}[htbp]
\begin{center}
\epsfysize=2.5in \epsfbox{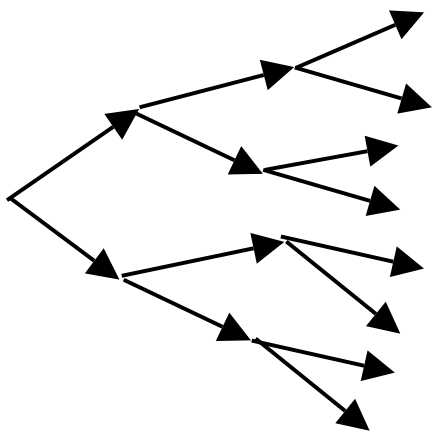}
\end{center}
\caption{Occurring time.}\label{occtime}
\end{figure}

A similar situation arises in decision
theory~\cite{dupuy2,dupuy3,dupuy4}. The familiar commonsense
temporality is encoded in a decision tree which we call
\textit{occurring time} (Figure~\ref{occtime}). Occurring time is
the linear time that embodies the commonsense understanding that
the future is open and the past is fixed. The agent has no causal
power over the past, but also no counterfactual power; on the
contrary, with regard to the future the agent has both causal and
counterfactual power. Decision theory employing this temporality
leads to many paradoxes, i.e. such cases where action prescribed
by the theory as the rational choice seems to be completely
bizarre and is practically never chosen by the real human decision
makers. Such paradoxes arise in a variety of settings, from simple
Take-or-Leave games to the nuclear deterrence problem and the
Newcomb paradox.

\begin{figure}[htbp]
\begin{center}
\epsfysize=2.5in \epsfbox{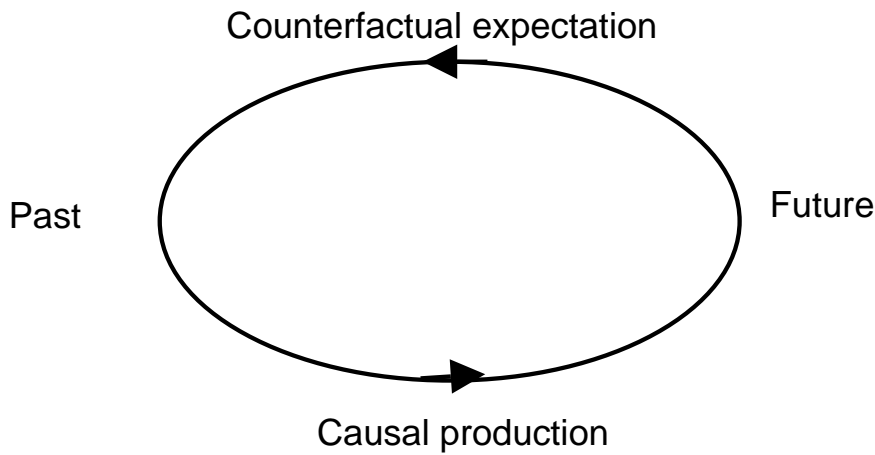}
\end{center}
\caption{Projected time.}\label{projtime}
\end{figure}

To avoid the paradoxes of decision theory in the occurring time,
Dupuy proposed a different temporality that he called
\textit{projected time}. Projected time is the time in which
reasoning of the agent takes place, and it is very different from
the linear occurring time: in fact, it takes the form of the loop
(Figure~\ref{projtime}). In the projected time future has
counterfactual power over the past, while the only causal power
is, as before, the power of the past over the future. To find a
decision-theoretic equilibrium in projected time, it is necessary
to seek a fixed point of the loop, where an expectation (on the
part of the past with regard to the future) and causal production
(of the future by the past) coincide. The agent, knowing that his
prediction is going to produce causal effects in the world, must
take account of this fact if he wants the future to confirm what
he has foretold.

Circular temporality of the projected time gives rise to a full
new decision theory drastically different from the old decision
theory that made use of the occurring time. Indeed,
\textit{decision} belongs in the kind of temporality in which
\textit{reasoning} is done, and this temporality is the one of the
circular projected time. Linear time, so to say, ceased to be the
interesting time. Projected time, which is not linear, raised to
become an upfront temporal decision-theoretic notion.

Whether there are or there are not good grounds to claim a
parallel between the two temporalities in the information-based
physical theory and in the decision theory, as of now we are not
yet ready to say. It is certainly tempting to seek an analogy
between the two: in the information-theoretic approach one speaks
about the temporality of facts being externally given to the
physical theory in the loop cut of Figure~\ref{loop01}, and this
is not far from the temporality of reasoning in the
decision-theoretic context. After all, facts are acts of bringing
about information, and reasoning is just the analysis of
information. So does the non-necessarily linear state-independent
notion of time have anything to do with the circular (i.e.
non-linear) temporality of projected time? To answer in the
affirmative would amount to an ambitious hypothesis that we can
only leave as subject to a future investigation.

\section{Philosophy and information technology}

As we repeatedly said in this dissertation, foundations of the
modern theory of information were laid out by von Neumann and
other scientists whose work initially belonged in the theoretical,
rather than applied, science. But these very people were also
among the pioneers of the construction of computers and what was
later called the field of information technology. Nowadays
information technology is a vast domain causing public excitement
and fascination and in which are employed thousands of
professionals most of whom have never given any attention to the
problems that interested the founding fathers of their discipline.
A software engineer does not need to think about thermodynamics
and its link with information. Chip maker does not need to worry
about advanced programming languages or web browsers that will be
run on computers using his chips. As many others, the field of
information technology is divided into numerous cells to each of
which are assigned hundreds of narrow specialists. Such is also
the situation in physics since 1970s, and today this situation
seems to be slowly changing: Queen Philosophy is coming back to
her kingdom of physics. Will information technology sooner or
later undergo a similar return to the fundamental questions?
Probably yes.

One prospective direction that information technology may take if
it decides to look back at the notions that lie in its foundation
is the route shown by Clifton, Bub and Halvorson, whose results we
discussed in Section~\ref{cbhsect}. Quantum information developed
powerful and beautiful theorems that are now used to serve as
foundation of the physical theory itself. Metaphorically, the
situation is like the one when a man for the first time looks in
the binoculars in the wrong direction: before this man used to
believe uncritically that the road is one-way only and that it
leads from quantum physics to quantum information, until one day,
out of curiosity, he looked in the binoculars from the wrong end,
and the view of the world has changed. It will never be the same:
we now know that quantum theory can be viewed as based on
information. Will information technology take the challenge to
produce for the world a new philosophy based on its values and its
fundamental notions? Will information technology, with the
development of the field of quantum information, install a clear
demarkation line between the superfluous ontological and the
efficient epistemological arguments? We are still living in the
days when articles by important information scientists speak about
``ontic states''~\cite{spekknctx}. Perhaps it is with the future
return of the interest toward its own fundamental concepts that
information technology will consistently and insistingly teach
other disciplines the language of information.

\singlespacing

\bibliographystyle{myplain}
\newpage\addcontentsline{toc}{chapter}{Bibliography}
\bibliography{inf}
\cleardoublepage
\end{document}